\documentclass[12pt]{report}
\usepackage[online]{suthesis-2e}
\usepackage{amsmath,amsfonts,bbm,subfigure,amsthm,dsfont}
\usepackage{algorithm,algorithmic,enumitem,comment,array}

\usepackage{tikz}
\usetikzlibrary{arrows}
\usepackage[colorlinks, linkcolor=blue, citecolor=blue, urlcolor=magenta, linktocpage, plainpages=false, pdftex]{hyperref}
\usepackage[htt]{hyphenat} % Allow linebreaks in texttt mode

\newtheorem{corollary}{Corollary}

\newtheorem{lemma}{Lemma}

\newtheorem{theorem}{Theorem}

\def\EE{{\mathbb{E}}}
\def\PP{{\mathbb{P}}}

\newcommand{\fall}{\,\forall\,}

\newcommand{\R}{{\mathbb R}}

\newcommand{\PR}{\pi}

\def\vecE{\mathbf{\underline{e}}}
\def\p{\mathbf{p}}
\def\r{\mathbf{r}}
\def\vecS{s}
\def\S{V}
\newcommand{\dbar}{\overline{d}}

\newcommand{\inlineheading}[1]{\textbf{#1}}
\newcommand{\outneighbors}[1]{\mathcal{N}^{out}(#1)}
\newcommand{\inneighbors}[1]{\mathcal{N}^{in}(#1)}
\newcommand{\outdegree}[1]{d^{out}(#1)}
\newcommand{\indegree}[1]{d^{in}(#1)}
\newcommand{\epr}{r_{\text{max}}}
\newcommand{\rmax}{\epr}

\newcommand{\abs}[1]{\left \lvert #1 \right \rvert}
\newcommand{\size}[1]{\left \lvert #1 \right \rvert}
\newcommand{\pn}[1]{\left ( #1 \right )}
\newcommand{\bk}[1]{\left [ #1 \right ]}
\newcommand{\pfail}{p_{\text{fail}}}
\newcommand{\norm}[1]{\left\lVert #1 \right\rVert}

\def\indicator{{\mathbbm{1}}}
\newcommand{\pluseq}{\mathrel{+}=}
\newcommand{\nsample}{n_{s}}
\newcommand{\pEst}{p}

\newcommand{\rpush}{\texttt{ReversePush}}
\newcommand{\fpush}{\texttt{ForwardPush}}
\newcommand{\bippr}{\texttt{BidirectionalPPR}}
\newcommand{\ubippr}{\texttt{UndirectedBiPPR}}
\newcommand{\mc}{\texttt{Monte Carlo}}

\newcommand{\bipprgrouped}{\texttt{BiPPR-Precomp-Grouped}}
\newcommand{\bipprsampling}{\texttt{BiPPR-Precomp-Sampling}}
\newcommand{\bipprprecomp}{\texttt{BiPPR-Precomp}}
\newcommand{\bipprbasic}{\texttt{BiPPR-Basic}}

\newcommand{\precompSamplers}{\texttt{PrecomputePathSamplers}}
\newcommand{\samplePath}{\texttt{SamplePathToTarget}}

\newcommand{\dmax}{d_{\text{max}}}
\newcommand{\rmaxr}{r_{\text{max}}^{\text{r}}}
\newcommand{\rmaxf}{r_{\text{max}}^{\text{f}}}
\newcommand{\nwalk}{n_{\text{w}}}
\newcommand{\nshard}{n_s}

\title{Efficient Algorithms for Personalized PageRank}
\author{Peter Lofgren}
\principaladviser{Ashish Goel}
\firstreader{Hector Garcia-Molina}
\secondreader{Jure Leskovec}

\begin{document}

% first the preface sections.  

% this includes the file preface.tex which should include the
% following commands
% \beforepreface
% \prefacesection{preface}
% body of the preface
\beforepreface
\prefacesection{Abstract}
We present new, more efficient algorithms for estimating random walk scores such as Personalized PageRank from a given source node to one or several target nodes.  These scores are useful for personalized search and recommendations on networks including social networks, user-item networks, and the web. Past work has proposed using Monte Carlo or using linear algebra to estimate scores from a single source to every target, making them inefficient for a single pair.  Our contribution is a new bidirectional algorithm which combines linear algebra and Monte Carlo to achieve significant speed improvements.  On a diverse set of six graphs, our algorithm is 70x faster than past state-of-the-art algorithms.  We also present theoretical analysis: while past algorithms require $\Omega(n)$ time to estimate a random walk score of typical size $\frac{1}{n}$ on an $n$-node graph to a given constant accuracy, our algorithm requires only $O(\sqrt{m})$ expected time for an average target, where $m$ is the number of edges, and is provably accurate.  %Our algorithm can also estimate the global PageRank of a single node in $O(\sqrt{m})$ time, 

In addition to our core bidirectional estimator for personalized PageRank, we present an alternative algorithm for undirected graphs, a generalization to arbitrary walk lengths and Markov Chains, an algorithm for personalized search ranking, and an algorithm for sampling random paths from a given source to a given set of targets.  We expect our bidirectional methods can be extended in other ways and will be useful subroutines in other graph analysis problems.

\newpage

% any other preface sections

% the last preface section (e.g., acknowledgement.tex)
% should look like
% \prefacesection{Acknowledgement}
% body 
% \afterpreface
\prefacesection{Acknowledgements}

I would like to acknowledge the help of the many people who helped me reach this place.  First I thank my PhD advisor Ashish Goel for his help, both in technical advice and in mentoring.  Ashish recognized that PPR was an important problem where progress could be made, helped me get started by suggesting sub-problems, and gave excellent advice at every step.  I'm always amazed by the breadth of his understanding, from theoretical Chernoff bounds to systems-level caching to the business applications of recommender systems.  He also helped me find excellent internships,  was very gracious when I took a quarter off to try working at a start-up, and generously hosted parties at his house for his students.  

I also thank my other reading committee members, Hector and Jure.  Hector gave great advice on two crowd-sourcing papers (independent of my thesis) and is remarkably nice, making the InfoLab a positive community and hosting frequent movie nights at his house.  Jure taught me much when I CAed his data mining and social networking courses and has amazing energy and enthusiasm.

I thank postdoc (now Cornell professor) Sid Banerjee for amazing collaboration in the last two years.  Sid brought my research to a completely new level, and everything in this thesis was co-developed with Sid and Ashish.  I thank my other collaborators during grad school for helping me learn to do research and write papers: Pankaj Gupta, C.~Seshadhri, Vasilis Verroios, Qi He, Jaewon Yang, Mukund Sundararajan, and Steven Whang.   I thank Dominic Hughes for his excellent advice and mentorship  when I interned at his recommendation algorithms start-up and in the years after.  

Going back further, I thank my undergraduate professors and mentors, including Nick Hopper, Gopalan Nadathur, Toni Bluher, Jon Weissman, and Paul Garrett.  Going back even further, I thank all the teachers who encouraged me in classes and math or computer science competitions, including Fred Almer, Brenda Kellen, and Brenda Leier.  

I thank my family for raising me to value learning and for always loving me unconditionally.  I thank all my friends (you know who you are) who made grad school fun through hiking, camping, playing board games, acting, dancing, discussing politics, discussing grad school life, and everything else.  I look forward to many more years enjoying life with you.  

Finally I thank everyone I forgot to list here who also gave me their time, advice, help, or kindness.  It takes a village to raise a PhD student.

%% afterpreface produces a table of contents and any other tables
%% wanted. At the end pagenumbering changes from roman to arabic and
%% is restarted
\afterpreface

\chapter{Introduction}
\section{Motivation from Personalized Search}
As the amount of information available to us grows exponentially, we need better ways of finding the information that is relevant to us.  For general information queries like ``What is the population of India?'' the same results are relevant to any user.  However, for context-specific queries like ``Help me connect with the person named Adam I met at the party yesterday,'' or ``Where is a good place to eat lunch?'' giving different results to different users (personalization) is essential.  There are many approaches to personalization, including using the user's search history, their attributes like location, or their social network, which we discuss in Section \ref{sec:other_personalization}.  In this thesis we focus on one approach to personalization, using random walks as described in Sections \ref{sec:def_ppr} and \ref{sec:other_scores}, which has proven useful in a variety of applications listed in Section \ref{sec:other_applications}.  We present new, dramatically more efficient algorithms for computing random walk scores, %including the Heat Kernel and personalized SALSA (see Section \ref{sec:other_scores}), 
and for concreteness we focus on computing the most well-known random walk score, Personalized PageRank.

%Personalization is important for many web applications.  For example, if a user searches for a common name like ``Adam'' on an interest-based social network like Twitter, the results should reflect the interests of the searching user.  A user interested in musicians should see different results than a user interested in technology.  Such personalization is even more critical on a friendship-based social network like Facebook.  If a user searches for ``Adam'' on Facebook, the user expects results based on their personal friendships, and the search must be personalized.  

 Personalized PageRank (PPR) \cite{Page1999} is the personalized version of the PageRank algorithm which was important to Google's initial success.  On any graph, given a starting node $s$ whose point of view we take, Personalized PageRank assigns a score to every node $t$ of the graph. This score models how much the user $s$ is in interested in $t$, or how much $s$ trusts $t$.  More generally we can personalize to a distribution over starting nodes, for example in web search we can create a distribution with equal probability mass on the web pages the searching user has bookmarked.  If we personalize to the uniform distribution over all nodes, the score is no longer personalized, and we recover the standard (global) PageRank score.  

\begin{figure}
  \centering
\subfigure[Ranked results on twitter.com in 2014 for query ``Adam'' from a logged-in user who followed people in technology and politics.  Notice that results appear un-personalized.]{
  \fbox{\includegraphics[height=2.5in]{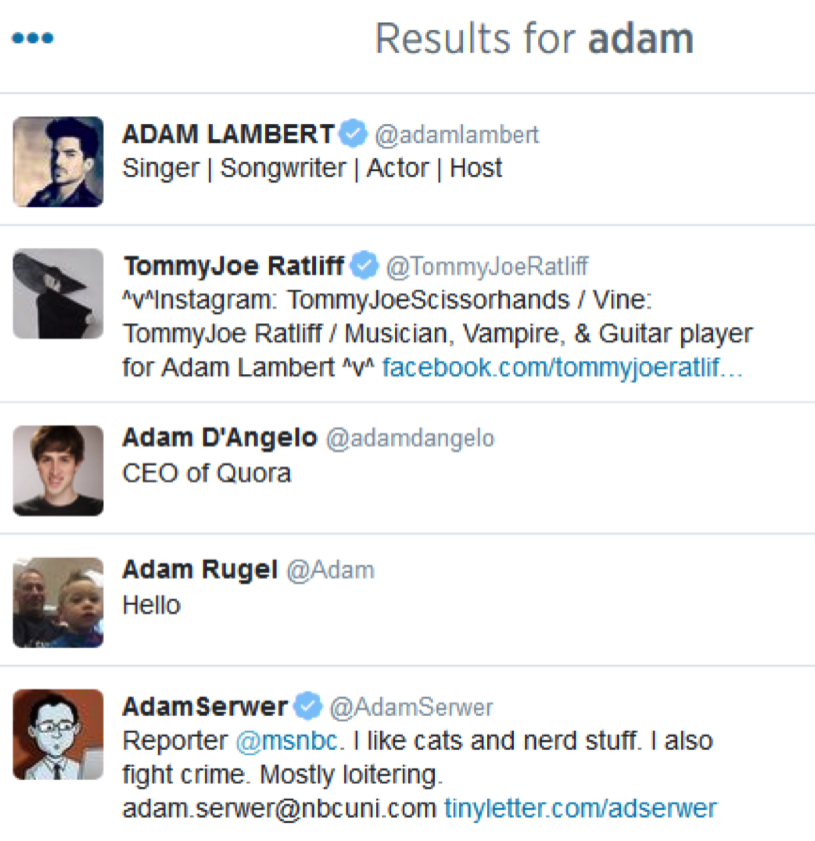}}
  \label{fig:twitter_adam_search}
}
\hfill
\subfigure[Results for the same query, re-ranked by Personalized PageRank from the searching user's account.  Notice that the top results are now in technology, reflecting the user's personalized interests.]{
\footnotesize
  \begin{tabular}[b]{|l|m{1.4in}|}
\hline
    Name & Description \\
\hline
Adam Messinger & CTO \@twitter \\
\hline
Adam D'Angelo & CEO of Quora \\
\hline
Adam Satariano & Technology Reporter, Bloomberg News \\
\hline
Adam Steltzner & Rocket scientist, intermittent gardener, master of mars, and dangerous dinner guest.  Co-founder of Adam and Trisha's dog and baby farm. \\
\hline
Adam Rugel & Hello \\
\hline
  \end{tabular}
  \label{fig:ppr_adam_search}
}
\caption{A comparison of PPR with a production search algorithm on Twitter in 2014.}
\end{figure}

As an example of how PPR can personalize search results, in Figure \ref{fig:twitter_adam_search} 
we show a search for ``Adam'' done on Twitter in 2014 by the author, who was signed in to an account which followed people in technology and politics.  However, the top result is a singer-songwriter, and the second result is a guitar player--results with much global popularity, but which are not personalized based on who the searcher follows.  If we re-rank these results using Personalized PageRank from the same account, we get the results in Figure \ref{fig:ppr_adam_search}.  Now the top two results are both in technology, and the results are more relevant overall.

\begin{figure}
  \centering
  \includegraphics[width=0.9\columnwidth]{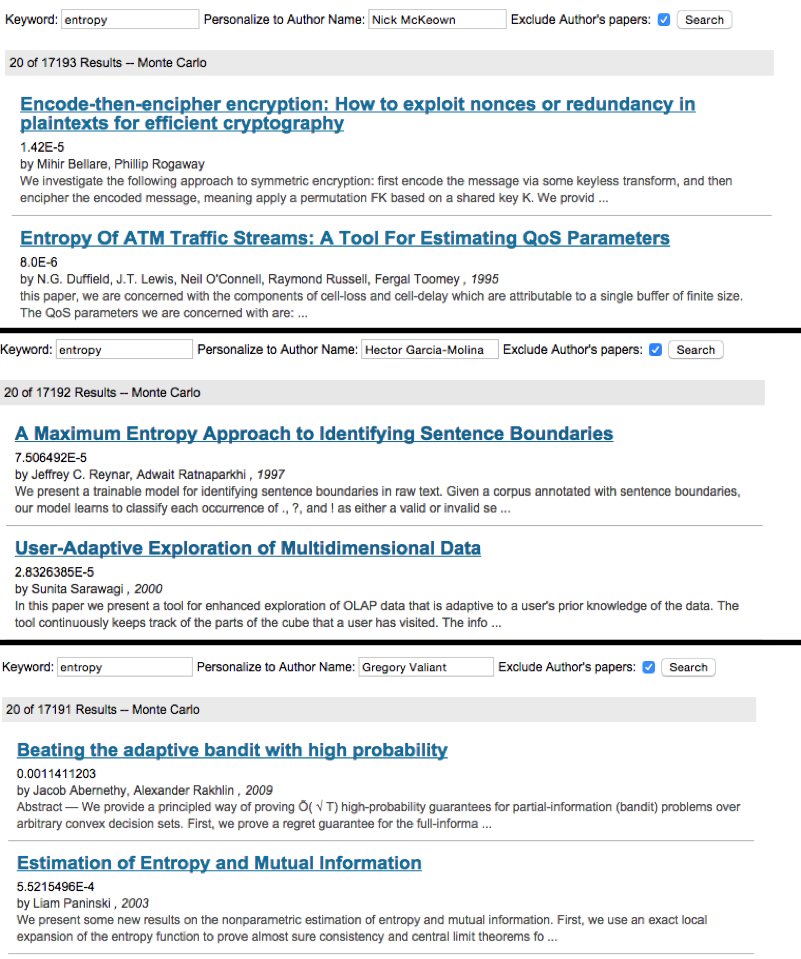}
  \caption[Academic Search Example]{The top two computer science papers containing the multi-faceted keyword ``entropy'' personalized to three different researchers' point of view: a networking researcher, a database/data-mining researcher, and a learning-theory researcher.  Notice how the ranking is very different for different researchers, and PPR on the citation graph captures the interests of each researcher.}
  \label{fig:entropy_search}
\end{figure}

As an example of how changing the source $s$ of the PPR algorithm results in different rankings, we consider personalized search on a citation graph.  On a citation graph provided by Cite-seer, we created a demo which given a keyword and researcher name, finds all papers containing that keyword, and ranks them from the point of view of the given researcher.  For a given researcher, we find all papers by that researcher, and define a source distribution $s$ giving equal weight to those papers. We then use Personalized PageRank from $s$ on the citation graph to rank papers matching the given keyword.  As an example, the keyword ``entropy'' means different things to different researchers, so we compare the top results for keyword ``entropy'' from different points of view.  In Figure \ref{fig:entropy_search}
we show how the search results change when personalized for different researchers: a networking researcher (McKeown), a database/data-mining researcher (Garcia-Molina), and a learning-theory researcher (Valiant).  These different rankings demonstrate that by changing the source distribution, we can use Personalized PageRank to understand the importance of nodes in a graph from different points of view.

\section{Other Applications}
\label{sec:other_applications}
The importance of Personalized PageRank extends beyond search on social networks.  PPR has found use in many other domains, including friend recommendation on Facebook \cite{backstrom2011supervised}, who to follow on Twitter \cite{gupta2013wtf}, graph partitioning \cite{Andersen2006},  community detection \cite{yang2012defining}, and other applications \cite{tong2006fast}.  On the web graph there is significant work on using Personalized PageRank to rank web pages (e.g. \cite{Jeh2003,haveliwala2002topic}).  PPR is also useful on heterogeneous graphs; for example it can be used to rank items in a
bipartite user-item graph, in which there is an edge from a user to
an item if the user has liked that item.  Random walks on a user-item graph have proven useful for YouTube when recommending videos \cite{baluja2008video}.  Gleich's survey \cite{Gleich-preprint-pagerank-beyond} lists a variety of applications of PageRank beyond the web, in fields from biology to chemistry to civil engineering.

%Social search protocols find widespread use -- from personalization of general web searches \cite{Page1999,Jeh2003,yin2010}, to more specific applications like collaborative tagging networks \cite{yahia2008}, ranking name search results on social networks \cite{vieira2007}, social Q\&A sites \cite{horowitz2010}, etc.

\section{Preliminaries}
We work on a directed graph $G=(V, E)$ with $n = \size{V}$ nodes and $m = \size{E}$ edges.  Our algorithms all generalize naturally to weighted graphs with weight matrix $W$ having entry $w_{u, v} \geq 0$ for each edge $(u, v) \in E$, although we sometimes present algorithms and proofs for the unweighted case to simplify the exposition.  For simplicity we assume the weights are normalized such that for all $u$, $\sum_v w_{u,v} = 1$.  

Define the out-neighbors of a node $u$ by $\outneighbors{u}=\{v: (u, v) \in E\}$ and let $\outdegree{u} = \size{\outneighbors{u}}$; define $\inneighbors{u}$ and $\indegree{u}$ similarly.  For example, in the graph in Figure \ref{fig:ppr_example}, we have $\outneighbors{c}=\{v, x, w\}$ and $\inneighbors{c}=\{s, x, w\}$. Define the average degree of nodes $\bar{d} = \frac{m}{n}$.

For a vector $x \in \R^n$, we denote the $i$th entry as $x[i] \in \R$.  

As alluded to before, we denote the starting node or distribution by $s$.  We overload notation to denote either a single node $s \in V$ or a distribution $s \in \R^n$ assigning weight $s[v] \geq 0$ to node $v \in V$.  We can interpret a node $s \in V$  as a distribution giving probability mass 1 to $s$ and 0 to other nodes.  Our algorithms generalize naturally to a start distribution, but for exposition we usually assume $s$ is a single node in $V$.

\section{Defining Personalized PageRank}
\label{sec:def_ppr}
Personalized PageRank has two equivalent definitions.  We define both here because they are both useful way of thinking about PPR.  In the first definition, Personalized PageRank recursively models the importance of nodes on a directed graph.  At a high level, given a start node $s$ whose point of view we take, we say that $s$ is important, and in addition we say a node is important if its in-neighbors are important.  To keep the importance scores bounded, we normalize the importance given from a node $u$ to a node $v$ though an edge $(u, v)$ by dividing by $u$'s out-degree.  In addition, we choose a decay term $\alpha$, and transfer a fraction $1-\alpha$ of each node $u$'s importance to $u$'s out-neighbors.  Formally, given normalized edge weight matrix $W$ (as defined earlier, entry $w_{u, v}$ is the weight of edge $(u, v)$ and $\forall u \sum_v w_{u, v}=1$), the Personalized PageRank vector $\PR_s$ with respect to source node $s$ is the solution to the recursive equation
\begin{align}
\label{eq:prrec}
\PR_s = \alpha e_s + (1-\alpha)\PR_s W.
\end{align}
If $s$ is a distribution (viewed as a column vector where entry $s[i]$ is the weight $s$ gives to node $i$) the recursive equation becomes
\[ \PR_s = \alpha s + (1-\alpha)\PR_s W. \]

An equivalent definition is in terms of the terminal node of a random walk starting from $s$. 
Let $(X_0,X_1,\ldots, X_L)$ be a random walk starting from $X_0 = s$ of length $L\sim Geometric(\alpha)$.  Here by $L\sim Geometric(\alpha)$ we mean $\Pr[L = \ell] = (1-\alpha)^\ell \alpha$.  This walk starts at $s$ and does the following at each step: with probability $\alpha$, terminate; and with the remaining probability $1-\alpha$, continue to a random out-neighbor of the current node.  Here if the current node is $u$, the random neighbor $v \in \outneighbors{u}$ is chosen with probability $w_{u, v}$ if the graph is weighted or with uniform probability $1 / \outdegree{u}$ if the graph is unweighted.
Then the PPR of any node $t$ is the probability that this walk stops at $t$:
\begin{align}
\label{eq:defmc}
\pi_s[t] = \PP[X_L = t].
\end{align}
Notice in this definition there is a single random walk\footnote{
Yet another definition of $\pi_s[t]$ to be aware of is the fraction of time spent at $t$ in the stationary distribution of the following random process over nodes:  at each step, with probability $\alpha$ we transition (``teleport'') back to $s$, and with remaining probability we transition from the current node $u$ to a random out-neighbor $v \in \outneighbors{u}$.  Because this chain is ergodic there is a unique stationary distribution.}.

The equivalence of these two definitions can be seen by solving Equation \ref{eq:prrec} for $\pi_s$ and then using a power series expansion~\cite{Avrachenkov2007}:
\[ \PR_s = (I - (\alpha e_s + (1-\alpha)W))^{-1} = \sum_{\ell=0}^{\infty} \alpha (1 - \alpha)^\ell e_s W^\ell. \]
In this expression we see the probability a walk has length $\ell$, $\alpha (1 - \alpha)^\ell$, and the probability distribution after $\ell$ steps from $s$, $e_s W^\ell$.
  For understanding the algorithms in this thesis, the second definition ($\pi_s[t]$ is the probability of stopping at $t$ on a single random walk from $s$) is the more useful one.

If some node $u$ has no out-neighbors, different conventions are possible.  For concreteness, we choose the convention of adding an artificial sink state to the graph, adding an edge from any such node $u$ to the sink, and adding a self-loop from the sink to itself.  In his survey~\cite{Gleich-preprint-pagerank-beyond}, Gleich describes other conventions for dealing with such dangling nodes.

\begin{figure}
  \centering
  \includegraphics[width=0.7\columnwidth]{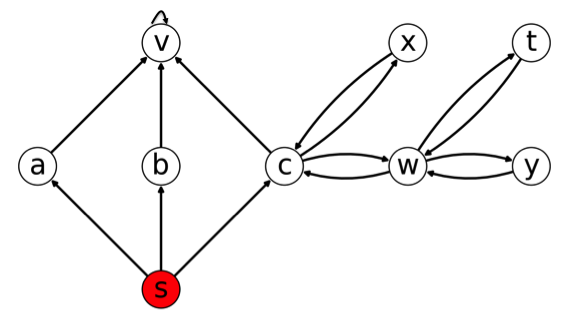}
  \caption[Example Graph]{A graph with 9 nodes.  We will compute PPR starting from $s$.}
  \label{fig:ppr_example}
\end{figure}

\begin{figure}
  \centering
  \includegraphics[width=0.7\columnwidth]{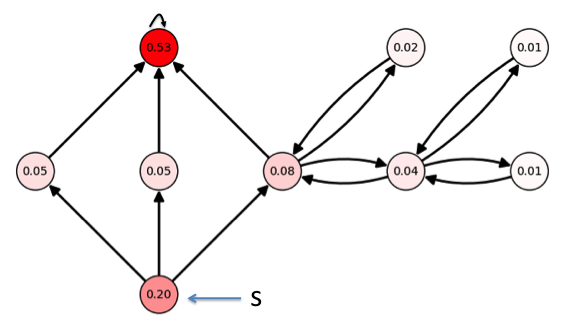}
  \caption[PPR Values on Example Graph]{The PPR values from $s$ to each other node in the graph of Figure \ref{fig:ppr_example}.  Nodes are also shaded by their PPR value. Notice that the top left node has the highest score.  This is because all of $s$'s out-neighbors point to $t$, and it matches our intuition that $s$ should be interested in the top-left node, or trust it.  On the other hand, the lowest score is for the two nodes on the right.  This is because they are far from $s$ and separated from $s$ by several nodes with other neighbors, and it matches our intuition that $s$ is likely less interested in these nodes.}
  \label{fig:ppr_example_values}
\end{figure}

%\subsection{Personalized PageRank Example}
As an example of the scores personalized PageRank assigns, consider the graph in Figure \ref{fig:ppr_example}.
If we compute PPR values from $s$ to all other nodes, we get the numbers shown in Figure \ref{fig:ppr_example_values}

\section{Other Random Walk Scores}
\label{sec:other_scores}
Personalized PageRank uses a geometric distribution for random walk length, assigning weight $\alpha (1 - \alpha)^\ell$ to walks of length $\ell$ for some parameter $\alpha$, but there is nothing magical about that length distribution\footnote{One unique property of a geometric random variable is that it is memoryless: if we have taken $\ell$ steps of a random walk with geometrically distributed length, then the expected number of remaining steps is the same as the initial number of steps.  Expressed another way: if after each step of the walk we stop with probability $\alpha$, we do not need to keep track of how long we've been walking, and can ``forget'' how many steps we've taken.  This gives PPR the simpler recursive Equation \ref{eq:prrec}.}.  If we instead use a Poisson length distribution, giving weight $e^{-\alpha}\alpha^\ell / \ell!$ to paths of length $\ell$, we get the the Heat Kernel which is useful for community detection \cite{Kloster2014}. In general, we can choose any weights $(\alpha_{\ell})_{l=0}^{\infty}$ such that $\sum_{\ell=0}^{\infty} \alpha_\ell = 1$ and assign weight $\ell$ to walks of length $\ell$. If $W$ is the random walk matrix, normalized so every row sums to 1, then there resulting score from $s$ to $t$ is
\[\sum_{\ell=0}^{\infty} \alpha_{\ell} s W^\ell e_t. \]

It is also not necessary to do a standard random walk to get a useful personalization score.  In personalized SALSA \cite{Bahmani2010} we alternate between taking forward and reverse walks along edges.  For example, on Twitter if two users Alice and Bob both follow Barack Obama, and Bob follows an opposing political candidate like Jeb Bush, then an alternating random walk from Alice might go to Obama, then to Bob, then to Bush.  On the other hand, a standard random walk would go from Alice to Obama to someone Obama follows (which might not include politicians of the opposite party).  For an alternating random walk, even length walks tend to find similar users to the source user, since if two users both follow the same people they are likely similar.  Odd length random walks tend to find users the source might want to follow (since they are followed by users similar the to source user).  See the section on SALSA in \cite{gupta2013wtf} for a fuller discussion of the meaning of SALSA. In an experiment at Twitter on real users,  SALSA actually worked better than PPR at generating who-to-follow candidates which the user subsequently followed \cite{gupta2013wtf}.

Our PPR estimation algorithms can also compute SALSA because SALSA on a graph $G$ is actually equivalent to PPR on a transformed graph $G'$.  To transform a given directed graph, for each node $u$ in the original graph $G$, we create two nodes, a ``consumer-node'' $u'$ and a ``producer-node'' $u''$ in $G'$.  Any directed edge $(u, v)$ in the original graph is then converted into an undirected edge $(u', v'')$ from $u$'s consumer node to $v$'s producer node.  Since personalized SALSA from $s$ to $t$ is defined as the probability that a random walk from $s$ which alternates between following forwards and reverse edges stops at $t$, we see that personalized SALSA on the original graph is equal to PPR on the transformed graph.

%Another recursively defined score is SimRank \cite{jeh2002simrank}.  For brevity we don't define it or its applications here, but we note that there is a complex reduction from SimRank to PPR \cite{sarlos2006randomize}, so it is plausible that with additional work our methods could be applied to compute SimRank.

\section{Problem Statements}

For personalized recommendations, such as who to follow on Twitter or friend recommendation on Facebook, a natural problem to solve is the $\delta$-significant Personalized PageRank problem: Given source node or distribution $s$, find the the targets $t$ with Personalized PageRank $\pi_s[t] \geq \delta$.  For this problem, previous algorithms like Monte Carlo (see section \ref{sec:monte_carlo}) run in $O \pn{\frac{1}{\delta}}$ time, and in the worst-case there are $\Omega \pn{\frac{1}{\delta}}$ results to return, so there is not much improvement to be made.  

In search-related applications, however, we are often interested in a small number of targets which are relevant to the search.  This motivates the \emph{Single Pair Personalized PageRank Estimation Problem}:
%\begin{center}
Given source node (or distribution) $s$ and target node $t$, estimate the Personalized PageRank $\PR_s[t]$ up to a small relative error.
%\end{center}
Since smaller values of $\PR_s[t]$ are more difficult to detect, we parameterize the problem by threshold $\delta$, requiring small relative error only if $\PR_s[t] \geq \delta$. 

We generalize the Single Pair PPR Estimation Problem to any given random walk length. Let a weighted, directed graph $G=(V, E)$ with normalized weight matrix $W$ be given (or equivalently let a Markov Chain with state space $V$ and transition matrix $W$ be given).  The \emph{$\ell$-Step Transition Probability Estimation Problem} is the following:
given an initial source distribution $s$ over $V$, a target state $t\in V$ and a fixed length $\ell$, compute the probability that a random walk starting at a state sampled from $s$ stops at $t$ after $\ell$ steps.   This probability is equal to the $t$th coordinate of $s W^\ell$.  We parameterize by a minimum probability $\delta$ and require a small relative error if this probability is at least $\delta$.

 In a typical personalized search application, we are given a set of candidate results $T$ for a query, and we want to find the top results ranked by Personalized PageRank.  This leads us to study the \emph{Personalized PageRank Search Problem}: Given 
\begin{itemize}
\item  a graph with nodes $V$ (each associated with a set of keywords) and edges $E$ (optionally weighted and directed),
\item a keyword inducing a set of targets \\$T = \{t \in V: \text{$t$ is relevant to the keyword}\}$, and
\item a node $s \in V$ or a distribution $s$ over starting nodes, based on the user performing the search
\end{itemize}
return the top-$k$ targets $t_1, \ldots, t_k \in T$ ranked by Personalized PageRank $\pi_s(t_i)$.

In future applications, it may be useful to not only score target nodes, but also compute paths to them.  For example, in personalized search on a social network, we could show the user performing the search some paths between them and each search result.  This motivates the \emph{Random Path Sampling Problem}: Given a graph, a set of targets $T$, and a source node or distribution $s$, sample a random path from $s$ of geometric length conditioned on the path ending in $T$.  %This problem could be a useful subroutine in future graph algorithms.

\inlineheading{Limitations of Prior Work}
Despite a rich body of existing work \cite{Jeh2003,Avrachenkov2007,Andersen2006,Andersen2007,Bahmani2010,Borgs2013,Sarma2013}, past algorithms for PPR estimation or search are too slow for real-time search.  For example, in Section \ref{sec:bippr_experimental_time}, we show that on a Twitter graph with 1.5 billion edges, past algorithms take five minutes to compute PPR from a given source.  Five minutes is far too long to wait for a user performing an interactive search.  Analytically we can understand this slowness through the assymptotic running time of past algorithms: On a graph with $n$ nodes, they take $\Theta(n)$ time to estimate PPR scores of average size $1/n$.  These algorithms requiring $\Theta(n)$ operations (more precicely, random memory accesses) are too slow for graphs with hundreds of millions of nodes.  

\section{High Level Algorithm}
Our PPR estimators are much faster than previous estimators because they are bidirectional; they work from the target backwards and from the source forwards.  %For intuition, consider the running time of breadth first search compared to bidirectional search.  
For intuition, suppose we want to detect if there is a path of length $\ell$ between two nodes $s$ and $t$ in a given directed graph, and suppose for ease of analysis that every node has in-degree and out-degree $d$.  Then finding a length-$\ell$ path using breadth first search forwards from $s$ requires us to expand $O(d^\ell)$ neighbors, and similarly breadth first search backwards from $t$ requires $O(d^\ell)$ time.  In contrast, if we expand length $\ell / 2$ from $s$ and expand length $\ell / 2$ from $t$ and intersect the set of visited nodes, we can detect a length $\ell$ path in time $O(\sqrt{d^l})$.  This is a huge difference; for example if $d=100$ and $\ell=4$ we are comparing order 10,000 node visits with order 100,000,000 node visits.  Our algorithm achieves a similar improvement: past PPR estimation algorithms take time $\Theta(n)$ to estimate a PPR score of size $1/n$ within constant relative error with constant probability, while our algorithm on average takes time $O(\sqrt{m})$.

The challenge for us designing a bidirectional estimator is that in graphs with large degrees, doing a breadth first search of length $\ell/2$, where $\ell$ is the maximum walk length, will take too much time, so a bidirectional search based on path length is not fast enough.  (The running time $O(\sqrt{d^\ell})$ is still exponential in $\ell$, whereas our bidirectional algorithm takes time polynomial in $\ell$.)  Instead we go ``half-way'' back from the target in the sense of probability: if we're estimating a probability $\pi_s[t]$ of size $1/n$, we first find all nodes with probability greater than $\sqrt{\frac{\dbar}{n}}$ of reaching the target on a random walk, where $\dbar = m/n$ is the average degree.  Then we take $O(\sqrt{m})$ random walks from $s$ forwards.  In Section \ref{sec:bippr} we describe how intersecting these random walks with the in-neighbors of nodes close to the target leads to a provably accurate estimator.

\section{Contributions}
The main result of this thesis is a bidirectional estimator \bippr{} (Section \ref{sec:bippr}) for the Single Pair Personalized PageRank Estimation Problem which is significantly faster than past algorithms in both theory and practice.  In practice, we find that on six diverse graphs, our algorithm is 70x faster than past algorithms.  For example, on the Twitter graph from 2010 with 1.5 billion edges, past algorithms take 5 minutes to estimate PPR from a random source to a random\footnote{This is for target sampled in proportion to their PageRank, since in practice users are more likely to search for popular targets.  If targets are chosen uniformly at random, past algorithms take 2 minutes, while our algorithm takes just 30 milliseconds, and improvement of 3000x.} target, while our algorithm takes less than three seconds.  Our algorithm is not just a heuristic that seems to work well on some graphs, but we prove it is accurate on any graph with any edge weights.  

We also analytically bound the running time of \bippr{} on arbitrary graphs.  First note that for a worst-case target, the target might have a large number of in-neighbors, making the reverse part of our bidirectional estimator very slow.  For example, on Twitter, Barack Obama has more than 60 million followers, so even one reverse step iterating over his in-neighbors would take significant time.  Analytically, if the target $t$ has $\Theta(n)$ followers, then our algorithm will take $\Theta(n)$ time to estimate a score of size $1/n$.  Since worst case analysis leads to weak running time guarantees for \bippr{}, we present an average case analysis.  In Theorem \ref{thm:fastpprtime} of Section \ref{sec:bippr_running_time}, we show that for a uniform random target, we can estimate a score of size at least $1/n$ to within a given accuracy $\epsilon$ with constant probability\footnote{We state results here for achieving relative error $\epsilon$ with probability $2/3$, but we can guarantee relative error $\epsilon$ with probability $1 - \pfail$ by multiplying the running time by $\log \frac{1}{\pfail}$.} in time
\[O \pn{\frac{\sqrt{m}}{\epsilon}}.  \]
We prove this running time bound without making any assumptions on the graph.  In contrast \mc{} (Section \ref{sec:monte_carlo}) takes $O \pn{\frac{n}{\epsilon^2}}$ time for the same problem and \rpush{} (Section \ref{sec:reverse_push}) takes $O \pn{\frac{n}{\epsilon}}$ time.  To make these numbers concrete, on Twitter-2010 (a crawl of Twitter from 2010), there are $n = 40 \text{ million}$ nodes and $m = 1.5 \text{ billion}$ edges, so the difference between $n$ (past algorithms) and $\sqrt{m}$ (our algorithm) is a factor of 1000.  In fact on this graph, when we choose targets uniformly at random and compare running times, our algorithm is 3000x faster than past algorithms, so this analysis explains why our algorithm is much faster than past algorithms.  This estimator also applies to standard (global) PageRank, giving the best known running time for estimating the global PageRank of a single node.

For undirected graphs we propose an alternative estimator \ubippr{} (Section \ref{sec:ubippr}) for the Single Pair PPR Estimation Problem which has worst-case running time guarantees.  %Whereas \bippr{} does local linear algebra ``push'' operations from the target (todo: Ashish wants me to change this to be less algorithm detailed) and random walks forwards, \ubippr{} does local push operations from the source and random walks from the target.  
Before we state the running time, note that on undirected graphs, the natural global importance of a node $t$ is $\frac{d_t}{2m}$, since this is the probability of stopping at $t$ as the walk length goes to infinity, so only Personalized PageRank scores larger than $\frac{d_t}{2m}$ are interesting. \ubippr{} can  estimate a Personalized PageRank score $\pi_s[t]$ of size greater than $\frac{d_t}{2m}$  within relative error $\epsilon$ in expected time 
\[O \pn{ \frac{\sqrt{m}}{\epsilon}}.\]
This running time bound applies for a worst-case target $t$ and worst-case undirected graph.  There is a nice symmetry between \bippr{} and \ubippr{}, and in one preliminary experiment (Section \ref{sec:ubippr_experiment}) \ubippr{} and \bippr{} have about the same running time.  

Our results are not specific to PPR, but actually generalize to a bidirectional estimator (Chapter \ref{sec:mc_chapter}) that solves the $\ell$-Step Transition Probability Estimation Problem on any weighted graph (or Markov Chain) from any source distribution to any target.  This estimator has relative error $\epsilon$ with constant probability. For a uniform random target, if the estimated probability is greater than $\delta$, the expected running time is 
\[O \pn{\frac{\ell^{3/2}}{\epsilon} \sqrt{\dbar/\delta}} \]
where $\dbar = m/n$ is the average degree of a node.
Compare this to the running time of Monte Carlo, $O \pn{\frac{\ell}{\epsilon^2 \delta}}$ and note that in practice $\ell$, $1/\epsilon$, and $\dbar$ are relatively small, say less than 100, whereas $\delta$ is tiny, perhaps $10^{-8}$ if we are estimating a very small probability, so the dependence on $\delta$ is crucial.  This algorithm has applications in simulating stochastic systems and in estimating other random walk scores like the Heat Kernel which has been useful for community detection \cite{Kloster2014}.

In practice, the \bippr{} algorithm is slower when the target node is very popular (having many in-neighbors or high PageRank) than for average targets.  In Chapter \ref{precompute_ppr} we present a method of precomputing some data and distributing it among a set of servers to enable fast Single Pair PPR Estimation even when the target is popular.  % to make PPR estimation efficient even for popular targets.%Since in practice users are likely to search for popular targets like Barack Obama which are challenging to \bippr{} due to their high in-degree,  in Chapter \ref{precompute_ppr} we propose a variation of \bippr{} which uses pre-computation to achieve worst-case guarantees for any target.  
Analytically, with pre-computation time and space of  $ O \pn{n \frac{\sqrt{m}}{\epsilon}}$ we can achieve worst-case expected running time 
\[O \pn{\frac{\sqrt{m}}{\epsilon}} \]
for any target, even targets with very large in-degree.  In practice, the analysis in Section \ref{pprStorageAnalysis}
estimates that the storage needed on the Twitter-2010 graph to pre-compute \bippr{} is 8 TB, or 1.4 TB after an optimization.  This is large, but not in-feasible, storage, and allows for very fast PPR computation for any target.  For comparison, if we pre-computed Monte-Carlo walks from every source (as proposed in \cite{Fogaras2005}), we would need $\Omega(n^2)$ storage, or 1600 TB.  

%In practice, the \bippr{} algorithm is slower when the target node is very popular (having many in-neighbors or high PageRank) than for average targets.  To address this, in Chapter \ref{precompute_ppr} we present a method of precomputing some data and distributing it among a set of servers to make PPR estimation efficient even for popular targets.

When doing a personalized search, we typically do not have a single target, but have many candidate targets and want to find the most relevant ones, motivating the Personalized PageRank Search Problem.  In Chapter \ref{sec:search_chapter} we describe a personalized search algorithm which can use precomputation to efficiently rank targets matching a query. In experiments on a 1.5 billion edge Twitter graph, this algorithm is more efficient than past algorithms, taking less than 0.2 seconds per query, while past algorithms take 10 seconds for some target set sizes.  However, this improved running time comes at the expense of significant pre-computation, roughly 16 TB for the 2010 Twitter graph.  Analytically, given $O(n \sqrt{m})$ precomputation time and space, this algorithm can respond to name search queries with rigorous approximate accuracy guarantees in time $O(\sqrt{m})$.

Finally, as another extension of these ideas, in Section \ref{sec:path_chapter} we present an algorithm for the Random Path Sampling Problem.  Given a given start $s$ and target $t$ (or a set of targets) this algorithm can sample a random path from $s$ conditioned on stopping at $t$.  Its average expected running time is
\[O \pn{\sqrt{m}}\]
for a uniform random target.  In contrast, Monte Carlo would take $\Omega(n)$ time for this task. %This problem  could be used to show a user $s$ why a target $t$ is relevant to them

%  LocalWords:  personalization PageRank

\chapter{Background and Prior Work}
\label{sec:relwork}
In this chapter we briefly consider other approaches to personalized search, and then, focusing on Personalized PageRank, we describe the prior algorithms for computing PPR which our bidirectional estimators build upon.  Our bidirectional estimator uses these past estimators as subroutines, so understanding them is important for understanding our bidirectional algorithm.
% algorithms which we later both as baselines for our work and as sub-routines in our bidirectional algorithm.

\section{Other Approaches to Personalization}
\label{sec:other_personalization}
The field of personalized search is large, but here we briefly describe some of the methods of personalizing search and cite a few representative papers.  One method is using topic modeling based on the search history of a user \cite{speretta2005personalized} to improve the rank of pages whose topic matches the topic profile of a user.  For example, if a user has a search history of computer science terms, while another has a search history of travel terms, that informs what to return for the ambiguous term ``Java.''  A simpler method of using history, compared to topic-modeling methods in \cite{dou2007large}, is to boost the rank of results the user has clicked on in the past (for example, if the user has searched for ``Java'' before and clicked on the Wikipedia result for the language, then the Wikipedia result could be moved to the top in future searches for ``Java'' by that user), or to boost the rank of results similar users have clicked on in the past, where user similarity is measured using topic models.  A related method is to personalize based on a user's bookmarks \cite{kim2006personalized}.

Other past work has proposed using a user's social network to personalize searches. The authors of \cite{carmel2009personalized} score candidate results based not just on the searching user's history of document access, but also based on similar user's history of document access, where users are said to be similar if they use similar tags, share communities, or interact with the same documents.  Considering the problem of name search on the Orkut social network, \cite{vieira2007} proposes ranking results by shortest path distance from the searching user.  %Our intuition is that PPR should provide a better ranking for name search than shortest path ranking, since PPR can distinguish between the following two cases: 

\section{Power Iteration}
\label{sec:power_iter}
The original method of computing PageRank and PPR proposed in \cite{Page1999} is known as power iteration.  This method finds a fixed-point of the recursive definition Equation \ref{eq:prrec} by starting with an initial guess, say $p^0_s = (\frac{1}{n}, \ldots, \frac{1}{n}) \in \R^n$, and repeatedly setting
\[ p^{i+1}_s = \alpha e_s + (1 - \alpha) p^i_s W \]
until $\norm{p^{i + 1} - p^{i}}_{\infty}$ falls below some threshold.  Standard power iteration analysis \cite{ilprints582} shows that to achieve additive  error $\delta$, $O \pn{\log \frac{1}{\delta}}$ iterations are sufficient, so the running time is $O \pn{ m \log \frac{1}{\delta}}$.  On large graphs, $m$ is more than a billion, so this technique works fine for computing global PageRank (which can be computed once and stored), but using it to personalize to every node is  computationally impractical.  This motivates local algorithms for PPR which focus on regions of the graph close to the source or target node and avoid iterating over the entire graph.

\section{Reverse Push}
\label{sec:reverse_push}
One local variation on Power Iteration starts at a given target node $t$ and works backwards, computing an estimate $p^t(s)$ of $\pi_s(t)$ from every source $s$ to the given target. This technique was first proposed by Jeh and Widom \cite{Jeh2003}, and subsequently improved by other researchers \cite{Fogaras2005,Andersen2006}. The algorithms are primarily based on the following recurrence relation for $\PR_u$:
\begin{align}
\label{eq:reverse_recurrence}
\PR_s(t)=\alpha e_s+\frac{(1-\alpha)}{|\mathcal{N}^{out}(s)|}.\sum_{v\in\mathcal{N}^{out}(s)}\PR_v(t) .
\end{align}
Intuitively, this equation says that for $s$ to decide how important $t$ is, first $s$ gives score $\alpha$ to itself, then adds the average opinion of its out-neighbors, scaled by $1 - \alpha$.

\begin{figure}[tbph]
  \centering
  \includegraphics[width=0.4\columnwidth]{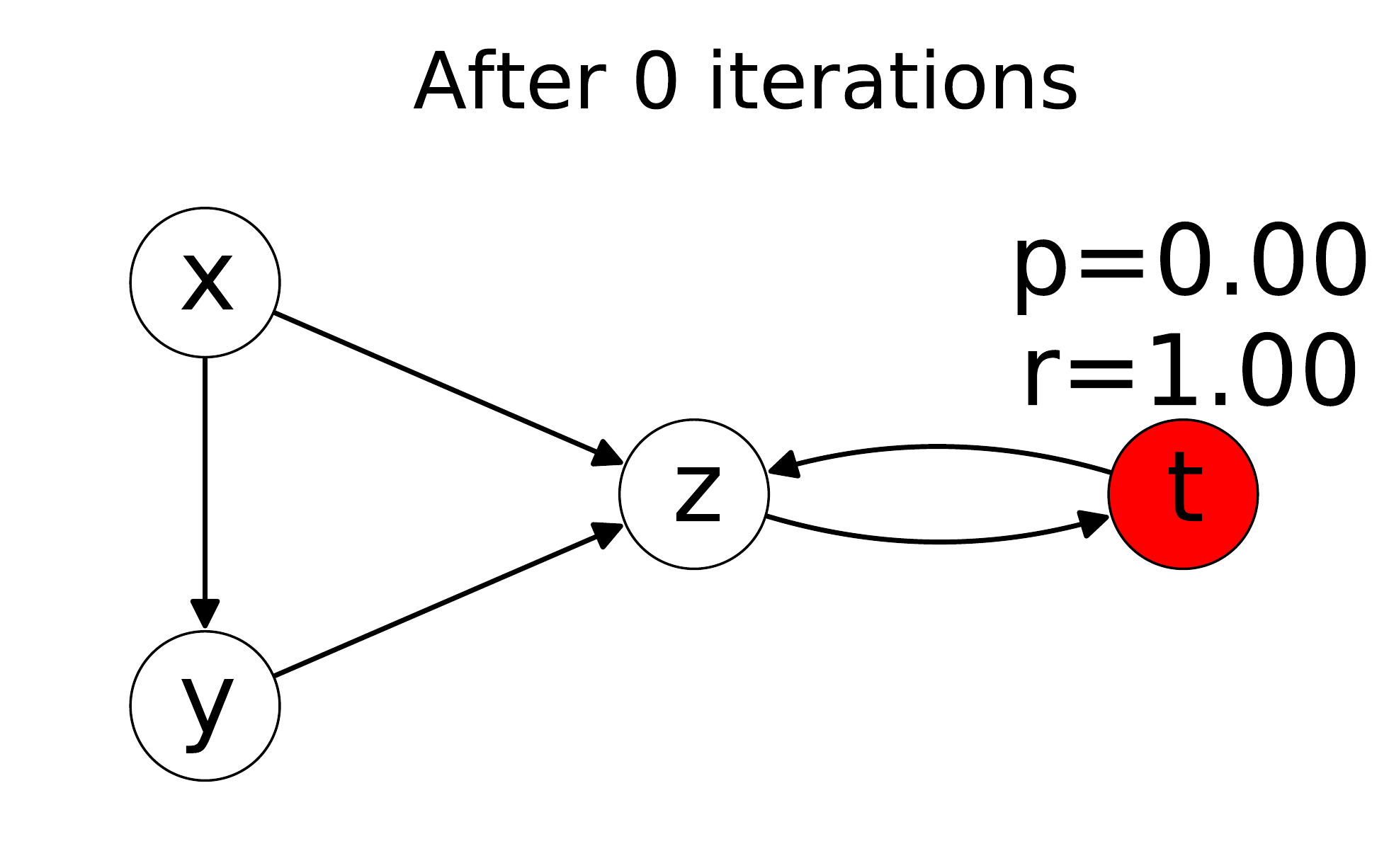}
  \includegraphics[width=0.4\columnwidth]{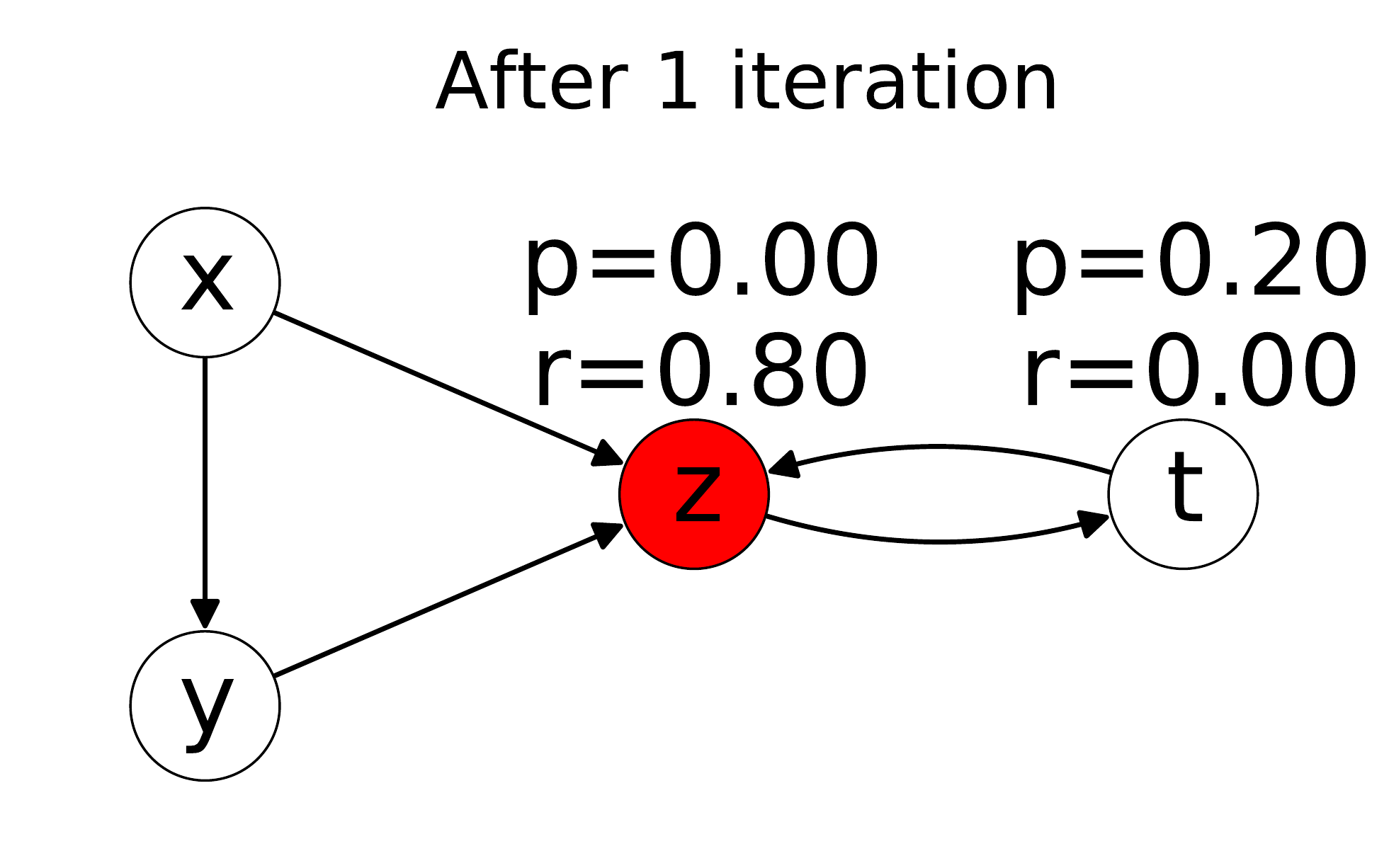}
  \includegraphics[width=0.4\columnwidth]{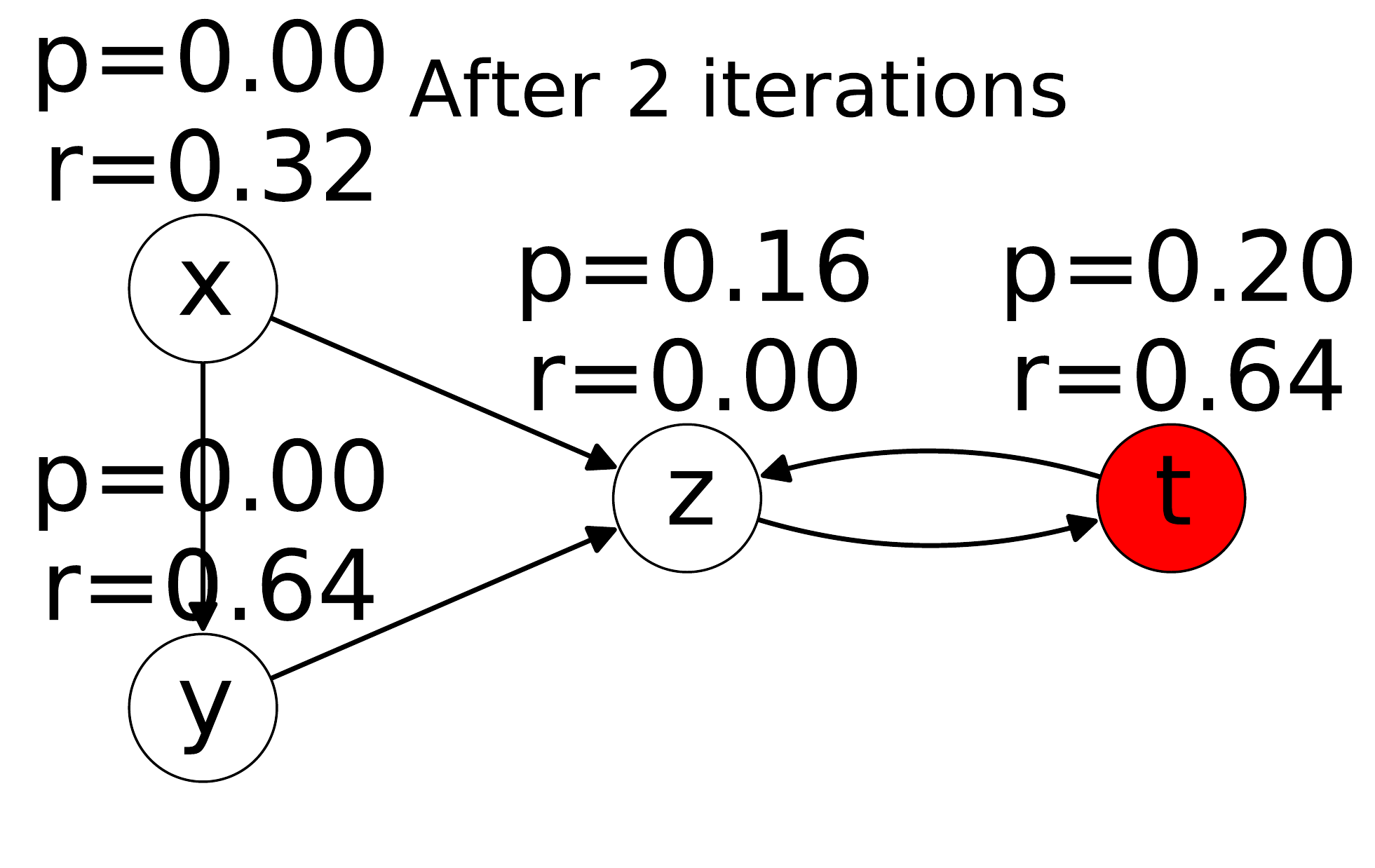}
  \includegraphics[width=0.4\columnwidth]{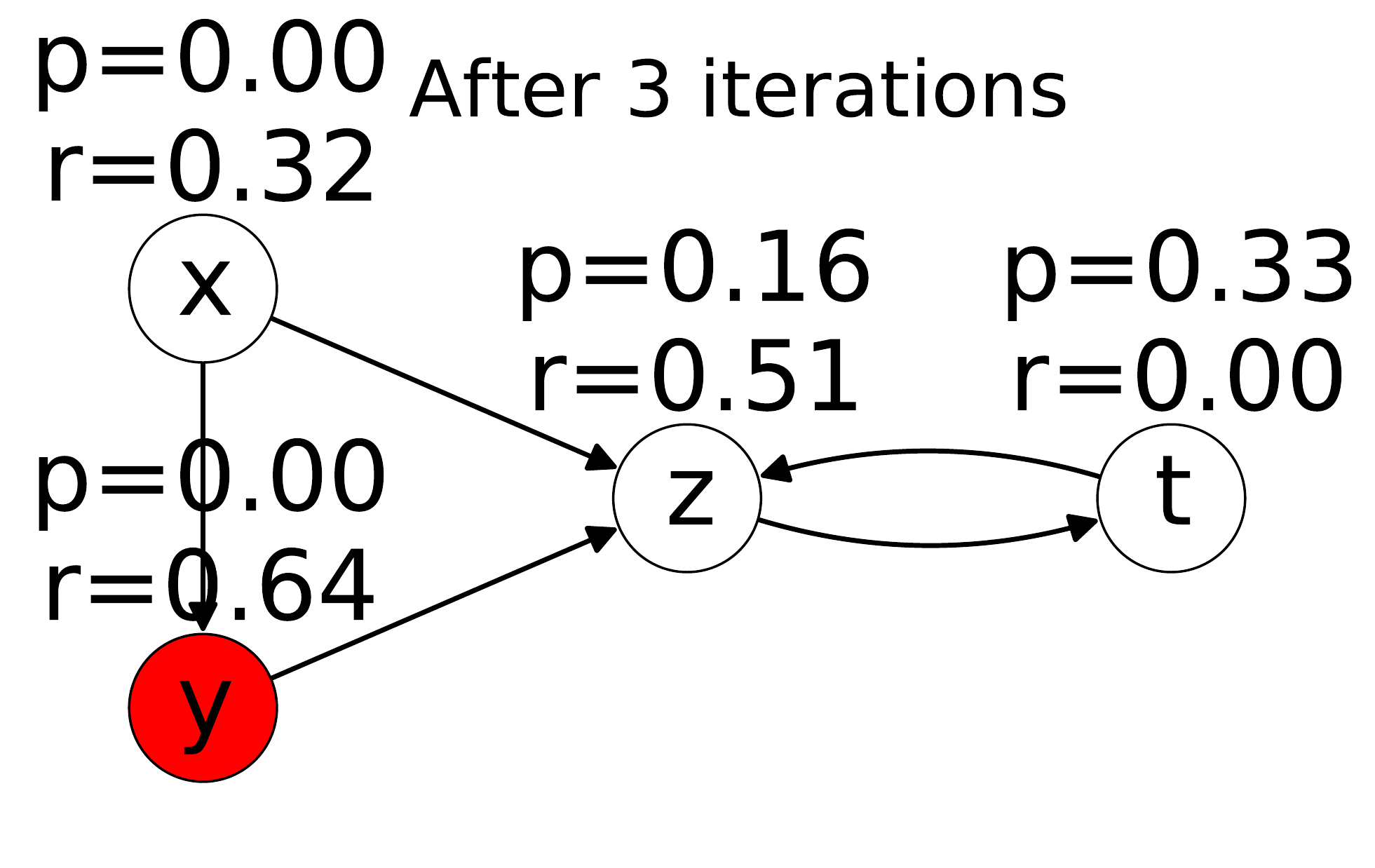}
  \includegraphics[width=0.4\columnwidth]{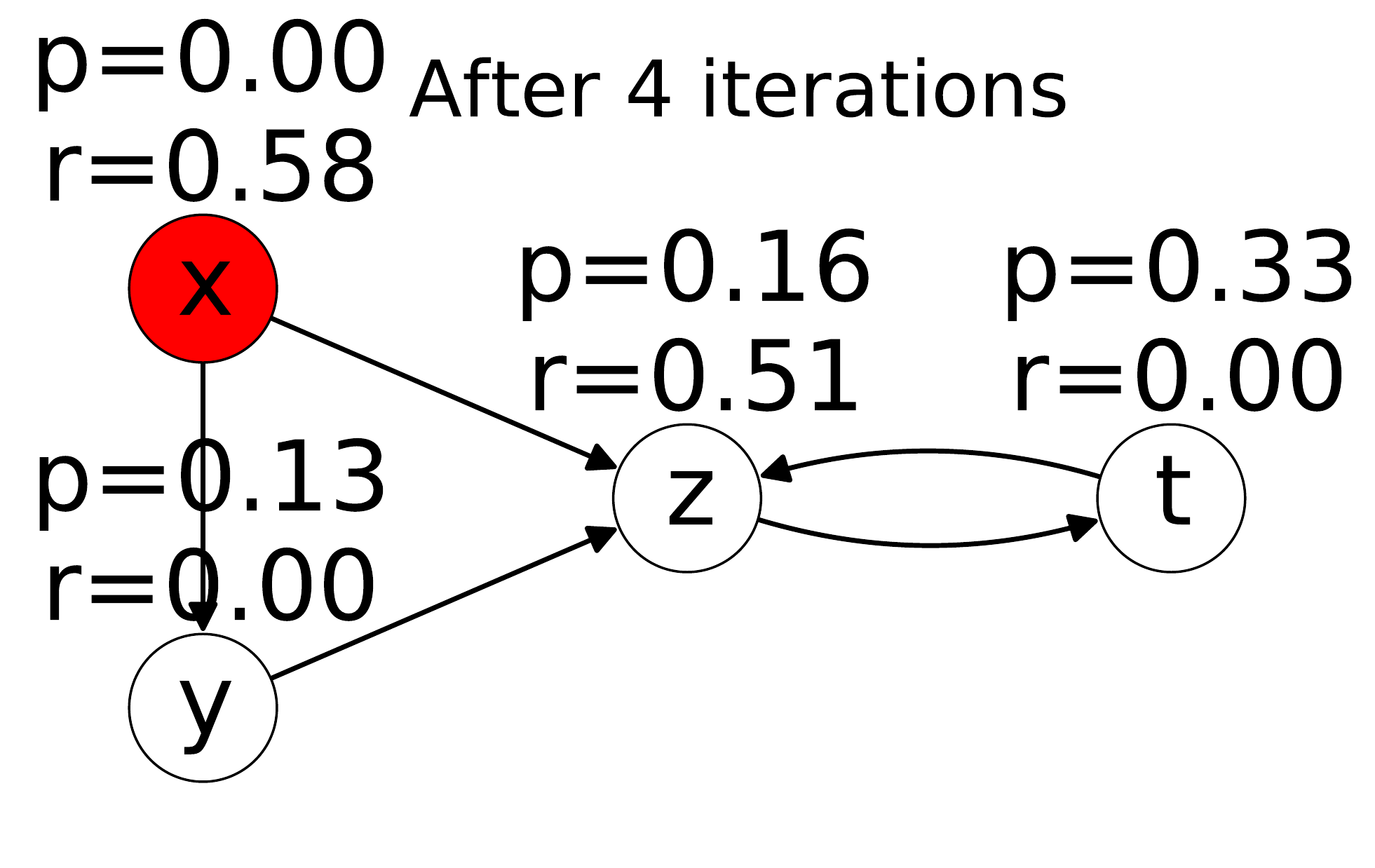}
  \includegraphics[width=0.4\columnwidth]{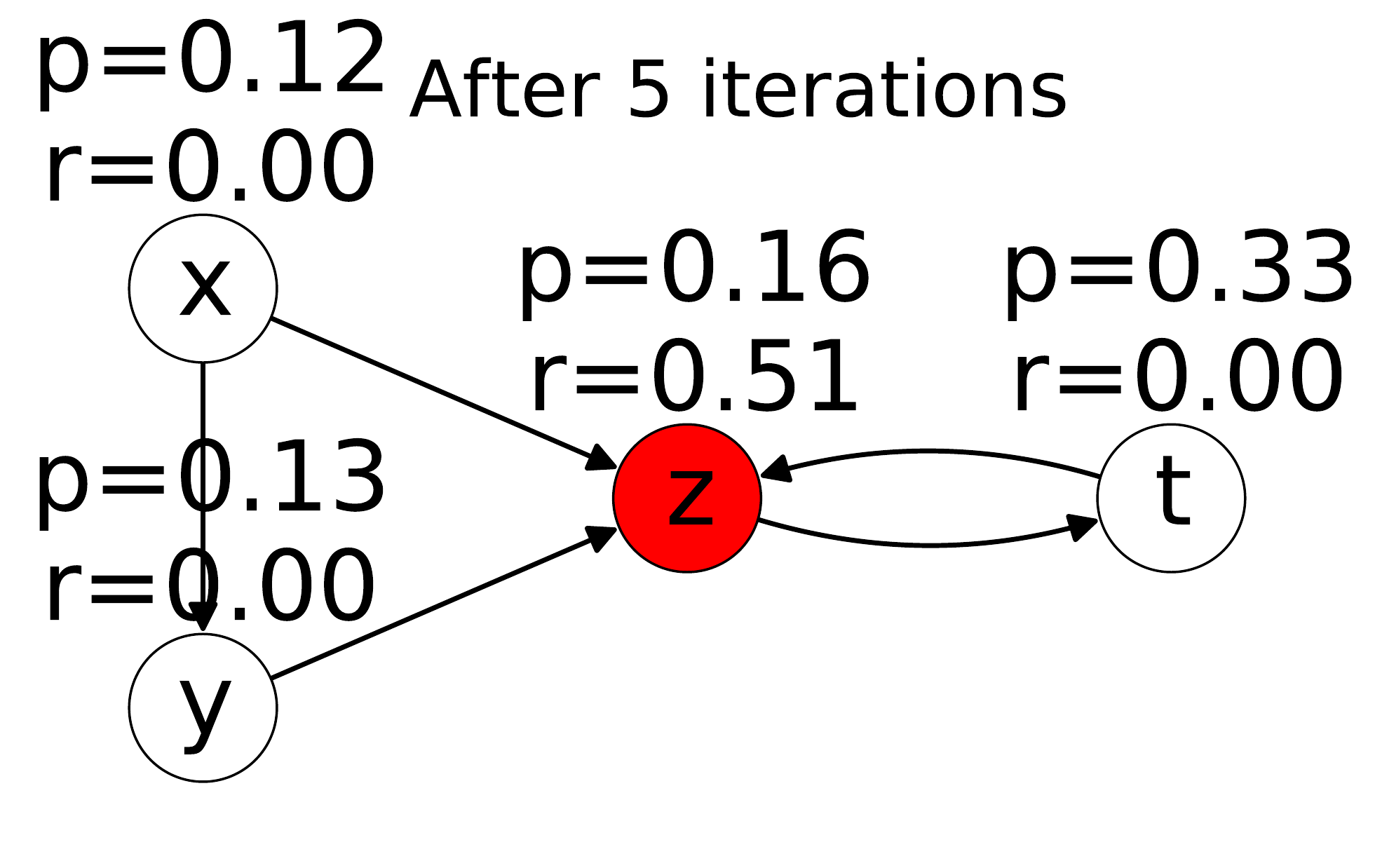}
  \includegraphics[width=0.4\columnwidth]{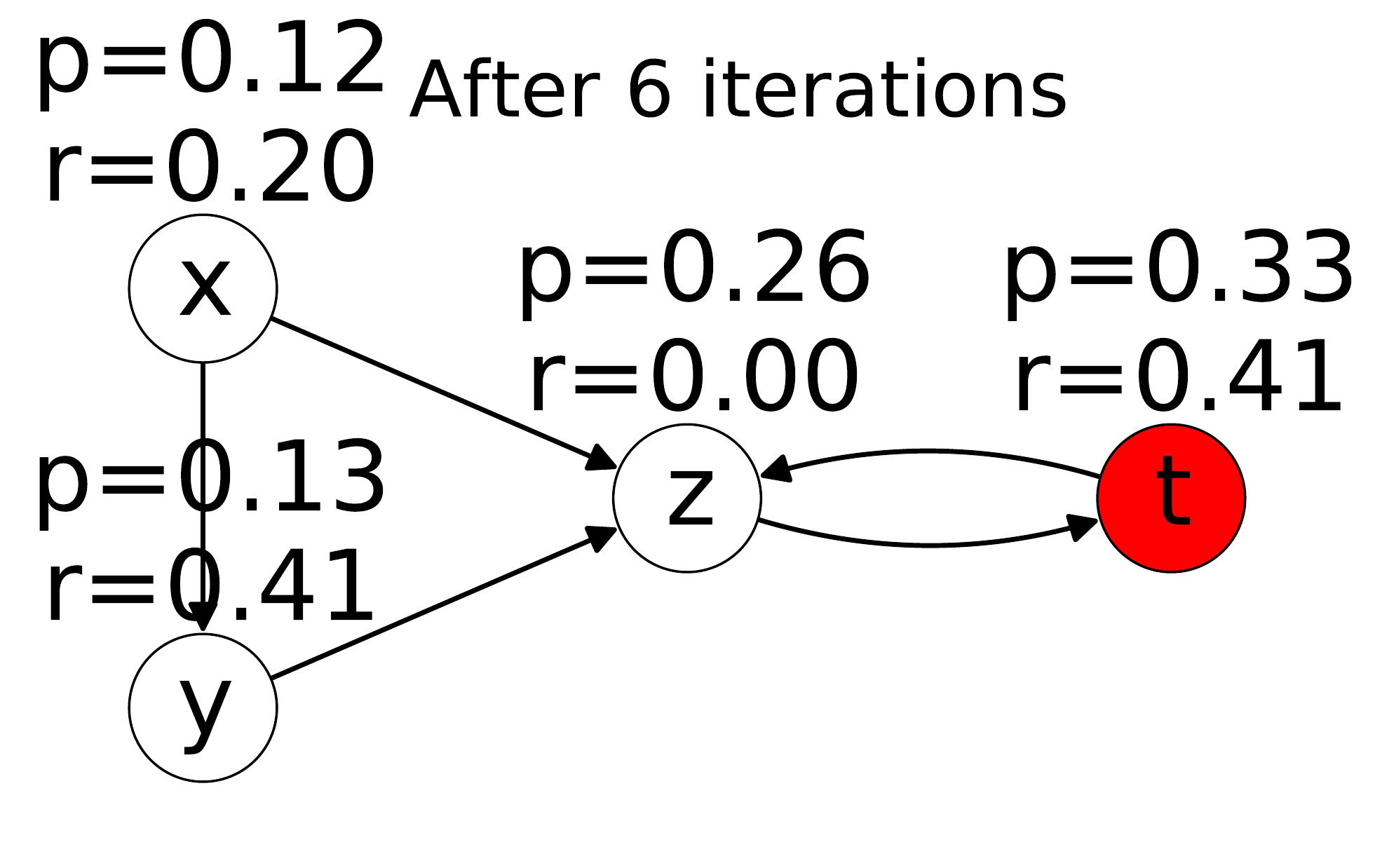}
  \includegraphics[width=0.4\columnwidth]{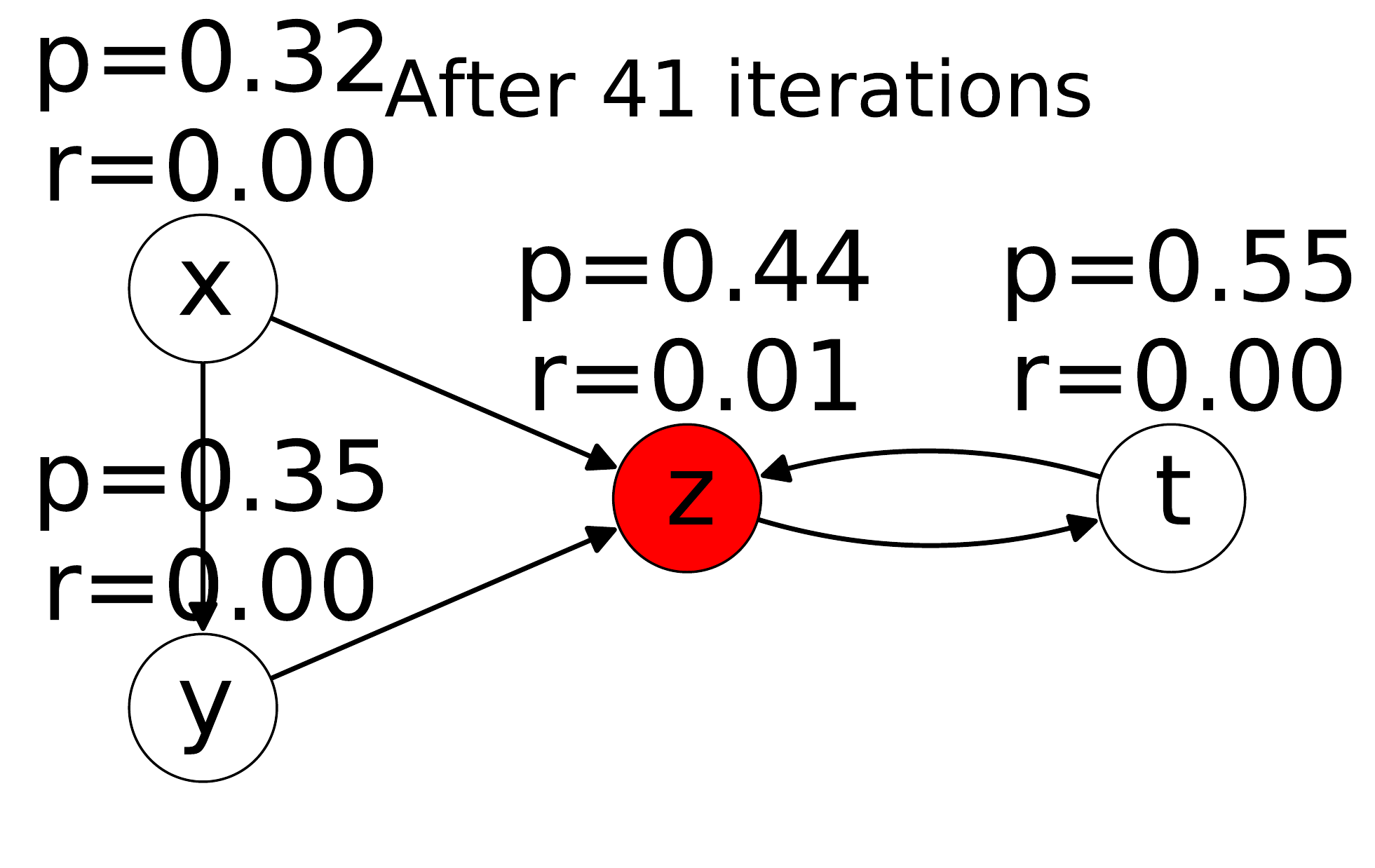}
  \caption[\rpush{} Example]{The first 7 iterations of the \rpush{} algorithm to target $t$ on a simple four-node graph.  Each node $v$ is annotated with its current residual $r^t[v]$ (mass that needs to be pushed back) and estimate $p^t[v]$ (estimate of $\pi_v[t])$. At each iteration, the node about to be pushed from is highlighted in red.  We also show the result after 41 iterations, when all of the residuals have fallen below $0.01$.  If we set $\rmax = 0.01$ the algorithm would terminate after 41 iterations.  At this point, all estimates $p^t[v] \in [\pi_v[t] - 0.01, \pi_v[t]]$. } 
  \label{fig:reverse_push_example}
\end{figure}

Andersen et.~al.~\cite{Andersen2007} present and analyze a local algorithm for PPR based on this recurrence.  This algorithm can be viewed as a message passing algorithm which starts with a message at the target.  Each local push operation involves taking the message value (or ``residual'') $r^t[v]$ at some node $v$, incorporating $r^t[v]$ into an estimate $p^t[v]$ of $\pi_v[t]$, and sending a message to each in-neighbor $u \in \inneighbors{v}$, informing them that $p^t[v]$ has increased.  Because we use it in our bidirectional algorithm, we give the full pseudo-code here as  Algorithm \ref{alg:invPPR}.
\begin{algorithm}[!ht]
\caption{\rpush{}$(t,\epr, \alpha)$~\cite{Andersen2007}}
\label{alg:invPPR}
\begin{algorithmic}[1]
\REQUIRE graph $G$ with edge weights $(w_{u,v})_{u,v \in V}$, target node $t$, maximum residual $\epr$, teleport probability $\alpha$
% multiplicative factor $\beta$ (default= $1/6$)
\STATE Initialize (sparse) estimate-vector $p_t = \vec{0}$ and (sparse) residual-vector $r_t = e_t$ (i.e.~$r_t(v) = 1$ if $v=t$; else $0$)

\WHILE{$\exists v\in V\,s.t.\,r_t(v)> \epr$}
\FOR{$u\in \inneighbors{v}$}
   \STATE $r_t(u) \pluseq (1 - \alpha) w_{u,v} r_t(v)$
\ENDFOR
\STATE      $p_t(v) \pluseq \alpha r_t(v)$
\STATE      $r_t(v)=0$
\ENDWHILE
\RETURN $(p_t, r_t)$
\end{algorithmic}
\end{algorithm}    
An example on a small four-node graph is shown in Figure \ref{fig:reverse_push_example}.

Andersen et.~al.~prove that after this algorithm completes, for all $s$ the additive error $\abs{\pi_s(t) - p^t(s)} \leq \rmax$.  Their proof uses the following loop invariant (Lemma $1$ in \cite{Andersen2007})
\begin{equation}
  \label{eq:ppr_dot_product_rel_work}
  \pi_s[t] = p^t[s] + \sum_{v \in V} \pi_s[v] r^t[v].
\end{equation}
This loop invariant can be interpreted as saying that the true value of $\pi_s[t]$ is equal to the estimate from messages processed at $s$ plus the sum over all nodes $v$ of their un-processed message value $r^t[v]$, weighted by their proximity to $s$, $\pi_s[t]$, since nearby messages are more relevant than distant messages.
%This summation is exactly a dot product (we later show this more explicitly).  
\rpush{} terminates once every residual value $r^t[v] < \epr$.  Since $\sum_v \pi_s[v] = 1$, the convex combination  $\sum_{v \in V} \pi_s[v] r^t[v] < \rmax$.  Viewing this as an error term, Andersen et.~al.~observe that $p^t[s]$ estimates $\pi_s[t]$ from every $s \in V$ up to a maximum additive error of $\epr$.  For a given $(s, t)$ pair, in Section \ref{sec:bippr} we will show how to use these residuals to get a more accurate estimate of $\pi_s(t)$.

Andersen et.~al.~also show that the number of push operations is at most $\frac{n \pi(t)}{\rmax}$, and the running time of pushing from some node $v$ is proportional to its in-degree $\indegree{v}$, so the total running time is proportional to $\frac{n \pi(t)}{\rmax}$ times the average in-degree of nodes pushed.  

In \cite{Lofgren2013}, Lofgren and Goel show that if $t$ is chosen uniformly at random, the average running time is 
\[ O \pn{ \frac{\bar{d}}{\rmax}} \]
where $\bar{d} = m/n$ is the average in-degree of a node.  They also analyze the running time of a variant algorithm where a priority queue is used in the while loop, so the node $v$ with the greatest residual value $r^t(v)$ is pushed on each iteration.  For this variant, as $\delta \to 0$, the running time is $O \pn{ m \log \frac{1}{\delta}}$.  Finally, they empirically measure the running time, and roughly find that for $\delta > \frac{1}{n}$, when running on a Twitter-based graph the running time roughly scales with $\frac{1}{\delta}$.
%Another use of such local update algorithms is for estimating the inverse-PPR vector for a target node. Local-update algorithms for inverse-PageRank are given in \cite{Andersen2007} (where inverse-PPR is referred to as the `contribution PageRank vector'), and \cite{Lofgren2013} (where it is called `susceptibility'). However, one can exhibit graphs where these algorithms need a running-time of $O(1/\delta)$ to get additive guarantees on the order of $\delta$.

\section{Forward Push}
\label{sec:forward_push}
An alternative local version of power iteration starts from the start node $s$ and works forward along edges.  Variations on this were proposed in \cite{berkhin2006bookmark,Jeh2003} and others, but the variation most useful for our work is in Andersen et.~al.~\cite{Andersen2006} because of the analysis they give.  Because we use it a variation of our bidirectional algorithm, we give the full pseudo-code here as Algorithm \ref{alg:forwardPush}.
\begin{algorithm}[tbph]
\caption{\fpush{}$(G,\alpha, s,\epr)$~\cite{Andersen2006}}
\label{alg:forwardPush}
\begin{algorithmic}[1]
\REQUIRE undirected graph $G$, teleport probability $\alpha$, start node $s$, maximum residual $\epr$
\STATE Initialize (sparse) estimate-vector $p_s = \vec{0}$ and (sparse) residual-vector $r_s = e_s$ (i.e.~$r_s[v] = 1$ if $v=s$; else $0$)

\WHILE{$\exists u\in V\,s.t.\, \frac{r_s[u]}{d_u}> \epr$}
\FOR{$v\in \mathcal{N}[u]$}
   \STATE $r_s[v] \pluseq (1 - \alpha) r_s[u] / d_u$
\ENDFOR
\STATE      $p_s[u] \pluseq \alpha r_s[u]$
\STATE      $r_s[u]=0$
\ENDWHILE
\RETURN $(p_s, r_s)$
\end{algorithmic}
\end{algorithm}    

To our knowledge, there is no clean bound on the error $\norm{p_s - \pi_s}$ as a function of $\rmax$ for a useful choice of norm.  The difficulty is illustrated by the following graph: we have $n$ nodes, $\{s, t, v_1, \ldots, v_{n-2}\}$, and $2(n-2)$ edges, $(s, v_i)$ and $(v_i, t)$ for each $i$.  If we run \fpush{} on this graph starting from $s$ with $\rmax= \frac{1}{n-2}$, then after pushing from $s$, the algorithm halts, with estimate at $t$ of $p_s(t) = 0$.  However, $\pi_s(t) = \Omega(1)$, so the algorithm has a large error at $t$ even though $\rmax$ is getting arbitrarily small as $n \to \infty$.  

The loop invariant in \cite{Andersen2006} does give a bound on the error of \fpush{}, but it is somewhat subtle, involving the personalized PageRank personalized to the resulting residual vector. In Equation \eqref{eq:forward_push_error} of Section \ref{sec:undirected_bippr} we give an additive error bound in terms of $\rmax$ and the target's degree $d_t$ and PageRank  $\pi(t)$ which applies on undirected graphs.

\section{Monte Carlo}
\label{sec:monte_carlo}
The random walk definition of personalized PageRank, Equation \eqref{eq:defmc}, inspires a natural randomized algorithm for estimating PPR.  We simple sample some number $w$ of random walks from the start node (or distribution) $s$ and see what empirical fraction of them end at each target.  This technique of simulating a random process and empirically measuring probabilities or expectations is known as Monte Carlo.  More formally, for each $i \in [w]$ we sample length $L_i \sim \text{Geometric}(1/\alpha)$, take a random walk of length $L_i$ starting from $s$, and and let $V_i$ be the endpoint.  Then we define 
\[ \tilde{\pi}_s(t) = \frac{\# \{i: V_i = t\}}{w}. \]
It is immediate from Equation \ref{eq:defmc} that for any $t$, $\EE [ \tilde{\pi}_s(t)] = \pi_s[t]$.  If we want to estimate a probability $\pi_s(t)$ of size $\delta$ to within relative error $\epsilon$ with constant probability, then standard Chernoff bounds like those used in Theorem \ref{thm:bidirmain} imply that 
\[ w = O \pn{\frac{1}{\epsilon^2 \delta}} \]
is a sufficient number of random walks.  These walks can be computed as needed or computed in advance, and various papers 
\cite{Avrachenkov2007,Bahmani2010,Borgs2013,Sarma2013} consider different variations. Such estimates are easy to update in dynamic settings \cite{Bahmani2010}. However, for estimating PPR values close to the desired threshold $\delta$, these algorithms need $\Omega(1/\delta)$ random-walk samples, which makes them slow for estimating the PPR between a random pair of nodes, since the expected PPR score is $1/n$, so their running time is $\Omega(n)$.

\section{Algorithms using Precomputation and Hubs}
In \cite{berkhin2006bookmark}, Berkhin builds upon the previous work by Jeh and Widom \cite{Jeh2003} and proposes efficient ways to compute the personalized PageRank vector $\pi_s$ at run time by combining pre-computed PPR vectors in a query-specific way.  In particular, they identify ``hub'' nodes in advance, using heuristics such as global PageRank, and precompute approximate PPR vectors $\hat{\pi}_h$ for each hub node using a local forward-push algorithm called the Bookmark Coloring Algorithm (BCA).  
%At runtime, given personalization vector $s$, Berkhin proposes a modified BCA algorithm which uses the precomputed vector $\hat{\pi}_h$ whenever the forward-push algorithm encounters a hub node $h$.   
Chakrabarti \cite{chakrabarti2007dynamic} proposes a variant of this approach, where Monte-Carlo is used to pre-compute the hub vectors $\hat{\pi}_h$ rather than BCA.

Both approaches differ from our work in that they construct complete approximations to $\pi_s$, then pick out entries relevant to the query. 
This requires a high-accuracy estimate for $\pi_s$ even though only a few entries are important.  
In contrast, our bidirectional approach allows us compute only the entries $\pi_s(t_i)$ relevant to the query.  

\chapter{Bidirectional PPR Algorithms}
In this chapter we present our bidirectional estimators for Personalized PageRank.  We briefly describe the first estimator we developed, FAST-PPR, since its main idea gives useful intuition.  Then we fully describe a simpler, more efficient estimator, \bippr{}, as well an alternative estimator for undirected graphs \ubippr{}.  Finally we demonstrate the speed and accuracy of our estimators in experiments on real graphs.  %Finally we describe a method of using pre-computation to improve the speed of \bippr{}.

\section{FAST-PPR}
\begin{figure}
\centering
\includegraphics[width=0.9\columnwidth]{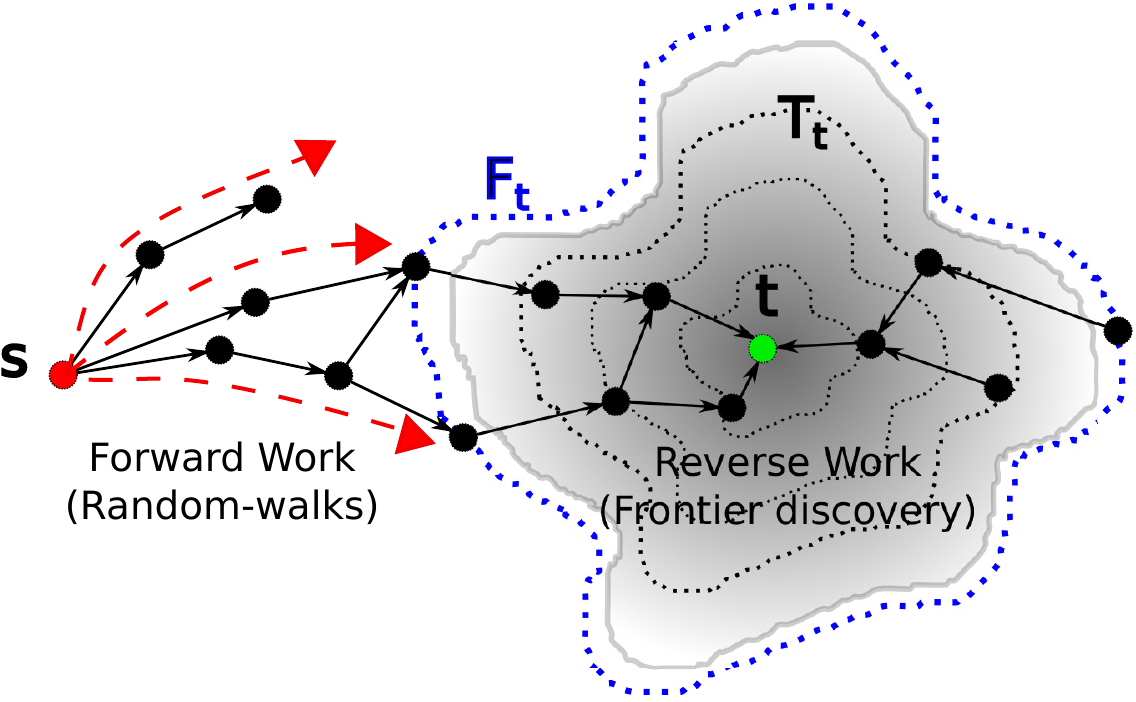}
\caption[FAST-PPR Algorithm]{The FAST-PPR Algorithm: We first work backward from the target $t$ to find the frontier $F_t$ (with inverse-PPR estimates). Next we work forward from the source $s$, generating random-walks and testing for hitting the frontier.}
\label{fig:fastppr_diag}
\end{figure}

The first bidirectional estimator we proposed was FAST-PPR \cite{fastppr}. Its main idea is that the personalized PageRank from $s$ to $t$ can be expressed as a sum
\begin{equation}
  \label{eq:ppr-decomposition}
 \PR_s[t]=\sum_{w\in F}\PP\left[\text{a walk from $s$ first hits $F$ at $w$} \right] \cdot \PR_w[t]  
\end{equation}
where $F$ is any set which intercepts every path from $s$ to $t$. Think of $F$ as some set surrounding the set of nodes ``close'' to the target, as shown in Figure \ref{fig:fastppr_diag}.
In FAST-PPR, we say a node $w$ is ``close'' to the target if $\pi_w[t]$ is above some threshold, and  $F$ is the set of in-neighbors of nodes ``close'' to the target.  FAST-PPR uses the \rpush{} algorithm (Section \ref{sec:reverse_push}) to estimate the values $\pi_w[t]$ for $w \in F$, then uses Monte Carlo random walks to estimate the value of Equation \ref{eq:ppr-decomposition}.  FAST-PPR has been superseeded \footnote{
%FAST-PPR achieved significant efficiency improvements over past algorithms, but 
The difficulty in FAST-PPR is how to estimate the values $\pi_w[t]$ for $w \in F$ accuratly.  The \rpush{} algorithm gives an additive error guarantee for nodes in $F$, whereas the analysis needs a relative error guarantee.  This issue makes FAST-PPR slower in our experiments than \bippr{} to achieve the same accuracy and complicates FAST-PPR's theoretical analysis.  
%\bippr{} improves these difficulties by using a different decomposition of Personalized PageRank based on the residuals of the \rpush{} algorithm.  It achieve significant efficiency improvements over past algorithms, but it has been superseded by a newer algorithm, \bippr{} which is more efficient, has a simpler analysis, and has a simpler linear structure which makes it easier to generalize to other length distributions or extend to use in personalized search.  Thus we do not present FAST-PPR in this thesis.
}
by \bippr{} so we now move on to describing that estimator.
 
\section{Bidirectional-PPR}
\label{sec:bippr}
In this section we present our new bidirectional algorithm for PageRank estimation \cite{bippr}. %We describe the algorithm, given an example, prove accuracy and efficiency guarantees, and finally describe a practical running time improvement.
At a high level, our algorithm estimates $\pi_s[t]$ by first working backwards from $t$ to find a set of intermediate nodes `near' $t$ and then generating random walks forwards from $s$ to detect this set.  Our method of combining reverse estimates with forward random walks is based on the residual vector returned by the \rpush{} algorithm (given in Section \ref{sec:reverse_push} as Algorithm \ref{alg:invPPR}) of Andersen et.~al.~\cite{Andersen2007}.  Our \bippr{} algorithm is based on the observation that in order to estimate $\pi_s[t]$ for a \emph{particular} $(s, t)$ pair, we can boost the accuracy of \rpush{} by sampling random walks from $s$ and incorporating the residual values at their endpoints.

\subsection{The \bippr{} Algorithm}
The reverse work from $t$ is done via the \rpush{}$(t, \rmax, \alpha)$, where $\rmax$ is a parameter we discuss more in Section \ref{sec:balancing}.  Recall from Section \ref{sec:reverse_push} that this algorithm produces two non-negative vectors $p^t \in \R^n$ (estimates) and $r^t \in \R^n$ (residual messages) which satisfy the following invariant (Lemma $1$ in \cite{Andersen2007})
\begin{equation}
  \label{eq:ppr_dot_product}
  \pi_s[t] = p^t[s] + \sum_{v \in V} \pi_s[v] r^t[v].
\end{equation}
The central idea is to reinterpret Equation \eqref{eq:ppr_dot_product} as an expectation:
\begin{equation}
  \label{eq:ppr_expectation}
  \pi_s[t] = p^t[s] + \EE_{v \sim \pi_s}[r^t[v]]. \end{equation}
Now, since $\max_v r^t[v] \leq \epr$, the expectation $\EE_{v \sim \pi_s[v]}[r^t[v]]$ can be efficiently estimated using Monte Carlo.  
To do so, we generate $w = c \frac{\epr}{\delta}$ random walks of length $Geometric(\alpha)$ from start node $s$. This choice of $w$ is  based on Theorem \ref{thm:bidirmain}, which provides a value for $c$ based on the desired accuracy, and here $\delta$ is the minimum PPR value we want to accurately estimate.  
Let $V_i$ be the final node of the $i$th random walk. From Equation \ref{eq:defmc}, $\Pr[V_i=v] = \pi_s[v]$.  Our algorithm returns the natural emperical estimate for Equation \ref{eq:ppr_expectation}:
%Let $X_i = r^t[V_i]$ denote the residual from the final node of the $i$th random walk, and $\bar{X} = \frac{1}{w} \sum_{i=1}^w X_i$. 
%Then \bippr{} returns as an estimate of $\pi_s[t]$: 
\[\widehat{\pi}_s[t] = p^t[s] + \frac{1}{w}\sum_{i=1}^w r^t[V_i]
\]
The complete pseudocode is given in Algorithm \ref{alg:BIPPR}.

\begin{algorithm}[!ht]
\caption{\bippr{}$(s,t,\delta)$}
\label{alg:BIPPR}
\begin{algorithmic}[1] 
\REQUIRE graph $G$, teleport probability $\alpha$, start node $s$, target node $t$, minimum probability $\delta$, accuracy parameter $c$ (in our experiments we use $c = 7$)

\STATE $(p_t, r_t)$ =  \rpush{}$(t,\epr, \alpha)$

\STATE Set number of walks $w=c \epr/\delta$ \quad(cf. Theorem \ref{thm:bidirmain})
\FOR{index $i\in [w]$}
\STATE Sample a random walk starting from $s$ (sampling a start from $s$ if $s$ is a distribution), stopping after each step with probability $\alpha$, and let $V_i$ be the endpoint
%\STATE Set $X_i = r_t[V_i]$
\ENDFOR 
\RETURN $\widehat{\PR}_s[t]=p_t[s] + (1/w)\sum_{i\in[w]} r_t[V_i]$
\end{algorithmic}
\end{algorithm}

\begin{figure}[tbph]
  \centering
  \includegraphics[width=0.9\columnwidth]{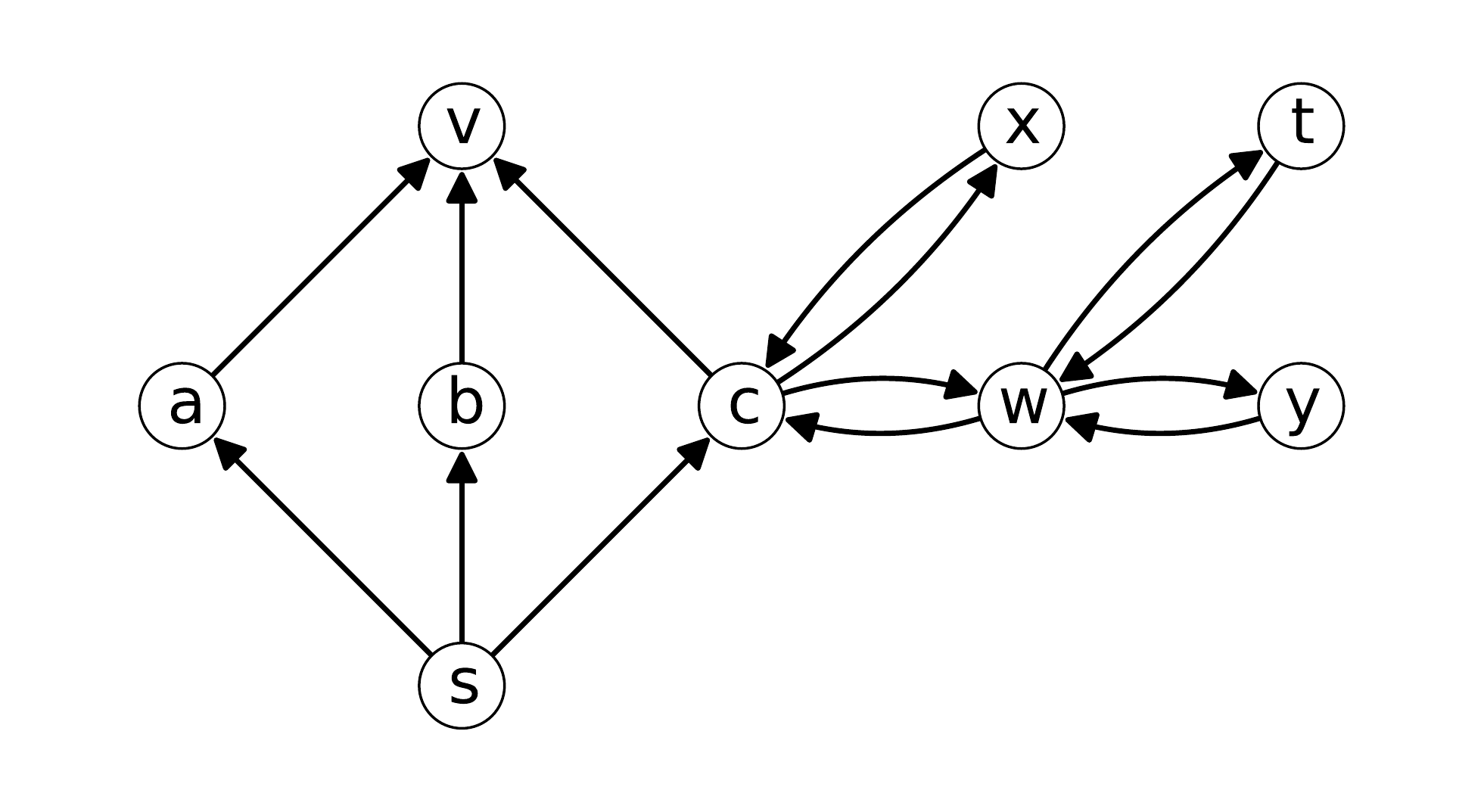}
  \caption{An example graph.  We will estimate $\pi_s[t]$.}
  \label{fig:example_9node_graph}
\end{figure}

\begin{figure}[tbph]
  \centering
  \includegraphics[width=0.9\columnwidth]{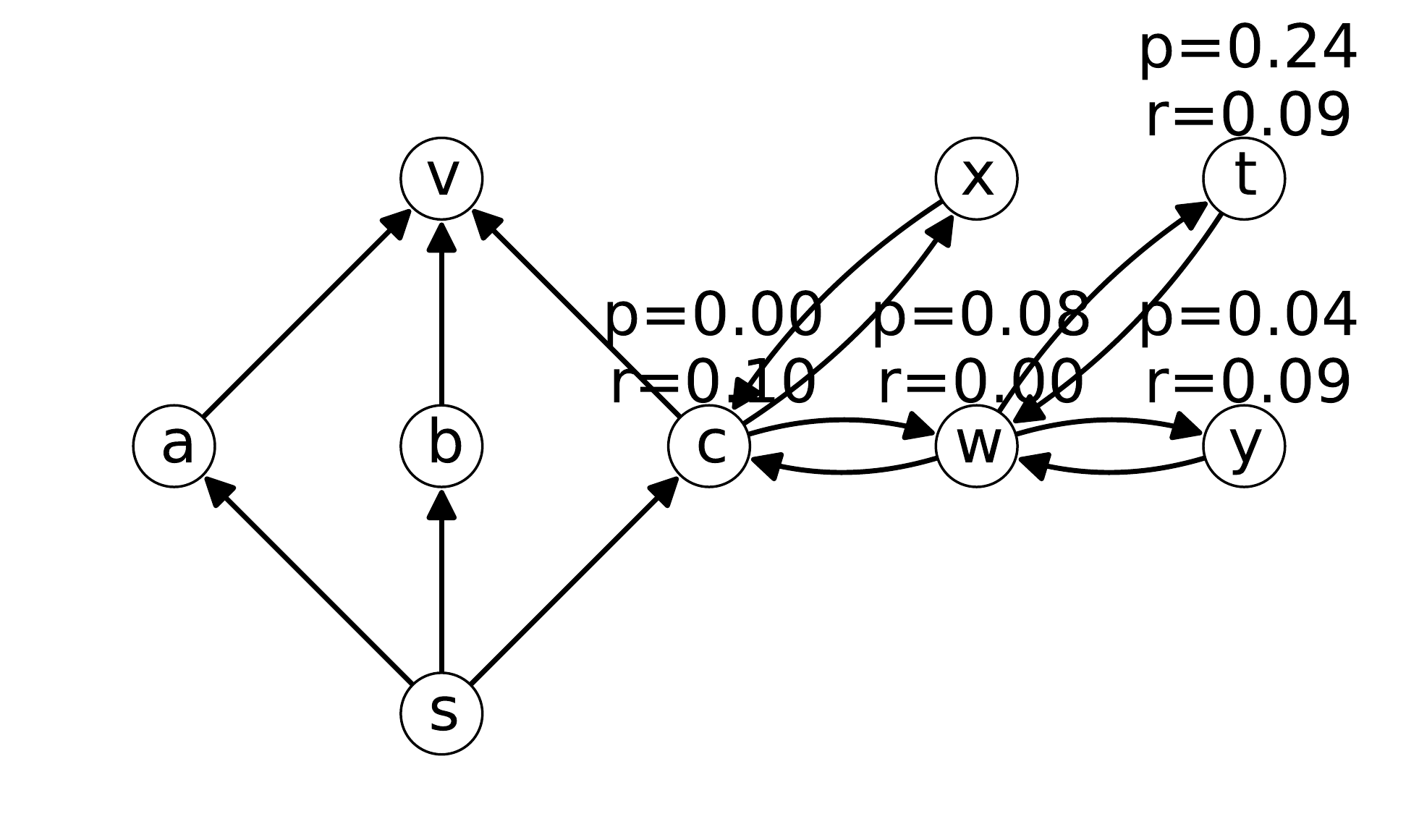}
  \caption{An example graph.  We will estimate $\pi_s[t]$.}
  \label{fig:example_9node_iteration5}
\end{figure}

\begin{figure}[tbph]
  \centering
  \includegraphics[width=0.9\columnwidth]{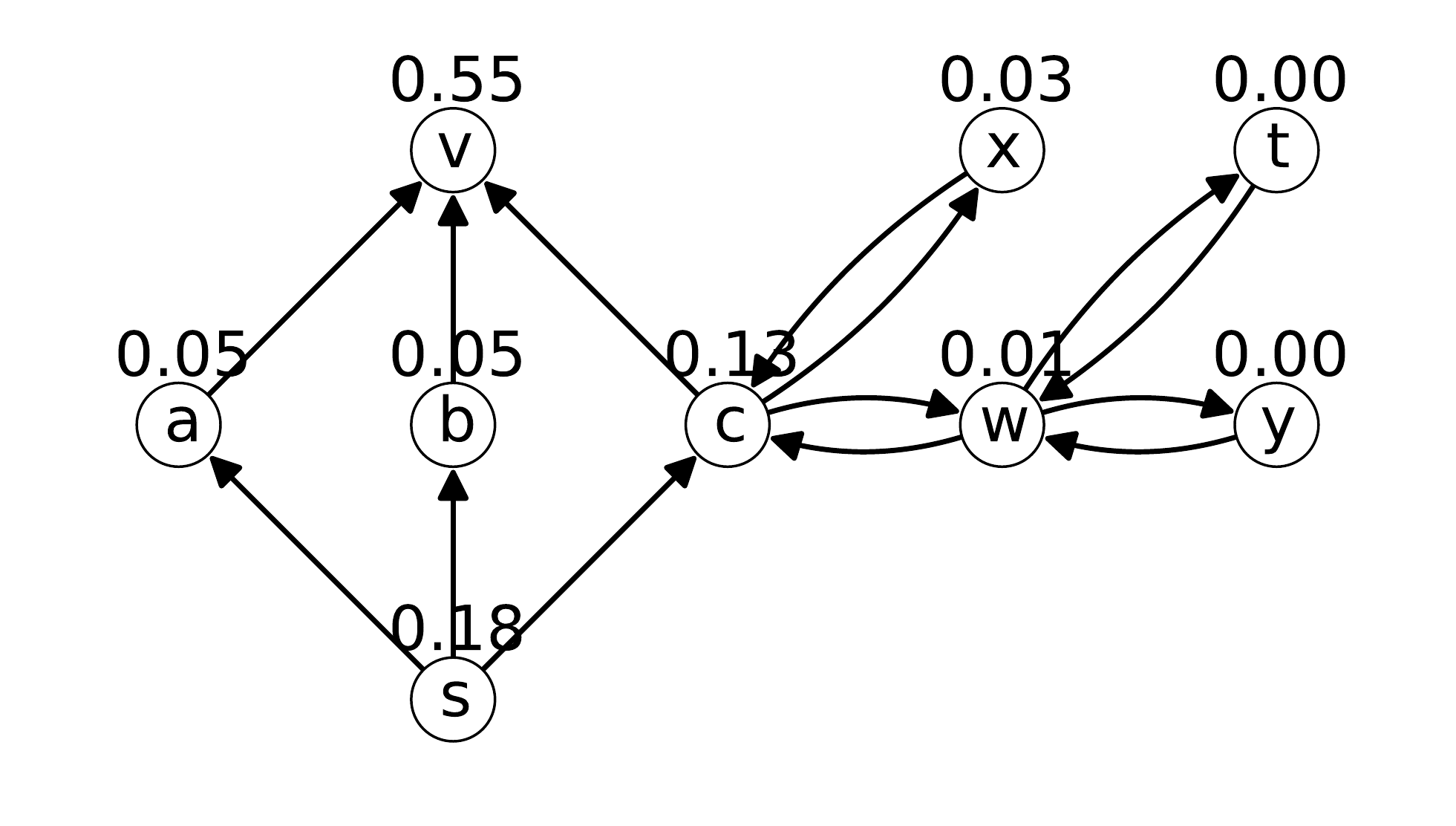}
  \caption{The empirical probability of stopping at each node after 77 walks from $s$ for some choice of random seed.  Note there is an implicit self loop at $v$.}
  \label{fig:example_9node_pp}
\end{figure}

As an example, consider the graph in Figure \ref{fig:example_9node_graph}.
Suppose we run \rpush{} on this graph with target $t$ and teleport probability $\alpha$ to accuracy $\rmax=0.11$.  The resulting PPR estimates $p$ and residuals $r$ are shown in Figure \ref{fig:example_9node_iteration5}.
In the second phase of the algorithm we take random walks forwards. If we set minimum PPR $\delta=0.01$ and chernoff constant, $c=7$, the number of walks we need to do is 
$ w = c \frac{\rmax}{\delta} = 77$
If we perform 77 walks, we might get the empirical forward probabilities $\tilde{\pi}_s$ shown in Figure \ref{fig:example_9node_pp}.
If we look at the residuals $r$ from Figure \ref{fig:example_9node_iteration5} and the forward probabilities $\tilde{\pi}_s$ in Figure \ref{fig:example_9node_pp} we see the only node with nonzero value on both happens to be $c$, so our estimate for this run of the algorithm is
\[\hat{\pi}_s[t] = p^t[s] + \sum_v \tilde{\pi}_s[v] r^t[v] = 0 + 0.13 \cdot 0.1 = 0.013. \]

%\subsection{Analysis}
\subsection{Accuracy Analysis}
We first prove that  high probability, \bippr{} returns an estimate with small relative error if $\pi_s[t]$ is above a given threshold, and returns an estimate with small additive error otherwise.
\begin{theorem}
\label{thm:bidirmain} 
Given start node $s$ (or source distribution $\sigma$), target $t$, minimum PPR $\delta$, maximum residual $\epr > \frac{2e \delta}{\alpha \epsilon}$, relative error $\epsilon \leq 1$, and failure probability $\pfail$, \bippr{} outputs an estimate $\widehat{\PR}_s[t]$ such
that with probability at least $1- \pfail$ the following hold: 
\begin{itemize}[nosep,leftmargin=*]
\item If $\pi_s[t] \geq \delta$: \hspace{1cm} $\abs{\pi_s[t]-\hat{\pi}_s[t]} \leq \epsilon \pi_s[t]$. 
\item If $\pi_s[t] \leq \delta$: \hspace{1cm}
%$\abs{\pi_s[t]-\hat{\pi}_s[t]} \leq \frac{\epsilon^2}{3 \ln(2)} \delta$.
$\abs{\pi_s[t]-\hat{\pi}_s[t]} \leq 2e\delta$.
\end{itemize}
\end{theorem}
The above result shows that the estimate $\hat{\pi}_s[t]$ can be used to distinguish between `significant' and `insignificant' PPR pairs: for pair $(s,t)$, Theorem \ref{thm:bidirmain} guarantees that if $\pi_s[t] \geq \frac{(1+2e)\delta}{(1-\epsilon)}$, then the estimate is greater than $(1+2e)\delta$, whereas if $\pi_s[t] < \delta$, then the estimate is less than $(1+2e)\delta$. 
The assumption $\epr > \frac{2e \delta}{\alpha \epsilon}$ is easily satisfied, as typically $\delta=O\pn{\frac{1}{n}}$ and $\rmax =\Omega\pn{\frac{1}{\sqrt{m}}}$.

\begin{proof} 
%(Based on the proof in \cite{generalized_fastppr}).
As shown in Algorithm \ref{alg:BIPPR}, we will average over 
\[ w = c \frac{\epr}{\delta}, \]
walks, where $c$ is a parameter we choose later.
Each walk is of length $Geometric(\alpha)$, and we denote $V_i$ as the last node visited by the $i^{th}$ walk so that $V_i\sim\pi_s$.
Now let $X_i = r^t[V_i]$.  The estimate returned by \bippr{} is:
\[\widehat{\pi}_s[t] = p^t[s] + \frac{1}{w} \sum_{i=1}^w X_i. \]
First, from Equation \eqref{eq:ppr_dot_product}, we have that $\EE[\widehat{\pi}_s[t]]= \pi_s[t]$, that is, our estimate is unbiased.   
Since \rpush{} guarantees that for all $v$, $r^t[v]<\epr$, each $X_i$ is bounded in $[0,\epr]$.  Before we apply Chernoff bounds, we rescale $X_i$ by defining $Y_i = \frac{1}{\epr} X_i \in [0,1]$. We also define $Y = \sum_{i=1}^w Y_i$

We show concentration of the estimate via the following two Chernoff bounds (see Theorem $1.1$ in~\cite{DuPa09}):
\begin{enumerate}
\item $\PP[|Y - \EE[Y]| > \epsilon \EE[Y]] < 2 \exp(-\frac{\epsilon^2}{3}\EE[Y])$
\item $\textrm{For any } b > 2e\EE[Y], \PP[Y > b] \leq 2^{-b}$
\end{enumerate}
We perform a case analysis based on whether $\EE[X_i] \geq \delta$ or $\EE[X_i] < \delta$. 

First suppose $\EE[X_i] \geq \delta$.  Note that this implies that $\pi_s[t] \geq \delta$ so we will prove a relative error bound of $\epsilon$.
Now we have $\EE[Y] = \frac{w}{\epr} \EE[X_i] = \frac{c}{\delta} \EE[X_i] \geq c$, and thus:
\begin{align*}
\PP[\abs{\widehat{\pi}_s[t] - \pi_s[t]} > \epsilon \pi_s[t]] 
 &\leq \PP[\abs{\bar{X} - \EE[X_i]} > \epsilon \EE[X_i]] \\
  &= \PP[\abs{Y - \EE[Y]} > \epsilon \EE[Y]] \\
  &\leq 2 \exp\pn{-\frac{\epsilon^2}{3}\EE[Y]} \\
  &\leq 2 \exp\pn{-\frac{\epsilon^2}{3} c}
  \leq \pfail,
\end{align*}
where the last line holds as long as we choose
\[ c \geq \frac{3}{\epsilon^2} \ln \pn{\frac{2}{\pfail}}.\]
%This case is concluded because if $\abs{\bar{X} - \EE[X_i]} \leq \epsilon \EE[X_i]$ then $\abs{\widehat{\pi_s[t]} - \pi_s^t} \leq \epsilon \pi_s^t$.

Suppose alternatively that $\EE[X_i] < \delta$.  Then
\begin{align*}
\PP[\abs{\hat{\pi}_s[t] - \pi_s[t]} > 2e\delta]
&= \PP[\abs{\bar{X} - \EE[X_i]} > 2e\delta] \\
%&=  \PP\bk{\abs{\frac{\epr}{w} Y - \frac{\epr}{w} \EE[Y]} > 2e\delta} \\
&=  \PP\bk{\abs{Y -  \EE[Y]} > \frac{w}{\epr} 2e\delta} \\
&\leq  \PP\bk{Y > \frac{w}{\epr}2e\delta} .
\end{align*}
At this point we set $b = \frac{w}{\epr} 2e\delta=2ec$ and apply the second Hoeffding bound.  Note that $\EE[Y] = \frac{c}{\delta} \EE[X_i] < c$, and hence we satisfy $b > 2e \EE[Y]$.
The second bound implies that
\begin{equation}
  \label{eq:1}
   \PP[\abs{\hat{\pi}_s[t] - \pi_s[t]} > 2e\delta] \leq 2^{-b} \leq \pfail
\end{equation}
as long as we choose $c$ such that:
\[ c \geq \frac{1}{2e} \log_2 \frac{1}{\pfail}. \]
If  $\pi_s[t] \leq \delta$, then equation \ref{eq:1} completes our proof. 

% where the last line holds if we choose
% \[ c = \frac{3}{\epsilon^2} \ln \pn{\frac{2}{\pfail}}.\]
% %This case is concluded because if $\abs{\bar{X} - \EE[X_i]} \leq \epsilon \EE[X_i]$ then $\abs{\widehat{\pi_s[t]} - \pi_s^t} \leq \epsilon \pi_s^t$.

% Suppose alternatively that $\EE[X_i] < \delta$.  Then
% \begin{align*}
% \PP[\abs{\hat{\pi}_s[t] - \pi_s[t]} > \frac{\epsilon^2}{3 \ln(2)} \delta]
% &= \PP[\abs{\bar{X} - \EE[X_i]} > \frac{\epsilon^2}{3 \ln(2)} \delta] \\
% %&=  \PP\bk{\abs{\frac{\epr}{w} Y - \frac{\epr}{w} \EE[Y]} > 2e\delta} \\
% &=  \PP\bk{\abs{Y -  \EE[Y]} > \frac{w}{\rmax}\frac{\epsilon^2}{3 \ln(2)} \delta} \\
% &\leq  \PP\bk{Y > \frac{w}{\rmax}\frac{\epsilon^2}{3 \ln(2)} \delta } .
% \end{align*}
% At this point we set $b = \frac{w}{\rmax}\frac{\epsilon^2}{3 \ln(2)} \delta = \frac{c \epsilon^2}{3 \ln(2)} = \frac{\ln \pn{\frac{2}{\pfail}}}{\ln(2)} \geq -\log_2(\pfail)$ and apply the second Hoeffding bound.  Note that (todo) $\EE[Y] = \frac{c}{\delta} \EE[X_i] < c$, and hence we satisfy $b > 2e \EE[Y]$.
% The second bound implies that
% \begin{equation}
%   \label{eq:1}
%    \PP[\abs{\hat{\pi}_s[t] - \pi_s[t]} > \frac{\epsilon^2}{3 \ln(2)} \delta ] \leq 2^{-b} \leq \pfail
% \end{equation}
% as long as we choose $c$ such that:
% \[ c \geq \frac{1}{2e} \log_2 \frac{1}{\pfail}. \]
%If  $\pi_s[t] \leq \delta$, then equation \ref{eq:1} completes our proof. 

The only remaining case is when $\pi_s[t] > \delta$ but $\EE[X_i] < \delta$. This implies that $p^t[s] > 0$ since $\pi_s[t] = p^t[s] + \EE[X_i]$.  In the \rpush{} algorithm when we increase $p^t[s]$, we always increase it by at least $\alpha \rmax$, so we have $\pi_s[t] \geq \alpha \rmax$.  We have that
\[ \frac{\abs{\hat{\pi}_s[t] - \pi_s[t]}}{\pi_s[t]} \leq \frac{\abs{\hat{\pi}_s[t] - \pi_s[t]}}{\alpha \rmax}. \]
By assumption, $\frac{2e \delta}{\alpha \rmax} < \epsilon$, so by equation \ref{eq:1},
\[ \PP\bk{\frac{\abs{\hat{\pi}_s[t] - \pi_s[t]}}{\pi_s[t]} > \epsilon} \leq \pfail \]

The proof is completed by combining all cases and choosing $c = \frac{3}{\epsilon^2} \ln \pn{\frac{2}{\pfail}}$.  
We note that the constants are not optimized; in experiments we find that $c=7$ gives mean relative error less than 8\% on a variety of graphs.
\end{proof}

\subsection{Running Time Analysis}
\label{sec:bippr_running_time}

The runtime of \bippr{} depends on the target $t$: if $t$ has many in-neighbors and/or large global PageRank $\pi[t]$, then the running time will be slower because the reverse pushes will iterate over a large number of nodes.  For example, in the worst case, we might have $\indegree{t} = \Theta(n)$ and \bippr{} takes $\Theta(n)$ time.  However, we can give an average case analysis, and in Section \ref{sec:parameterized_bippr} we give a back-of-the-envelope  parameterized running time estimate.

\subsubsection{Average Case Running Time Analysis}
Alternatively, for \emph{a uniformly chosen target node}, we can prove the following:
\begin{theorem}
\label{thm:fastpprtime} 
For any start node $s$ (or source distribution $\sigma$), minimum PPR $\delta$, maximum residual $\epr$, relative error $\epsilon$, and failure probability $\pfail$, if the target $t$ is chosen uniformly at random, then \emph{\bippr{}} has expected running time
\[ O\pn{ \sqrt{\frac{\bar{d}}{\delta}} \frac{\sqrt{\log\pn{1/\pfail}}}{\alpha \epsilon}}.
 \]
\end{theorem}

In contrast, the running time for \texttt{Monte-Carlo} to achieve the same accuracy guarantee is $O\pn{\frac{1}{\delta} \frac{\log\pn{1/\pfail}}{\alpha \epsilon^2}}$, and the running time for \rpush{} is $O\pn{\frac{\bar{d}}{\delta \alpha}}$.  The \texttt{FAST-PPR} algorithm of \cite{fastppr} has an average running time bound of
$O\pn{\frac{1}{\alpha\epsilon^2}\sqrt{\frac{\bar{d}}{\delta}} \sqrt{ \frac{\log\pn{1/\pfail} \log\pn{1/\delta}}{\log\pn{1/(1-\alpha)}} }}$
 for uniformly chosen targets.  
The running time bound of \bippr{} is thus asymptotically better than \texttt{FAST-PPR}, and in our experiments (Section \ref{sec:bippr_experimental_time}) \bippr{} is significantly faster than past algorithms. %In experiments we find that \bippr{} is 3 to 8 times faster on a diverse set of graphs.
% Typical values might be $\pfail = 10^{-6}$, $\epsilon = \frac{1}{10}$ (since PPR values vary over a large range, even a constant relative error provides allows them to be ranked accurately), and $\delta = \frac{1}{n}$ (this is the average value of $\pi_s$, so if $\pi_s[t] < \frac{1}{n}$ then $t$ is not very relevant to $s$).  

\begin{proof} 
We first show that for a uniform random $t$, \rpush{} runs in average time $\frac{\bar{d}}{\alpha \epr}$ where $\bar{d}$ is the average degree of a node. This is because each time we push from a node $v$, we increase $p^t[v]$ by at least $\alpha \rmax$.  Since $p^t[v] \leq \pi_v[t]$ and $\sum_t \pi_v[t] = 1$, this lets us bound the average running time:
\begin{align*}
 &\frac{1}{n} \sum_{t \in V} \sum_{v \in V} \indegree{v}  [\text{\# times $v$ is pushed during \rpush{}($t, \rmax, \alpha$)}] \\
 &\leq \frac{1}{n} \sum_{t \in V} \sum_{v \in V} \indegree{v} \frac{\pi_v[t]}{\alpha \rmax} \\
 &\leq \frac{1}{n} \sum_{v \in V} \indegree{v} \frac{1}{\alpha \rmax} \\
& \leq \frac{\dbar}{\alpha \rmax}.
\end{align*}
  On the other hand, from Theorem \ref{thm:bidirmain}, the number of walks generated is
$O\pn{ \frac{\epr}{\delta\epsilon^2} \ln\pn{1/\pfail}}$ random walks, each of which can be sampled in average time $1/\alpha$.  Finally, we choose $\epr = \epsilon \sqrt{\frac{\bar{d} \delta}{ \ln\pn{2/\pfail}}}$ to minimize our running time bound and get the claimed result. 
\end{proof}

\subsubsection{A Practical Improvement: Balancing Forward and Reverse Running Time}
\label{sec:balancing}

The dynamic runtime-balancing heuristic proposed in \cite{fastppr} can improve the running time of Bidirectional-PageRank in practice.  In this technique, $\epr$ is chosen dynamically in order to balance the amount of time spent by \rpush{} and the amount of time spent generating random walks. 
To implement this, we modify \rpush{}  to use a priority queue in order to always push from the node $v$ with the largest value of $r_t[v]$.  
Then we change the while loop so that it terminates when the amount of time spent achieving the current value of $\epr$ first exceeds the predicted amount of time required for sampling random walks, $c_{\text{walk}}  \cdot c \cdot \frac{\epr}{\delta}$, where $c_{\text{walk}}$ is the average time it takes to sample a random walk.  For a plot showing how using a fixed value for $\rmax$ results in unbalanced running time for many targets, and how this variant empirically balances the reverse and forward times, see \cite{fastppr}.  
The complete pseudocode is given as Algorithm \ref{alg:invPPRBalanced} 
\begin{algorithm}[!ht]
\caption{\texttt{\rpush{}Balanced}$(G, t, \alpha, \delta)$~\cite{Andersen2007}}
\label{alg:invPPRBalanced}
\begin{algorithmic}[1]
\REQUIRE graph $G$ with edge weights $w_{u,v}$, teleport probability $\alpha$, target node $t$, minimum probability $\delta$, accuracy parameter $c$ (in our experiments we use $c = 7$)
\STATE Choose tuning constant $c_{\text{walk time}}$, the average time it takes to generate a walk.
\STATE Initialize (sparse) estimate-vector $p^t = \vec{0}$ and (sparse) residual-vector $r^t = e^t$ (i.e.~$r^t(v) = 1$ if $v=t$; else $0$)
\STATE Initial Max-Priority-Queue $q$ with priority $r^t$
\STATE Start stopwatch $w$
\STATE Define function forwardTime($\rmax$) = $ c_{\text{walk time}} c \frac{\rmax}{\delta}$
\WHILE{$Q$.nonEmpty() and $w\text{.elapsedTime()} < \text{forwardTime($Q$.maxPrioriority())}$}
\STATE Let $v = Q.\text{popMax()}$
\FOR{$u\in \inneighbors{v}$}
   \STATE $r^t(u) \pluseq (1 - \alpha) w_{u,v} r^t(v)$ (increasing $u$'s priority in $Q$)
\ENDFOR
\STATE      $p^t(v) \pluseq \alpha r^t(v)$
\STATE      $r^t(v)=0$ 
\ENDWHILE
\STATE $\rmax = Q$.maxPriority() if $Q$ is non-empty, otherwise $\rmax=0$
\RETURN $(p^t, r^t, \rmax)$
\end{algorithmic}
\end{algorithm}    
and Algorithm \ref{alg:BIPPR_balanced}.
\begin{algorithm}[!ht]
\caption{\texttt{Bidirectional-PPR-Balanced}$(s,t,\delta)$}
\label{alg:BIPPR_balanced}
\begin{algorithmic}[1] 
\REQUIRE graph $G$, teleport probability $\alpha$, start node $s$, target node $t$, minimum probability $\delta$, accuracy parameter $c$ (in our experiments we use $c = 7$)

\STATE $(p^t, r^t, \rmax)$ =  \texttt{ReversePushBalanced}$(G, t, \alpha, \delta)$

\STATE Set number of walks $w=c \epr/\delta$ \quad(for $c$'s value, see Theorem \ref{thm:bidirmain}; we use $c=7$ in experiments)
\FOR{index $i\in [w]$}
\STATE Sample a random walk starting from $s$ (sampling a start from $s$ if $s$ is a distribution), stopping after each step with probability $\alpha$; let $V_i$ be the endpoint
\STATE Set $X_i = r^t[V_i]$
\ENDFOR 
\RETURN $\widehat{\PR}_s[t]=p^t[s] + (1/w)\sum_{i\in[w]}X_i$
\end{algorithmic}
\end{algorithm}

\subsubsection{Choosing the Minimum Probability  $\delta$}
\label{sec:choosing_delta} 
%\inlineheading{Choosing $\delta$}
So far we have assumed that the minimum probability $\delta$ is either given or is $O\pn{\frac{1}{n}}$, but a natural choice in applications when estimating $\pi_s[t]$ is to set $\delta = \pi[t]$.  The motivation for this is that if $\pi_s[t] < \pi[t]$, then $s$ is less interested in $t$ than a random source is, so it is not useful to quantify just how (un)interested $s$ is in $t$.  %(More generally we can set $\delta = a \pi[t]$ for some constant $a$.)
%  If we set $\delta = \pi[t]$ we get running time
%\[O \pn{ \frac{1}{\alpha \epsilon}\sqrt{ c m }}. \]

\subsubsection{Generalization}
Bidirectional-PageRank extends naturally to generalized PageRank using a source distribution $\sigma$ rather than a single start node -- we simply sample an independent starting node from $s$ for each walk, and replace $p^t[s]$ with %the expected value of $p^t[s]$ when $s$ is sampled from the starting distribution
$\sum_{v \in V} p^t[v] s[v]$ where $s[v]$ is the weight of $v$ in the start distribution $s$.  It also generalizes naturally to weighted graphs if we simply use the weights when sampling neighbors during the random walks.  For a generalization to other walk distributions or to any given walk length, see Chapter \ref{sec:mc_chapter}.

\section{Undirected-Bidirectional-PPR}
\label{sec:ubippr}
Here we present an alternative bidirectional estimator for undirected graphs \cite{ubippr}.
\subsection{Overview} 
The \bippr{} presented above is based on the \rpush{} algorithm (introduced in Section \ref{sec:reverse_push}) from \cite{Andersen2007} which satisfies the loop invariant
\begin{equation}
  \label{eq:loop_invariant}
  \pi_s[t] = p^t[s] + \sum_{v \in V} \pi_s[v] r^t[v].
\end{equation}
However, an alternative local algorithm is the \fpush{} algorithm (see Section \ref{sec:forward_push}) of Andersen et al.~\cite{Andersen2006}, which returns estimates\footnote{We use the superscript $p^t \in \R^n$ vs subscript $p_s \in \R^n$ to indicate which type of estimate vector we are referring to (reverse or forward, respectively).  We use a similar convention for reverse residual vector $r^t \in \R^n$ to target $t \in V$ and forward residual vector $r^s \in \R^n$ from source $s \in V$.} $p_s$ and residuals $r_s$ which satisfy the invariant
\begin{equation}
  \label{eq:forward_push}
 \pi_s[t] = p_s[t] + \sum_{v \in V} r_s[v] \pi_v[t],\qquad\fall t\in V .  
\end{equation}
On undirected graphs this invariant is the basis for an alternative algorithm for personalized PageRank which we call \ubippr{}.  In a preliminary experiment on one graph, this algorithm has comparable running time to \bippr{}, but it is interesting because because of its symmetry to \bippr{} and because it enables a parameterized worst-case analysis.  We now present our new bidirectional algorithm for PageRank estimation in undirected graphs.

%\subsection{Preliminaries}

%We consider an undirected graph $G(V,E)$, with $n$ nodes and $m$ edges.
%For ease of notation, we henceforth consider unweighted graphs, and focus on the simple case where $\sigma = \vecE_s$ for some single node $s$. 
%We note however that all our results extend to weighted graphs and any source distribution $\sigma$ in a straightforward manner.% (with some additional notation and assumptions).

%Recall that in a weighted directed graph, the normalized adjacency matrix obeys $W=D^{-1}A$, where $A$ is the matrix of edge weights $w_{ij}$, and $D = diag(d_i)$ is a diagonal-matrix with $d_i:=\sum_jw_{ij}$. 
%In an un-weighted graph (i.e., if all edge-weights are $1$), then $A$ is the adjacency matrix and $d_i$ the out-degree of node $i$. The case where $A$ is symmetric corresponds to a (weighted) undirected graph. 

\subsection{A Symmetry for PPR in Undirected Graphs}

%Consider an undirected graph $G(V,E)$, with $n$ nodes and $m$ edges. The edges have associated edge-weights $w_{ij}\fall(i,j)\in E$ (and $w_{ij}=w_{ji}$); recall that we define $d_i=\sum_j w_{ij}\fall i\in V$. The probability that a random-walk transitions from node $i$ to node $j$ is then given by $(D^{-1}A)_{ij} = w_{ij}/d_i$, and from $j$ to $i$ by $w_{ij}/d_j$. 

The \ubippr{} Algorithm critically depends on an underlying \emph{reversibility property} exhibited by PPR vectors in undirected graphs. 
This property, stated before in several earlier works \cite{avrachenkov2013,Grolmusz2015}, is a direct consequence of the reversibility of random walks on undirected graphs.
To keep our presentation self-contained, we present this property, along with a simple probabilistic proof, in the form of the following lemma:
\begin{lemma} 
\label{undirected_ppr_lemma} 
Given any undirected graph $G$, for any teleport probability $\alpha \in (0,1)$ and for any node-pair $(s, t) \in V^2$, we have:
\[ \pi_s[t] =  \frac{d_t}{d_s} \pi_t[s]. \]  
\end{lemma}

\begin{proof}
For path $P=\{s,v_1,v_2,\ldots,v_k,t\}$ in $G$, we denote its length as $\ell(P)$ (here $\ell(P) = k+1$), and define its reverse path to be $\overline{P}=\{t,v_k,\ldots,v_2,v_1,s\}$ -- note that $\ell(P) = \ell(\overline{P})$. 
Moreover, we know that a random-walk starting from $s$ traverses path $P$ with probability 
$\PP[P]= \frac{1}{d_s}\cdot \frac{1}{d_{v_1}}\cdot \ldots \cdot \frac{1}{d_{v_k}}$,
and thus, it is easy to see that we have:
\begin{equation}
\label{eq:path}
\PP[P]\cdot d_s = \PP[\overline{P}]\cdot d_t 
\end{equation}
Now let $\mathcal{P}_{st}$ denote the set of paths in $G$ starting at $s$ and terminating at $t$. Then we can re-write Eqn. \eqref{eq:defmc} as:
\begin{align*}
	\PR_s[t] = \sum_{P\in \mathcal{P}_{st}}\alpha(1-\alpha)^{\ell(P)}\PP[P] 
	= \sum_{\overline{P}\in \mathcal{P}_{ts}}\alpha(1-\alpha)^{\ell(\overline{P})}\PP[\overline{P}]
	= \frac{d_t}{d_s} \PR_t[s]\quad\square
\end{align*}
\end{proof}

%\begin{proof}[of Lemma \ref{undirected_ppr_lemma}]
%We simply use the random walk interpretation of PageRank to write down the probability of a walk occurring as a sum over the set of all $s$-$t$ paths of a given length $l$.  Let $P_l = \{(p_0, p_1, \ldots, p_l) \in V^{l+1}: p_0=s, p_l=t, \forall i \in \{0,\ldots, l-1\} (p_i, p_{i+1}) \in E\}$. Then
%\begin{align*}
% \pi_v[t] &= \sum_{l=0}^{\infty} (1-\alpha)^l \alpha \sum_{p \in P_l} \prod_{i=0}^{l-1} \frac{1}{d_{p_i}}  \\
% &= \frac{d_t}{d_v} \sum_{l=0}^{\infty} (1-\alpha)^l \alpha \sum_{p \in P_l} \prod_{i=1}^{l} \frac{1}{d_{p_i}}  \\
%&=  \frac{d_t}{d_v} \pi_t[v].   \quad\square
%\end{align*}
%\end{proof}

\subsection{The \ubippr{} Algorithm} 
\label{sec:undirected_bippr}
At a high level, the \ubippr{} algorithm has two components:
\begin{itemize}[nosep,leftmargin=*]
\item \textbf{Forward-work}: Starting from source $s$, we first use a forward local-update algorithm, the \fpush{}$(G,\alpha,s,\epr)$ algorithm of Andersen et al.~\cite{Andersen2006} (given in section \ref{sec:forward_push} as Algorithm \ref{alg:forwardPush}). This procedure begins by placing one unit of ``residual'' probability-mass on $s$, then repeatedly selecting some node $u$, converting an $\alpha$-fraction of the residual mass at $u$ into probability mass, and pushing the remaining residual mass to $u$'s neighbors.  For any node $u$, it returns an estimate $p_s[u]$ of its PPR $\PR_s[u]$ from $s$ %with additive error at most $d_u\epr$.
 as well as a residual $r_s[u]$ which represents un-pushed mass at $u$.
\item \textbf{Reverse-work}:  
We next sample random walks of length $L\sim Geometric(\alpha)$ starting from $t$, and use the residual at the terminal nodes of these walks to compute our desired PPR estimate. Our use of random walks backwards from $t$ depends critically on the symmetry in undirected graphs presented in Lemma \ref{undirected_ppr_lemma}.
\end{itemize}
Note that this is in contrast to \texttt{FAST-PPR} and \bippr{}, which performs the local-update step in reverse from the target $t$, and generates random-walks forwards from the source $s$.

In more detail, %our forward-work component is based on the local update algorithm of \cite{Andersen2006} (shown here as Algorithm \ref{alg:forwardPush}) which works forwards from a start $s$ to compute estimates $p_s \in \R^n$ of $\pi_s$ and residuals $r_s \in \R^n$ which represent mass not yet pushed forward.
our algorithm will choose a maximum residual parameter $\epr$, and apply the local push operation in Algorithm \ref{alg:forwardPush} until for all $v$, $r_s[v]/d_v < \epr$. Andersen et al.~\cite{Andersen2006} prove that their local-push operation preserves the following invariant for vectors $(p_s,r_s)$:
\begin{equation}
  \label{eq:forward_invariant}
 \pi_s[t] = p_s[t] + \sum_{v \in V} r_s[v] \pi_v[t],\qquad\fall t\in V .  
\end{equation}
Since we ensure that $\forall v, r_s[v]/d_v < \epr$, it is natural at this point to use the symmetry Lemma \ref{undirected_ppr_lemma} and re-write this as:
\[ \pi_s[t] = p_s[t] + d_t \sum_{v \in V}  \frac{r_s[v]}{d_v} \pi_t[v]. \]
Now using the fact that $\sum_t \pi_v[t]=n\pi[t]$ get that $\fall t\in V$, 
\begin{equation}
  \label{eq:forward_push_error}
 \left|\pi_s[t] - p_s[t]\right|  \leq \epr d_t n \pi[t].  
\end{equation}

However, we can get a more accurate estimate by using the residuals.  
The key idea of our algorithm is to re-interpret this as an expectation:
\begin{align}
\label{eq:estimate}
\pi_s[t] = p_s[t] + d_t  \EE_{V \sim \pi_t} \bk{\frac{r_s[v]}{d_V} }.
\end{align}
We estimate the expectation using standard Monte-Carlo.  Let $V_i \sim \pi_t$ and $X_i = r_s[V_i] d_t/d_{V_i}$, so we have $\pi_s[t] = p_s[t] + \EE[X]$. Moreover, each sample $X_i$ is bounded by $d_t \epr$ (this is the stopping condition for \fpush{}), which allows us to efficiently estimate its expectation.
To this end, we generate $w$ random walks, where
\[ w = \frac{c}{\epsilon^2} \frac{\epr}{\delta / d_t}.
 \]
The choice of $c$ is specified in Theorem \ref{thm:accuracy}. Finally, we return the estimate:
\[ \widehat{\pi}_s[t] = p_t[s] + \frac{1}{w} \sum_{i=1}^w X_i. \]
The complete pseudocode is given in Algorithm \ref{alg:UBIPPR}.

\begin{algorithm}[!ht]
\caption{\ubippr{}$(s,t,\delta)$}
\label{alg:UBIPPR}
\begin{algorithmic}[1] 
\REQUIRE graph $G$, teleport probability $\alpha$, start node $s$, target node $t$, minimum probability $\delta$, accuracy parameter $c= 3\ln \pn{2/\pfail}$ \quad(cf. Theorem \ref{thm:accuracy})

\STATE $(p_s, r_s)$ =  \fpush{}$(s,\epr)$

\STATE Set number of walks $w=c d_t \epr/(\epsilon^2\delta)$ 
\FOR{index $i\in [w]$}
\STATE Sample a random walk starting from $t$, stopping after each step with probability $\alpha$; let $V_i$ be the endpoint
\STATE Set $X_i = r_s[V_i] / d_{V_i}$
\ENDFOR 
\RETURN $\widehat{\PR}_s[t]=p_s[t] + (1/w)\sum_{i\in[w]}X_i$
\end{algorithmic}
\end{algorithm}

\subsection{Analyzing the Performance of \ubippr{}}

\subsubsection{Accuracy Analysis}
We first prove that \ubippr{} returns an unbiased estimate with the desired accuracy:

\begin{theorem}
\label{thm:accuracy} 
In an undirected graph $G$, for any source node $s$, minimum threshold $\delta$, maximum residual $\epr$, relative error $\epsilon$, and failure probability $\pfail$, Algorithm \ref{alg:UBIPPR} outputs an estimate $\widehat{\PR}_s[t]$ such
that with probability at least $1- \pfail$ we have: $\qquad\abs{\pi_s[t]-\hat{\pi}_s[t]} \leq \max\{\epsilon\pi_s[t],2e\delta\}$. 

%\begin{itemize}[nosep,leftmargin=*]
%\item If $\pi_s[t] \geq \delta$: \hspace{1cm} $\abs{\pi_s[t]-\hat{\pi}_s[t]} \leq \epsilon \pi_s[t]$. 
%\item If $\pi_s[t] \leq \delta$: \hspace{1cm}
%$\abs{\pi_s[t]-\hat{\pi}_s[t]} \leq 2e\delta$.
%\end{itemize}
\end{theorem}

%The above result shows that the estimate $\hat{\pi}_s[t]$ can be used to distinguish between `significant' and `insignificant' PPR pairs: for pair $(s,t)$, Theorem \ref{thm:accuracy} guarantees that if $\pi_s[t] \geq \frac{(1+2e)\delta}{(1-\epsilon)}$, then the estimate is greater than $(1+2e)\delta$, whereas if $\pi_s[t] < \delta$, then the estimate is less than $(1+2e)\delta$. 

The proof follows a similar outline as the proof of Theorem \ref{thm:bidirmain}.
For completeness, we sketch the proof here:
\begin{proof} 
As stated in Algorithm \ref{alg:UBIPPR}, we average over $w= cd_t\epr/\epsilon^2\delta$ walks, where $c$ is a parameter we choose later.
Each walk is of length $Geometric(\alpha)$, and we denote $V_i$ as the last node visited by the $i^{th}$ walk; note that $V_i\sim\pi_t$.
As defined above, let $X_i = r_s[V_i] d_t/d_{V_i}$; the estimate returned by \ubippr{} is:
\[\widehat{\pi}_s[t] = p_t[s] + \frac{1}{w} \sum_{i=1}^w X_i. \]
First, from Eqn. \eqref{eq:estimate}, we have that $\EE[\widehat{\pi}_s[t]]= \pi_s[t]$.   
Also, \fpush{} guarantees that for all $v$, $r_s[v]<d_v\epr$, and so each $X_i$ is bounded in $[0,d_t\epr]$; for convenience, we rescale $X_i$ by defining $Y_i = \frac{1}{d_t\epr} X_i$.

We now show concentration of the estimates via the following Chernoff bounds (see Theorem $1.1$ in~\cite{DuPa09}):
\begin{enumerate}[nosep]
\item $\PP[|Y - \EE[Y]| > \epsilon \EE[Y]] < 2 \exp(-\frac{\epsilon^2}{3}\EE[Y])$
\item $\textrm{For any } b > 2e\EE[Y], \PP[Y > b] \leq 2^{-b}$
\end{enumerate}
We perform a case analysis based on whether $\EE[X_i] \geq \delta$ or $\EE[X_i] < \delta$. 
First, if $\EE[X_i] \geq \delta$, then we have $\EE[Y] = \frac{w}{d_t\epr} \EE[X_i] = \frac{c}{\epsilon^2\delta} \EE[X_i] \geq \frac{c}{\epsilon^2}$, and thus:
\begin{align*}
\PP\left[\abs{\widehat{\pi}_s[t] - \pi_s[t]} > \epsilon \pi_s[t]\right] 
 &\leq \PP\left[\abs{\bar{X} - \EE[X_i]} > \epsilon \EE[X_i]\right]
  = \PP\left[\abs{Y - \EE[Y]} > \epsilon \EE[Y]\right] \\
  &\leq 2 \exp\pn{-\frac{\epsilon^2}{3}\EE[Y]} 
  \leq 2 \exp\pn{-\frac{c}{3} }
  \leq \pfail,
\end{align*}
where the last line holds as long as we choose  $c \geq 3 \ln \pn{2/\pfail}$.
%This case is concluded because if $\abs{\bar{X} - \EE[X_i]} \leq \epsilon \EE[X_i]$ then $\abs{\widehat{\pi_s[t]} - \pi_s_t} \leq \epsilon \pi_s_t$.

Suppose alternatively that $\EE[X_i] < \delta$.  Then:
\begin{align*}
\PP[\abs{\hat{\pi}_s[t] - \pi_s[t]} > 2e\delta]
&= \PP[\abs{\bar{X} - \EE[X_i]} > 2e\delta] 
%&=  \PP\bk{\abs{\frac{\epr}{w} Y - \frac{\epr}{w} \EE[Y]} > 2e\delta} \\
=  \PP\bk{\abs{Y -  \EE[Y]} > \frac{w}{d_t\epr} 2e\delta} \\
&\leq  \PP\bk{Y > \frac{w}{d_t \epr}2e\delta} .
\end{align*}
At this point we set $b = 2e\delta w/d_t\epr=2ec/\epsilon^2$ and apply the second Chernoff bound.  
Note that $\EE[Y] = c\EE[X_i]/\epsilon^2\delta  < c/\epsilon^2$, and hence we satisfy $b > 2e \EE[Y]$.
We conclude that:
\[ \PP[\abs{\hat{\pi}_s[t] - \pi_s[t]} > 2e\delta] \leq 2^{-b} \leq \pfail \]
as long as we choose $c$ such that $c \geq \frac{\epsilon^2}{2e} \log_2 \frac{1}{\pfail}$. The proof is completed by combining both cases and choosing $c = 3 \ln \pn{2/\pfail}$.$\quad\square$
\end{proof}

\subsubsection{Running Time Analysis}
\label{sec:ubippr_running_time}
 The more interesting analysis is that of the running-time of \ubippr{} -- we now prove a worst-case running-time bound:
\begin{theorem}
\label{thm:worstcase}
  In an undirected graph, for any source node (or distribution) $s$, target $t$ with degree $d_t$, threshold $\delta$, maximum residual $\epr$, relative error $\epsilon$, and failure probability $\pfail$, \ubippr{} has a worst-case running-time of:
\[ O \pn{ \frac{\sqrt{\log{\frac{1}{\pfail}}}}{\alpha \epsilon} \sqrt{\frac{d_t}{\delta}}}. \]
\end{theorem}

Before proving this result, we first state a crucial lemma from \cite{Andersen2006}:
\begin{lemma}[Lemma $2$ in \cite{Andersen2006}]
	\label{lem:andersen}
Let $T$ be the total number of push operations performed by \fpush{}, and let $d_k$ be the degree of the vertex involved in the $k^{th}$ push. Then:
\begin{align*}
\sum_{k=1}^Td_k\leq\frac{1}{\alpha\epr}	
\end{align*}
\end{lemma}	
To keep this work self-contained, we also include a short proof from \cite{Andersen2006}
\begin{proof}
Let $v_k$ be the vertex pushed in the $k^{th}$ step -- then by definition, we have that $r_s(v_k)>\epr d_k$. Now after the local-push operation, the sum residual $||r_s||_1$ decreases by at least $\alpha\epr d_k$. However, we started with $||r_s||_1 = 1$, and thus we have $\sum_{k=1}^T\alpha\epr d_k\leq 1$. $\qquad\square$
\end{proof}
Note also that the amount of work done while pushing from a node $v$ is $d_v$.
\begin{proof}[of Theorem \ref{thm:worstcase}]
  As proven in Lemma \ref{lem:andersen}, the push forward step takes total time $O\pn{1/\alpha \epr}$ in the \emph{worst-case}.  The random walks take $O(w) = O\pn{\frac{1}{\alpha \epsilon^2} \frac{\epr}{\delta / d_t}}$ time.  Thus our total time is
\[ O\pn{\frac{1}{\alpha \epr} + \frac{\ln{\frac{1}{\pfail}}}{ \alpha \epsilon^2} \frac{\epr}{\delta / d_t}}. \]
Balancing this by choosing $r_{\max}=\frac{\epsilon}{\sqrt{\ln{\frac{1}{\pfail}}}} \sqrt{\delta/d_t}$, we get total running-time:
\[ O \pn{ \frac{\sqrt{\ln{\frac{1}{\pfail}}}}{\alpha \epsilon} \sqrt{\frac{d_t}{\delta}}}. \qquad\square\]
\end{proof}
We can get a cleaner worst-case running time bound if we make a natural assumption on $\pi_s[t]$.  In an undirected graph, if we let $\alpha = 0$ and take infinitely long walks, the stationary probability of being at any node $t$ is $\frac{d_t}{2 m}$.  Thus if $\pi_s[t] < \frac{d_t}{2 m}$, then $s$ actually has a lower PPR to $t$ than the non-personalized stationary probability of $t$, so it is natural to say $t$ is not significant for $s$.  If we set a significance threshold of $\delta = \frac{d_t}{2m}$, and apply the previous theorem, we immediately get the following:
\begin{corollary}
	\label{corr:worstcase}
  If $\pi_s[t] \geq \frac{d_t}{2m}$, we can estimate $\pi_s[t]$ within relative error $\epsilon$ with constant probability in worst-case time:
\[ O\pn{\frac{\sqrt{m}}{\alpha \epsilon} }. \]
\end{corollary}

%In contrast, the running time for \texttt{Monte-Carlo} to achieve the same accuracy guarantee is $O\pn{\frac{1}{\delta} \frac{\log\pn{1/\pfail}}{\alpha \epsilon^2}}$, and the running time for \texttt{ApproximatePageRank} is $O\pn{\frac{\bar{d}}{\delta \alpha}}$.  The \texttt{FAST-PPR} algorithm of \cite{fastppr} has an \emph{average case} running time of $O\pn{\frac{1}{\alpha\epsilon^2}\sqrt{\frac{\bar{d}}{\delta}} \sqrt{ \frac{\log\pn{1/\pfail} \log\pn{1/\delta}}{\log\pn{1/(1-\alpha)}} }}$ for uniformly chosen targets, but has no clean worst-case running time bound because its running time depends on the degree of nodes pushed from in the linear-algebraic part of the algorithm.  

\section{Experiments}
\subsection{PPR Estimation Running Time}
\label{sec:bippr_experimental_time}
We now compare \bippr{} to its predecessor algorithms (namely \texttt{FAST-PPR} \cite{Lofgren2013}, \texttt{Monte Carlo} \cite{Avrachenkov2007,Fogaras2005} and \rpush{} \cite{Andersen2007}). 
The experimental setup is identical to that in \cite{Lofgren2013}; for convenience, we describe it here in brief. 
We perform experiments on 6 diverse, real-world networks: two directed social networks (Pokec (31M edges) and Twitter-2010 (1.5 billion edges)), two undirected social network (Live-Journal (69M edges) and Orkut (117M edges)), a collaboration network (dblp (6.7M edges)), and a web-graph (UK-2007-05 (3.7 billion edges)).  
Since all algorithms have parameters that enable a trade-off between running time and accuracy, we first choose parameters such that the mean relative error of each algorithm is approximately 10\%.  
For bidirectional-PPR, we find that setting $c=7$ (i.e., generating $7 \cdot \frac{\epr}{\delta}$ random walks) results in a mean relative error less than 8\% on all graphs; for the other algorithms, we use the settings determined in \cite{Lofgren2013}. 
We then repeatedly sample a uniformly-random start node $s \in V$, and a random target $t \in T$ sampled either uniformly or from PageRank (to emphasize more important targets).  
For both \bippr{} and \texttt{FAST-PPR}, we used the dynamic-balancing heuristic described above.

\begin{figure*}[t]
\centering
\subfigure[Sampling targets uniformly]{
\label{fig:runtime_uniform}
\includegraphics[width=0.9\columnwidth]{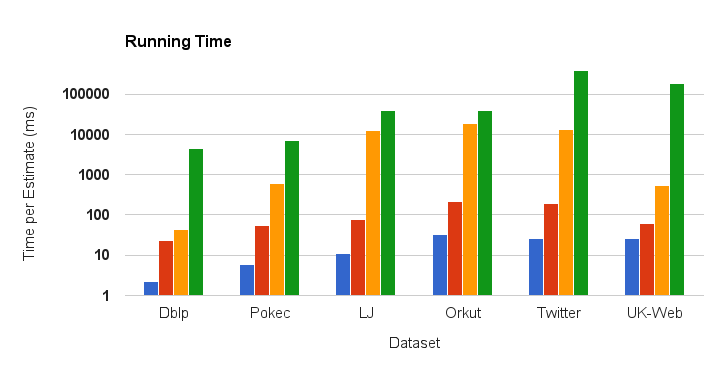}
}
\hfill
\subfigure[Sampling targets from PageRank distribution]{
\label{fig:runtime_pagerank}
\includegraphics[width=1.1\columnwidth]{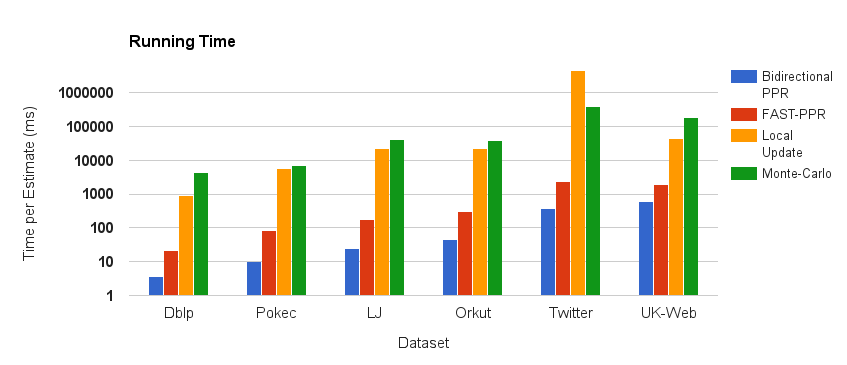}
}
%\subfigure[Other stuff]{
%\includegraphics[width=.65\textwidth]{somethingelse.png}}
\caption[Running-time Plots]{Average running-time (on log-scale) for different networks. We measure the time required for estimating PPR values $\pi_s[t]$ with threshold $\delta=\frac{4}{n}$ for $1000$ $(s,t)$ pairs. For each pair, the start node is sampled uniformly, while the target node is sampled uniformly in Figure \ref{fig:runtime_uniform}, or from the global PageRank distribution in Figure \ref{fig:runtime_pagerank}. In this plot we use teleport probability $\alpha=0.2$.}
\label{fig:ppr_runtime}
\end{figure*}

The results are shown in Figure \ref{fig:ppr_runtime}.  
We find that \bippr{} is at least 70x faster than \mc{} or \rpush{} on all six graphs. For example, on Orkut, past algorithms take 20 seconds per pair while \bippr{} takes just 50 milliseconds.  On Tiwtter-2010, past algorithms take 5 minutes, while \bippr{} takes just 3 seconds

Comparing \bippr{} to its predecessor $\texttt{FAST-PPR}$, we find it is $3$ to $8$ times faster than across all graphs.
In particular, \bippr{} only needs to sample $7 \frac{\epr}{\delta}$ random walks, while FAST-PPR needs $350 \frac{\epr}{\delta}$ walks to achieve the same mean relative error.  
This is because the estimator used by \bippr{} is unbiased, while the estimator used in the basic \texttt{FAST-PPR} implementation uses a biased estimate provided by \rpush{} for nodes on the Frontier of the target.  
%Also, \bippr{} runs Local-Update until the maximum residual is less than $\epr$, while FAST-PPR does extra reverse-work, using a maximum residual of $\beta \epr$, where in experiments $\beta=\frac{1}{6}$.  
%The provably unbiased way in which \bippr{} uses residuals to estimate PPR enables it to achieve accurate estimation at a fraction of the running time of FAST-PPR.

\subsection{PPR Estimation Running Time on Undirected Graphs}
\label{sec:ubippr_experiment}
We present a preliminary comparison between the running time of \bippr{} and \ubippr{}.  In this experiment, we find that \ubippr{} is not significantly faster than \bippr{}, but we report this for scientific completeness.  To obtain a realistic, large, undirected graph, we compute personalized SALSA \cite{Bahmani2010} on Twitter-2010, since personalized SALSA is equivalent to PPR on a transformed undirected graph (see Section \ref{sec:other_scores} for details).  We call the resulting graph Twitter-2010-SALSA.  It has 1.5 billion undirected edges and 80 million nodes (two nodes for each node in Twitter-2010).  

On this graph, we first compare the accuracy of \bippr{} and \ubippr{} using the following experiment.  We run 100 trials.  In each trial, we sample $s \in V$ uniformly and $t \in V$ from global PageRank (since users are more likely to search for popular targets).  We check if $\pi_s[t] \geq \delta$ (setting $\delta=1/n$) and if not repeatedly sample $s$ and $t$ in the same way until $\pi_s[t] \geq \delta$.  We use the same constant $c=7$ for both algorithms, and used the variations which balance forward and reverse running time.  We find that the mean relative error for both algorithms is a little below 10\%.  This makes sense, since both algorithms are unbiased estimators and their accuracy comes from the same Chernoff Bound argument.

Next we compare the running time of the two algorithms.  We run 1000 trials.  As above, in each trial we sample $s$ uniformly and $t$ from global PageRank.  We use $\delta=4/n$ (for consistency with the runtime experiments in \cite{fastppr}), Chernoff constant $c=7$ (for determining the number of walks), and the balanced version of both algorithms, tuned such that the forward and reverse time are similar (within 20\%).  For both algorithms, the mean running time on Twitter-2010-SALSA is around 0.7 seconds per pair.  This surprised us at first, since part of our motivation for developing \ubippr{} was getting a worst-case running time bound.  However, this result makes sense if we estimate the running time of \bippr{} under an assumption on the in-degree of the nodes pushed from.  

\subsubsection{An Informal Parameterized Running Time Estimate}
\label{sec:parameterized_bippr}
%Since the running time of \bippr{} depends significantly on the global PageRank $\pi[t]$ of the target, 
To understand the relationship between the running time of \bippr{} and \ubippr{} we find it useful to analyze the running time dependence of \bippr{} on $\pi[t]$ under the heuristic assumption that the set of nodes we push back from have average in-degree.  The analysis in this section is similar to the rigorous running time analysis of \ubippr{} in Section \ref{sec:ubippr_running_time}.

Theorem 1 of \cite{Andersen2007} states that \rpush{}$(t, \epr, \alpha)$ performs at most $\frac{ n \pi[t]}{\alpha \epr}$ pushback operations, and the exact running time is proportional to the sum of the in-degrees of all the nodes we push back from. If we heuristically assume the average in-degree of these nodes is the average degree of the graph, $\dbar=m/n$, then the running time of \rpush{}($t, \epr, \alpha$) is $O \pn{\dbar \frac{ n \pi[t]}{\alpha \epr}}$ and the total running time for \bippr{} is
\[ O \pn{ \frac{c_1 \rmax}{\alpha \epsilon^2 \delta} +  \dbar \frac{ n \pi[t]}{\alpha \epr}}.  \]
Now if we set $\rmax = \sqrt{\frac{n \pi[t] \dbar}{c_1 \epsilon^2 \delta}}$ to minimize this expression, we get running time
\[ O \pn{ \frac{1}{\alpha \epsilon}\sqrt{ \frac{c m \pi[t]}{\delta} }}. \]

Similarly, in Section \ref{sec:ubippr_running_time} we show the parameterized running time of \ubippr{} is
\[ O \pn{ \frac{1}{\alpha \epsilon} \sqrt{\frac{c d_t}{\delta}}}. \]
On an undirected graph, the PageRank $\pi[t]$ tends to $\frac{d_t}{2m}$ as $\alpha \to 0$, and we expect $\pi[t]$ to approximate $\frac{d_t}{2m}$ even for positive $\alpha$.  If we replace $d_t$ in the second equation with $\dbar n \pi[t]$, then we actually get identical running times.  This explains the similar empirical running times we observe.

%\[ O \pn{ \frac{1}{\alpha \epsilon}\sqrt{c d^*_t(\rmax) n}}. \]
%Note that this is somewhat heuristic, because $d^*_t(\rmax)$ is a complex function of $t$ and $\rmax$.  
%If we further assume that the nodes pushed in \rpush{} have average in-degree, this simplifies to $O \pn{ \frac{1}{\alpha \epsilon}\sqrt{ c m }}$.  %In Section \ref{sec:ubippr} we give a similar parameterized bound for an alternative algorithm on undirected graphs, but where we do not have any heuristic dependence on $d^*_t(\rmax)$.
%Here the term $d^*$ is defined as the average degree of node we pushed from.  Since whenever we push from a node $v$, we increase $p^t[v]$ by at least $\alpha \rmax$, we only push from nodes $v$ with $\pi_v[t] \geq \alpha \rmax$.  Thus we can can bound $d^*$ by
% \[ d^* \leq  \sum_{v: \pi_v[t] \geq \rmax} \indegree{v}.\]
%We obtain a similar result rigorously on undirected graphs in Corollary \ref{corr:worstcase} of Section \ref{sec:ubippr_running_time}.
%There, under the assumption that nodes pushed from have average in-degree, we show the parameterized running time of \bippr{} (for constant relative error) is 
%\[ O \pn{ \frac{1}{\alpha \epsilon}\sqrt{ \frac{c m \pi[t]}{\delta} }} .\]

\chapter{Random Walk Probability Estimation}
\label{sec:mc_chapter}
This chapter generalizes our results to any estimating random walk probabilities where the walks have a given length, and hence to arbitrary Markov Chains \cite{generalized_fastppr}.

\section{Overview}
We present a new bidirectional algorithm for estimating fixed length transition probabilities: given a weighted, directed graph (Markov chain), we want to estimate the probability of hitting a given target state in $\ell$ steps after starting from a given source distribution. Given the target state $t$, we use a (reverse) local power iteration to construct an `expanded target distribution', which has the same mean as the quantity we want to estimate, but a smaller variance -- this can then be sampled efficiently by a Monte Carlo algorithm. Our method extends to any Markov chain on a discrete (finite or countable) state-space, and can be extended to compute functions of multi-step transition probabilities such as PageRank, graph diffusions, hitting/return times, etc. Our main result is that in `sparse' Markov Chains -- wherein the number of transitions between states is  comparable to the number of states -- the running time of our algorithm for a uniform-random target node is \emph{order-wise smaller} than Monte Carlo and power iteration based algorithms. In particular, while Monte Carlo takes time $O\pn{\frac{\ell}{\epsilon^2 p}}$ to estimate a probability $p$ within relative error $\epsilon$ with constant probability on sparse Markov Chains, our algorithm's running time is only $O\pn{\frac{\ell^{3/2}}{\epsilon}\sqrt{p}}$.

\section{Problem Description}

Markov chains are one of the workhorses of stochastic modeling, finding use across a variety of applications -- MCMC algorithms for simulation and statistical inference; to compute network centrality metrics for data mining applications; statistical physics; operations management models for reliability, inventory and supply chains, etc. 
In this paper, we consider a fundamental problem associated with Markov chains, which we refer to as the \emph{multi-step transition probability estimation} (or MSTP-estimation) problem: 
given a Markov Chain on state space $\S$ with transition matrix $W$, an initial source distribution $\vecS$ over $\S$, a target state $t\in\S$ and a fixed length $\ell$, we are interested in computing the \emph{$\ell$-step transition probability from $\vecS$ to $t$}. Formally, we want to estimate:
\begin{align}
\label{eq:MSTP}
\PR_{\vecS}^{\ell}[t]:= \langle\vecS W^{\ell},\vecE_t\rangle = \vecS W^{\ell}\vecE_t^T,
\end{align}
where $\vecE_t$ is the indicator vector of state $t$. A natural parametrization for the complexity of MSTP-estimation is in terms of the minimum transition probabilities we want to detect: given a desired minimum detection threshold $\delta$, we want algorithms that give estimates which guarantee small relative error for any $(\vecS,t,\ell)$ such that $\PR_{\vecS}^{\ell}[t]>\delta$.

Parametrizing in terms of the minimum detection threshold $\delta$ can be thought of as benchmarking against a standard Monte Carlo algorithm, which estimates $\PR_{\vecS}^{\ell}[t]$ by sampling independent $\ell$-step paths starting from states sampled from $\vecS$. An alternate technique for MSTP-estimation is based on linear algebraic iterations, in particular, the (local) power iteration. We discuss these in more detail in Section \ref{ssec:relwork}. Crucially, however, \emph{both these techniques have a running time of $\Omega(1/\delta)$} for testing if $\PR_{\vecS}^{\ell}[t]>\delta$ (cf. Section \ref{ssec:relwork}).

%Note however that neither Monte Carlo methods, nor linear algebraic techniques, are designed specifically for MSTP-estimation. In particular, neither make use of both the boundary conditions (source $\vecS$ and target $t$) -- instead, they proceed from one end ($\vecS$ for Monte Carlo in general chains, either $\vecS$ or $t$ for power iteration, or for Monte Carlo in reversible chains) and return an estimate for the entire $\ell$-step transition probability distribution (typically, with guarantees on the $\ell_1$-norm). Though in many applications this is precisely what is required, it raises the question if something better can be done for MSTP-estimation.

\subsection{Our Results}
\label{ssec:results}

To the best of our knowledge, our work gives {\em the first bidirectional algorithm for MSTP-estimation} which works for general discrete state-space Markov chains\footnote{Bidirectional estimators have been developed before for \emph{reversible} Markov chains~\cite{Goldreich2011}; our method however is not only more general, but conceptually and operationally simpler than these techniques (cf. Section \ref{ssec:relwork}).}.
The algorithm we develop is very simple, both in terms of implementation and analysis. Furthermore, we prove that in many settings, it is order-wise faster than existing techniques.

Our algorithm consists of two distinct \emph{forward} and \emph{reverse} components, which are executed sequentially. In brief, the two components proceed as follows:
\begin{itemize}[nosep,leftmargin=*]
\item \textbf{Reverse-work}: Starting from the target node $t$, we perform a sequence of reverse local power iterations -- in particular, we use the REVERSE-PUSH operation defined in Algorithm \ref{alg:push}.
%), which is based on the local-update procedure of Andersen et al.~\cite{Andersen2007}. 
%One novel feature is that in addition to storing estimates, we also keep track of the residual mass at each node. We perform sufficient REVERSE-PUSH operations to ensure all residues are less than a chosen {\em reverse threshold $\rmax$}.
\item \textbf{Forward-work}:  
%The crux of our algorithm is encoded in Equation \eqref{eq:mstpest} which expresses $\PR_{\vecS}^{\ell}[t]$ as a linear combination of estimates and residues. 
We next sample a number of random walks of length $\ell$, starting from $\vecS$ and transitioning according to $P$, and return the sum of residues on the walk as an estimate of $\PR_{\vecS}^{\ell}[t]$. 
%Since these residues are bounded by $\rmax$, the resulting samples have low variance and hence we get good concentrations for the estimates with very few samples.  
\end{itemize}

This full algorithm, which we refer to as the \texttt{Bidirectional-MSTP} estimator, is formalized in Algorithm \ref{alg:MSTP}. It works for all countable-state Markov chains, giving the following accuracy result:
\newtheorem*{thm:mc_accurate}{Theorem \ref{thm:MSTPmain}}
\begin{thm:mc_accurate}[For details, see Section \ref{ssec:analysis}]

Given any Markov chain $W$, source distribution $\vecS$, terminal state $t$, length $\ell$, threshold $\delta$ and relative error $\epsilon$, \texttt{Bidirectional-MSTP} (Algorithm \ref{alg:MSTP}) returns an unbiased estimate $\widehat{\PR}_{\vecS}^{\ell}[t]$ for $\PR_{\vecS}^{\ell}[t]$,
%. Moreover, if we use reverse threshold $\rmax\sim\sqrt{\delta}$ and number of trajectories $n_f\sim 1/\sqrt{\delta}$, then, 
which, with high probability, satisfies:
\begin{align*}
\left|\widehat{\PR}_{\vecS}^{\ell}[t]-\PR_{\vecS}^{\ell}[t]\right|<\max\left\{\epsilon\PR_{\vecS}^{\ell}[t],\delta\right\}.
\end{align*}
\end{thm:mc_accurate}

Since we dynamically adjust the number of REVERSE-PUSH operations to ensure that all residues are small, the proof of the above theorem follows from straightforward concentration bounds. 

Since \texttt{Bidirectional-MSTP} combines local power iteration and Monte Carlo techniques, a natural question is when the algorithm is faster than both. 
It is easy to to construct scenarios where the runtime of \texttt{Bidirectional-MSTP} is comparable to its two constituent algorithms -- for example, if $t$ has more than $1/\delta$ in-neighbors. 
Surprisingly, however, we show that in \emph{sparse Markov chains} and for \emph{random target states}, \texttt{Bidirectional-MSTP} is order-wise faster:

\newtheorem*{thm:mc_fast}{Theorem \ref{thm:runtime}}
\begin{thm:mc_fast}
[For details, see Section \ref{ssec:analysis}]
Given any Markov chain $W$, source distribution $\vecS$, length $\ell$, threshold $\delta$ and desired accuracy $\epsilon$; then for a uniform random choice of $t\in\S$, the \texttt{Bidirectional-MSTP} algorithm has a running time of $\widetilde{O}(\ell^{3/2}\sqrt{\dbar/\delta})$, where $\dbar$ is the average number of neighbors of nodes in $\S$.
\end{thm:mc_fast}

Thus, for random targets, {\em our running time dependence on $\delta$ is $1/\sqrt{\delta}$}. Note that we do not need for every state that the number of neighboring states is small, but rather, that they are small on average -- for example, this is true in `power-law' networks, where some nodes have very high degree, but the average degree is small. The proof of this result in  Section \ref{ssec:analysis} is similar to the analysis of \bippr{} in Section \ref{sec:bippr_running_time}. 

Estimating transition probabilities to a target state is one of the fundamental primitives in Markov chain models -- hence, we believe that our algorithm can prove useful in a variety of application domains. In Section \ref{sec:apps}, we briefly describe how to adapt our method for some of these applications -- estimating hitting/return times and stationary probabilities, extensions to non-homogenous Markov chains (in particular, for estimating graph diffusions and heat kernels), connections to local algorithms and expansion testing. In addition, our MSTP-estimator could be useful in several other applications -- estimating ruin probabilities in reliability models, buffer overflows in queueing systems, in statistical physics simulations, etc.

\subsection{Existing Approaches for MSTP-Estimation}
\label{ssec:relwork}

There are two main techniques used for MSTP-estimation, similar to the techniques described in Chapter \ref{sec:relwork} for PPR estimation. The first is a natural Monte Carlo algorithm (see Section \ref{sec:monte_carlo}): we estimate $\PR_{\vecS}^{\ell}[t]$ by sampling independent $\ell$-step paths, each starting from a random state sampled from $\vecS$. 
A simple concentration argument shows that for a given value of $\delta$, we need $\widetilde{\Theta}(1/\delta)$ samples to get an accurate estimate of $\PR_{\vecS}^{\ell}[t]$, irrespective of the choice of $t$, and the structure of $W$. 
Note that this algorithm is agnostic of the terminal state $t$; it  gives an accurate estimate for any $t$ such that $\PR_{\vecS}^{\ell}[t]>\delta$. 
%This is useful for many applications -- however in settings where we are interested in a small number of terminal states, then our algorithm is much faster.

On the other hand, the problem also admits a natural linear algebraic solution, using the standard power iteration (see Section \ref{sec:power_iter}) starting with $\vecS$, or the reverse power iteration starting with $\vecE_t$  which is obtained by re-writing Equation \eqref{eq:MSTP} as $\PR_{\vecS}^{\ell}[t]:= \vecS(\vecE_t(W^T)^{\ell})^T$. When the state space is large, performing a direct power iteration is infeasible -- however, there are localized versions of the power iteration that are still efficient. Such algorithms have been developed, among other applications, for PageRank estimation (see Section \ref{sec:forward_push} and \ref{sec:reverse_push}) and for heat kernel estimation~\cite{Kloster2014}. Although slow in the worst case~\footnote{In particular, local power iterations are slow if a state has a very large out-neighborhood (for the forward iteration) or in-neighborhood (for the reverse update).}, such local update algorithms are often fast in practice, as unlike Monte Carlo methods they exploit the local structure of the chain. However even in sparse Markov chains and for a large fraction of target states, their running time can be $\Omega(1/\delta)$. For example, consider a random walk on a random $d$-regular graph and let $\delta= o(1/n)$ -- then for $\ell\sim\log_d(1/\delta)$, verifying $\PR_{\vecE_s}^{\ell}[t]>\delta$ is equivalent to uncovering the entire $\log_d(1/\delta)$ neighborhood of $s$. Since a large random $d$-regular graph is (whp) an expander, this neighborhood has $\Omega(1/\delta)$ distinct nodes. Finally, note that as with Monte Carlo, power iterations do not only return probabilities between a single pair, but return probabilities from a single source to all targets, or from all sources to a single target.%be adapted to either the source or terminal state, but not both. 

For \emph{reversible Markov chains}, one can get a bidirectional algorithms for estimating $\PR_{\vecE_s}^{\ell}[t]$ based on colliding random walks. 
For example, consider the problem of estimating length-$2\ell$ random walk transition probabilities in a \emph{regular undirected graph} $G(V,E)$ on $n$ vertices~\cite{Goldreich2011,Kale2008}. 
The main idea is that to test if a random walk goes from $s$ to $t$ in $2\ell$ steps with probability $\geq\delta$, we can generate two independent random walks of length $\ell$, starting from $s$ and $t$ respectively, and detect if they \emph{terminate at the same intermediate node}. 
Suppose $p_w,q_w$ are the probabilities that a length-$\ell$ walk from $s$ and $t$ respectively terminate at node $w$ -- then from the reversibility of the chain, we have that $\PR_{\vecS}^{2\ell}[t]=\sum_{w\in V}p_wq_w$; this is also the collision probability. 
The critical observation is that if we generate $\sqrt{1/\delta}$ walks from $s$ and $t$, then we get $1/\delta$ potential collisions, which is sufficient to detect if $\PR_{\vecS}^{2\ell}[t]>\delta$.
This argument forms the basis of the \emph{birthday-paradox}, and similar techniques used in a variety of estimation problems (eg., see~\cite{Motwani2007}).
Showing concentration for this estimator is tricky as the samples are not independent; moreover, to control the variance of the samples, the algorithms often need to separately deal with `heavy' intermediate nodes, where $p_w$ or $q_w$ are much larger than $O(1/n)$. 
Our proposed approach is much simpler both in terms of algorithm and analysis, and more significantly, it extends beyond reversible chains to any general discrete state-space Markov chain.

%The most similar approach to ours is the recent FAST-PPR algorithm of Lofgren et al.~\cite{Lofgren2014} for PageRank estimation; our algorithm borrows several ideas and techniques from that work. However, the FAST-PPR algorithm relies heavily on the structure of PageRank -- in particular, the fact that the PageRank walk has $Geometric(\alpha)$ length (and hence can be stopped and restarted due to the memoryless property).
%Our work provides an elegant and powerful generalization of the FAST-PPR algorithm, extending the approach to general Markov chains. 
%Moreover, although~\cite{Lofgren2014} presents a theoretical unbiased estimator, it is too complicated to implement in practice (the FAST-PPR algorithm itself is a simpler approximate estimator). In contrast, we give a simple unbiased estimator which is easy to implement, as demonstrated by our experiments.

\section{The Bidirectional MSTP-estimation Algorithm}
\label{sec:generalizedBiPPR}

Our approach is similar to our approach for bidirectional PPR estimation (Section \ref{sec:bippr}).  The main difference is that for PPR, the random walk is memoryless, so the \rpush{} algorithm returns a single residual value $r^t[v]$ for each node.  Here, we have a fixed walk length, so we a compute separate residual value $r_t^k[v]$ for each number of steps $k \leq \ell$ back from the target.  We then combine residual values $k$ steps back from the target with a random walk probability $\ell - k$ steps forward from the source, and sum over $k \leq \ell$ to estimate $\pi_s^\ell[t]$.

%Notation: For a Markov chain with finite state-space $\S$ and transition probability matrix $P$, we denote $|\S|=n$ and denote $m$ to be the number of non-zero elements of $P$; the {\em average} number of neighbors of states in $\S$ is thus $\dbar=\frac{m}{n}$. As per convention, throughout this work we use $n\times 1$ row vectors to represent probability distributions over $\mathcal{S}$. 

\subsection{Algorithm}
\label{sec:algo}

As described in Section \ref{ssec:results}, given a target state $t$, our bidirectional MSTP algorithm keeps track of a pair of vectors -- the estimate vector $\p_t^k \in \R^n$ and the residual vector $\r_t^k \in \R^n$ -- for each length $k\in\{0,1,2,\ldots,\ell\}$. The vectors are initially all set to $\underline{0}$ (i.e., the all-$0$ vector), except $r_t^0$ which is initialized as $\vecE_t$. They are updated using a \emph{reverse push} operation defined as Algorithm \ref{alg:push}.

\begin{algorithm}[ht]
\caption{REVERSE-PUSH-MSTP$(v,i)$}
\label{alg:push}
\begin{algorithmic}[1]
\REQUIRE Transition matrix $W$, estimate vector $\p_t^i$, residual vectors $\r_t^i,\r_t^{i+1}$
\RETURN New estimate vectors $\{\widetilde{\p}_t^i\}$ and residual-vectors $\{\widetilde{\r_t^i}\}$ computed as:
\begin{align*}
	\widetilde{\p}_t^i &\leftarrow \p_t^i + \langle\r_t^i,\vecE_v\rangle\vecE_v \\
	\widetilde{\r}_t^i &\leftarrow \r_t^i -\langle\r_t^i,\vecE_v\rangle\vecE_v\\
	\widetilde{\r}_t^{i+1} &\leftarrow \r_t^{i+1} + \langle\r_t^i,\vecE_v\rangle\left(\vecE_vW^T\right)
\end{align*}	
\end{algorithmic}
\end{algorithm}    

The main observation behind our algorithm is that we can re-write $\PR_{\vecS}^{\ell}[t]$ in terms of $\{\p_t^k,\r_t^k\}$ as an {\em expectation over random sample-paths of the Markov chain} as follows (cf. Equation \eqref{eq:pushinv}): 
\begin{equation}
\label{eq:mstpest}
\PR_{\vecS}^{\ell}[t] = \langle\vecS,\p_t^{\ell}\rangle + \sum_{k=0}^{\ell}\EE_{V_k\sim\vecS W^k}\left[\r_t^{\ell-k}(V_k)\right] 
%\EE_{V_0\sim \vecS}\left[\p_t^{\ell}[V_0]\right] + \sum_{k=1}^{\ell}\EE_{V_k\sim\vecS W^k}\left[\r_t^{\ell-k}(V_k)\right]	
\end{equation}	
In other words, given vectors $\{\p_t^k,\r_t^k\}$, we can get an unbiased estimator for $\PR_{\vecS}^{\ell}[t]$ by sampling a length-$\ell$ random trajectory $\{V_0,V_1,\ldots,V_{\ell}\}$ of the Markov chain $W$ starting at a random state $V_0$ sampled from the source distribution $\vecS$, and then adding the residuals along the trajectory as in Equation \eqref{eq:mstpest}.
We formalize this bidirectional MSTP algorithm in Algorithm \ref{alg:MSTP}. 
%The input-output behavior of FAST-PPR is as follows:
%\begin{itemize}[nosep,leftmargin=*]
%\item {\bf Inputs:} The primary inputs are the transition matrix $P$, starting distribution $\vecS$, target state $t\in\S$, maximum number of steps $\ell_{\max}$ and threshold $\delta$. We also choose a desired \emph{accuracy $\epsilon$, failure probability $p_{f}$} and \emph{reverse threshold $\rmax$} -- these affect the accuracy and running time of the algorithm (refer Theorems \ref{thm:MSTPmain} and \ref{thm:runtime} for details). 

%\item {\bf Output:} An estimate $\widehat{\PR}_{\vecS}^{\ell}[t]$ for the multi-step transition probability ${\PR}_{\vecS}^{\ell}[t]$ for {\em every} $\ell\leq\ell_{\max}$.
%\end{itemize}

\begin{algorithm}[!h]
\caption{Bidirectional-MSTP$(W,\vecS,t,\ell_{\max},\delta)$}
\label{alg:MSTP}
\begin{algorithmic}[1] 
\REQUIRE Transition matrix $W$, source distribution $\vecS$, target state $t$, maximum steps $\ell_{\max}$, minimum probability threshold $\delta$, relative error bound $\epsilon$, failure probability $p_{f}$
\STATE Set accuracy parameter $c$ based on $\epsilon$ and $p_f$ and set reverse threshold $\rmax$ (cf. Theorems \ref{thm:MSTPmain} and \ref{thm:runtime})  (in our experiments we use $c = 7$ and $\rmax = \sqrt{\delta/c}$)
\STATE Initialize: Estimate vectors $\p_t^k = \underline{0}\,,\, \fall k\in\{0,1,2,\ldots,\ell\}$,\\
\hspace{1.4cm}  Residual vectors $\r_t^0 = \vecE_t$ and $\r_t^k = \underline{0}\,,\, \fall k\in\{1,2,3,\ldots,\ell\}$
\FOR{$i\in\{0,1,\ldots,\ell_{\max}\}$}
\WHILE{$\exists \, v\in \S\quad s.t.\quad\r_t^i[v]>\rmax$}
\STATE Execute REVERSE-PUSH-MSTP$(v,i)$
\ENDWHILE
\ENDFOR
\STATE Set number of sample paths $n_f=c\ell_{\max}\rmax/\delta$ \quad(See Theorem \ref{thm:MSTPmain} for details)
\FOR{index $i\in \{1,2,\ldots,n_f\}$}
\STATE Sample starting node $V_i^0\sim\vecS$
\STATE Generate sample path $T_i = \{V_i^0,V_i^1,\ldots,V_i^{\ell_{\max}}\}$ of length $\ell_{\max}$ starting from $V_i^0$ 
\STATE For $\ell\in\{1,2,\ldots,\ell_{\max}\}$: sample $k\sim Uniform[0,\ell]$ and compute $S_{t,i}^{\ell}=\ell \r_t^{\ell-k}[V_i^{k}]$ \\(We reinterpret the sum over $k$ in Equation \ref{eq:mstpest} as an expectation and sample $k$ rather sum over $k \leq \ell$  for computational speed.)
%=\sum_{i=0}^{\ell}\r_t^{\ell-i}[V_k^i]$
\ENDFOR 
\RETURN $\{\widehat{\PR}_{\vecS}^{\ell}[t]\}_{\ell\in[\ell_{\max}]}$, where $\widehat{\PR}_{\vecS}^{\ell}[t]=\langle\vecS,\p_t^{\ell}\rangle + (1/n_f)\sum_{i=1}^{n_f}S_{t,i}^{\ell}$
\end{algorithmic}
\end{algorithm}

\subsection{Some Intuition Behind our Approach}
\label{ssec:intuition}

\begin{figure}[!b]
\begin{center}
\includegraphics[width=0.8\textwidth]{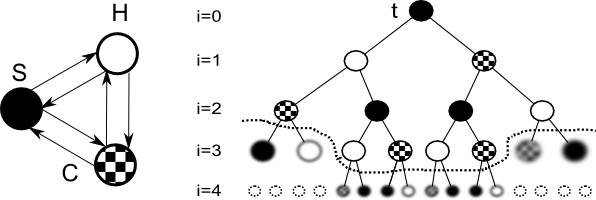}
\end{center}
\caption{Visualizing a sequence of REVERSE-PUSH-MSTP operations: Given the Markov chain on the left with $S$ as the target, we perform REVERSE-PUSH-MSTP operations $(S,0),(H,1),(C,1)$,$(S,2)$.
%Each state has multiple virtual nodes at any level $i$, each corresponding to a unique length-$i$ path to the target $R$. Edges correspond to completed push operations; colored virtual nodes are those where either the estimate or residue is non-zero.
% -- in particular, paths starting at solid nodes are captured in the corresponding estimate for that stage, while paths starting at blurred nodes are captured in the residue.
}
\label{fig:example}
\end{figure}

Before formally analyzing the performance of our MSTP-estimation algorithm, we first build some intuition as to why it works. 
In particular, we give an interpretation for the residual values $r_t^k[v]$ which is different from the message passing interpretation of the PPR residual values given in Section \ref{sec:reverse_push}. %which worked because of the memoryless length distribution of PPR random walks.% it is useful to interpret the estimates and residues in probabilistic/combinatorial terms. 
In Figure \ref{fig:example}, we have considered a simple Markov chain on three states -- Solid, Hollow and Checkered (henceforth $(S,H,C)$). On the right side, we have illustrated an intermediate stage of reverse work using $S$ as the target, after performing the REVERSE-PUSH-MSTP operations $(S,0),(H,1),(C,1)$ and $(S,2)$ in that order. Each push at level $i$ uncovers a collection of length-$(i+1)$ paths terminating at $S$ -- for example, in the figure, we have uncovered all length $2$ and $3$ paths, and several length $4$ paths. The crucial observation is that {\em each uncovered path of length $i$ starting from a node $v$ is accounted for in either $\p_v^{i}$ or $\r_v^{i}$}. In particular, in Figure \ref{fig:example}, all paths starting at solid nodes are stored in the estimates of the corresponding states, while those starting at blurred nodes are stored in the residue. Now we can use this set of pre-discovered paths to boost the estimate returned by Monte Carlo trajectories generated starting from the source distribution. The dotted line in the figure represents the current reverse-work \emph{frontier} -- it separates the fully uncovered neighborhood of $(S,0)$ from the remaining states $(v,i)$.

%In a sense, what the REVERSE-PUSH-MSTP operation does is construct a sequence of importance-sampling weights, which can then be used for Monte Carlo. An important novelty here is that the importance-sampling weights are: $(i)$ adapted to the target state, and $(ii)$ dynamically adjusted to ensure the Monte Carlo estimates have low variance. Viewed in this light, it is easy to see how the algorithm can be modified to applications beyond basic MSTP-estimation: for example, to non-homogenous Markov chains, or for estimating the probability of hitting a target state $t$ for the first time in $\ell$ steps (cf. Section \ref{sec:apps}). Essentially, we only need an appropriate reverse-push/dynamic programming update for the quantity of interest (with associated invariant, as in Equation \eqref{eq:mstpest}).  

\subsection{Performance Analysis}
\label{ssec:analysis}

We first formalize the critical invariant introduced in Equation \eqref{eq:mstpest}:
%; the proof follows the outline of a similar result in Andersen et al.~\cite{Andersen2007} for PageRank estimation:
\begin{lemma} \label{lemma:rev_push}
Given a terminal state $t$, suppose we initialize $\p_t^0=\underline{0}, \r_t^0=\vecE_t$ and $\p_t^k,\r_t^k = \underline{0}\,\forall\,k\geq 0$. 
Then for any source distribution $\vecS$ and length $\ell$, after any arbitrary sequence of REVERSE-PUSH-MSTP$(v,k)$ operations, the vectors $\{\p_t^k,\r_t^k\}$ satisfy the invariant:
\begin{equation}
\label{eq:pushinv}
\PR_{\vecS}^{\ell}[t] = \langle\vecS,\p_t^{\ell}\rangle + \sum_{k=0}^{\ell}\langle\vecS W^k,\r_t^{\ell-k}\rangle	
\end{equation}	
\end{lemma}

\begin{proof}
The proof follows the outline of a similar result in Andersen et al.~\cite{Andersen2007} for PageRank estimation. First, note that Equation \eqref{eq:pushinv} can be re-written as
%\begin{equation*}
$\PR_{\vecS}^{\ell}[t] = \left\langle\vecS,\p_t^{\ell} + \sum_{k=0}^{\ell}\r_t^{\ell-k}(W^k)^T\right\rangle
$.
%\end{equation*}	
Note that under the initial conditions specified in Algorithm \ref{alg:MSTP}, for any $\vecS,\ell$, the invariant reduces to:
%\begin{equation*}
$\PR_{\vecS}^{\ell}[t] = \left\langle\vecS,\p_t^{\ell} + \sum_{k=0}^{\ell}\r_t^{\ell-k}(W^k)^T\right\rangle	
= \langle\vecS,\vecE_t(W^{T})^{\ell}\rangle
= \langle\vecS W^{\ell},\vecE_t\rangle,
$
%\end{equation*}	
which is true by definition. We now prove the invariant by induction. Suppose at some stage, vectors $\{\p_t^k,\r_t^k\}$ satisfy Equation \eqref{eq:pushinv}; to complete the proof, we need to show that for any pair $(v,i)$, the REVERSE-PUSH-MSTP$(v,i)$ operation preserves the invariant. Let $\{\widetilde{\p}_t^k,\widetilde{\r}_t^k\}$ denote the estimate and residual vectors after the REVERSE-PUSH-MSTP$(v,i)$ operation is applied, and define:
\begin{equation*}
\Delta_v^i = \left(\widetilde{\p}_t^{\ell} + \sum_{k=0}^{\ell}\widetilde{\r}_t^{\ell-k}(W^k)^T\right) - \left(\p_t^{\ell} + \sum_{k=0}^{\ell}\r_t^{\ell-k}(W^k)^T\right)
\end{equation*}
Now, to prove the invariant, it is sufficient to show $\Delta_v^i=0$ for any choice of $(v,i)$. Clearly this is true if $\ell<i$; this leaves us with two cases:
\begin{itemize}[nosep,leftmargin=*]
\item If $\ell = i$ we have: $\Delta_v^i= \Big(\widetilde{\p_t}^i-\p_t^i\Big)+\Big(\widetilde{\r}_t^{i}-\r_t^{i}\Big)= \langle\r_t^k,\vecE_v\rangle\vecE_v-\langle\r_t^k,\vecE_v\rangle\vecE_v = 0
$
%\begin{align*}
%\Delta_v^i	&= \Big(\widetilde{\p_t}^i-\p_t^i\Big) +\Big(\widetilde{\r}_t^{i}-\r_t^{i}\Big) 
%%= \langle\vecS W^{\ell},\vecE_t\rangle 
%= \langle\r_t^k,\vecE_v\rangle\vecE_v-\langle\r_t^k,\vecE_v\rangle\vecE_v %= 0
%\end{align*}	
\item If $\ell> i$, we have:
\begin{align*}
\Delta_v^i
&= \Big(\widetilde{\r}_t^{i+1}-\r_t^{i+1}+ (\widetilde{\r}_t^i-\r_t^i)W^T\Big)\Big(W^T\Big)^{\ell-i-1} \\
&= \Big(\langle\r_t^k,\vecE_v\rangle(\vecE_tW^T)- (\langle\r_t^k,\vecE_v\rangle\vecE_v)W^T\Big)\Big(W^T\Big)^{\ell-i-1} = \underline{0}
\end{align*}	
\end{itemize}
This completes the proof of the lemma.
\end{proof}

Using this result, we can now characterize the accuracy of the \texttt{Bidirectional-MSTP} algorithm:

\begin{theorem}
\label{thm:MSTPmain}
We are given any Markov chain $W$, source distribution $\vecS$, terminal state $t$, maximum length $\ell_{\max}$ and also parameters $\delta, p_{f}$ and $\epsilon$ (i.e., the desired threshold, failure probability and relative error). Suppose we choose any reverse threshold $\rmax>\delta$, and set the number of sample-paths $n_f=c\rmax/\delta$, where $c=\max\left\{6e/\epsilon^2,1/\ln 2\right\}\ln\left(2\ell_{\max}/p_{f}\right)$.  Then for any length $\ell\leq\ell_{\max}$ with probability at least $1-p_{f}$, the estimate returned by \texttt{Bidirectional-MSTP} satisfies:
\begin{align*}
\left|\widehat{\PR}_{\vecS}^{\ell}[t]-\PR_{\vecS}^{\ell}[t]\right|<\max\left\{\epsilon\PR_{\vecS}^{\ell}[t],\delta\right\}.
\end{align*}
\end{theorem}

\begin{proof}
Given any Markov chain $W$ and terminal state $t$, note first that for a given length $\ell\leq\ell_{\max}$, Equation \eqref{eq:mstpest} shows that the estimate $\widehat{\PR}_{\vecS}^{\ell}[t]$ is an unbiased estimator.
Now, for any random-trajectory $T_k$, we have that the score $S_{t,k}^{\ell}$ obeys: $(i)\,\,\EE[S_{t,k}^{\ell}]\leq\PR_{\vecS}^{\ell}[t]$ and $(ii)\,\, S_{t,k}^{\ell}\in[0,\ell\rmax]$; the first inequality again follows from Equation \eqref{eq:mstpest}, while the second follows from the fact that we executed REVERSE-PUSH-MSTP operations until all residual values were less than $\rmax$.

Now consider the rescaled random variable $X_k=S_{t,k}^{\ell}/(\ell\rmax)$ and $X=\sum_{k\in[n_f]}X_k$; then we have that $X_k\in[0,1]$, $\EE[X]\leq (n_f/\ell\rmax)\PR_{\vecS}^{\ell}[t]$ and also $\left(X-\EE[X]\right)= (n_f/\ell\rmax)(\widehat{\PR}_{\vecS}^{\ell}[t]-\PR_{\vecS}^{\ell}[t])$. Moreover, using standard Chernoff bounds (cf. Theorem $1.1$ in~\cite{dubhashi2009}), we have that:
\begin{align*}
\PP\left[|X-\EE[X]|>\epsilon\EE[X]\right]&<2\exp{\left(-\frac{\epsilon^2\EE[X]}{3}\right)}\quad\mbox{ and }\quad
\PP[X>b]\leq 2^{-b}\mbox{ for any }b>2e\EE[X]
\end{align*}
Now we consider two cases:
\begin{enumerate}[nosep,leftmargin=*]

\item $\EE[S_{t,k}^{\ell}]>\delta/2e$ (i.e., $\EE[X]>n_f\delta/2e\ell\rmax=c/2e$): Here, we can use the first concentration bound to get:
\begin{align*}
\PP\left[\left|\widehat{\PR}_{\vecS}^{\ell}[t]-\PR_{\vecS}^{\ell}[t]\right|\geq \epsilon\PR_{\vecS}^{\ell}[t]\right]
&=\PP\left[\left|X-\EE[X]\right|\geq \frac{\epsilon n_f}{\ell\rmax}\PR_{\vecS}^{\ell}[t]\right]
\leq\PP\left[\left|X-\EE[X]\right|\geq \epsilon \EE[X]\right]\\
&\leq 2\exp{\left(-\frac{\epsilon^2\EE[X]}{3}\right)}
%&= 2\exp{\left(-\frac{\epsilon^2n_f\delta}{32e\ell\rmax}\right)}
\leq 2\exp{\left(-\frac{\epsilon^2 c}{6e}\right)},
\end{align*}
where we use that $n_f=c\ell_{\max}\rmax/\delta$ (cf. Algorithm \ref{alg:MSTP}). Moreover, by the union bound, we have:
\begin{align*}
\PP\left[\bigcup_{\ell\leq\ell_{\max}}\left\{\left|\widehat{\PR}_{\vecS}^{\ell}[t]-\PR_{\vecS}^{\ell}[t]\right|\geq \epsilon\PR_{\vecS}^{\ell}[t]\right\}\right]
&\leq 2\ell_{\max}\exp{\left(-\frac{\epsilon^2 c}{32e}\right)},
\end{align*}
Now as long as $c\geq \left(6e/\epsilon^2\right)\ln\left(2\ell_{\max}/p_{f}\right)$, we get the desired failure probability.

\item $\EE[S_{t,k}^{\ell}]<\delta/2e$ (i.e., $\EE[X]<c/2e$): In this case, note first that since $X>0$, we have that $\PR_{\vecS}^{\ell}[t]-\widehat{\PR}_{\vecS}^{\ell}[t]\leq (n_f/\ell\rmax)\EE[X]\leq \delta/2e<\delta$. On the other hand, we also have:
\begin{align*}
\PP\left[\widehat{\PR}_{\vecS}^{\ell}[t]-\PR_{\vecS}^{\ell}[t]\geq \delta \right]
&=\PP\left[X-\EE[X]\geq \frac{n_f\delta}{\ell\rmax}\right]
\leq\PP\left[X\geq c\right]\leq 2^{-c},
\end{align*}
where the last inequality follows from our second concentration bound, which holds since we have $c>2e\EE[X]$. Now as before, we can use the union bound to show that the failure probability is bounded by $p_{f}$ as long as $c\geq\log_2\left(\ell_{\max}/p_{f}\right)$. 

%, which is true by our assumption. Now using the concentration and the union bound, we have: 
%\begin{align*}
%\PP\left[\bigcup_{\ell\leq\ell_{\max}}\left\{\left|\widehat{\PR}_{\vecS}^{\ell}[t]-\PR_{\vecS}^{\ell}[t]\right|\geq\delta\right\}\right]
%&\leq \ell_{max}2^{-c}\leq p_{f},
%\end{align*}
%as long as $c\geq\log_2\left(\ell_{\max}/p_{f}\right)$. 
\end{enumerate}
Combining the two cases, we see that as long as $c\geq\max\left\{6e/\epsilon^2,1/\ln 2\right\}\ln\left(2\ell_{\max}/p_{f}\right)$, then we have $\PP\left[\bigcup_{\ell\leq\ell_{\max}}\left\{\left|\widehat{\PR}_{\vecS}^{\ell}[t]-\PR_{\vecS}^{\ell}[t]\right|\geq \max\{\delta,\epsilon\PR_{\vecS}^{\ell}[t]\}\right\}\right]
\leq p_{f}$.
\end{proof}

One aspect that is not obvious from the intuition in Section \ref{ssec:intuition} or the accuracy analysis is if using a bidirectional method actually improves the running time of MSTP-estimation. This is addressed by the following result, which shows that for typical targets, our algorithm achieves significant speedup:

\begin{theorem}
\label{thm:runtime}
Let any Markov chain $W$, source distribution $\vecS$, maximum length $\ell_{\max}$, minimum probability $\delta$, failure probability $p_{f}$, and relative error bound $\epsilon$ be given. 
Suppose we set $\rmax= \sqrt{\frac{\epsilon^2\delta}{\ell_{\max}\log(\ell_{\max}/p_{f})}}$.  Then for a uniform random choice of $t\in\S$, the \texttt{Bidirectional-MSTP} algorithm has a running time 
\[ O\left(\frac{\ell_{\max}^{3/2} \sqrt{\dbar \log(\ell_{\max}/p_f)}}{\epsilon} \frac{1}{\sqrt{\delta}} \right). \]
\end{theorem}

\begin{proof}
The runtime of Algorithm \ref{alg:MSTP} consists of two parts:\\
\noindent\textbf{Forward-work} (i.e., for generating trajectories): we generate $n_f=c\ell_{\max}\rmax/\delta$ sample trajectories, each of length $\ell_{\max}$ -- hence the running time is $O\left(c\delta\ell_{\max}^2/\delta\right)$ for any Markov chain $W$, source distribution $\vecS$ and target node $t$. Substituting for $c$ from Theorem \ref{thm:MSTPmain}, we get that the forward-work running time $T_f = O\left(\frac{\ell_{\max}^2\rmax\log(\ell_{\max}/p_{f})}{\epsilon^2\delta}\right)$. % \fall t\in\S

\noindent\textbf{Reverse-work} (i.e., for REVERSE-PUSH-MSTP operations): Let $T_r$ denote the reverse-work runtime for a \emph{uniform random choice of $t\in\S$}. Then we have:
\begin{align*}
\EE[T_r]&=\frac{1}{|\S|}\sum_{t\in\S}\sum_{k=0}^{\ell_{\max}}\sum_{v\in\S}(d^{in}(v)+1)\mathds{1}_{\left\{\mbox{REVERSE-PUSH-MSTP}(v,k)\mbox{ is executed}\right\}}
\end{align*}
Now for a given $t\in\S$ and $k\in\{0,1,\ldots,\ell_{\max}\}$, note that the only states $v\in\S$ on which we execute REVERSE-PUSH-MSTP$(v,k)$ are those with residual $\r_t^k(v)>\rmax$ -- consequently, for these states, we have that $\p_t^k(v)>\rmax$, and hence, by Equation \eqref{eq:pushinv}, we have that $\PR_{\vecE_v}^k[t]\geq\rmax$ (by setting $\vecS=\vecE_v$, i.e., starting from state $v$). Moreover, a REVERSE-PUSH-MSTP$(v,k)$ operation involves updating the residuals for $d^{in}(v)+1$ states. Note that $\sum_{t\in\S}\PR_{\vecE_v}^k[t]=1$ and hence, via a straightforward counting argument, we have that for any $v\in\S$,  $\sum_{t\in\S}\mathds{1}_{\{\PR_{\vecE_v}^k[t]\geq\rmax\}}\leq1/\rmax$. Thus, we have:
\begin{align*}
\EE[T_r]&\leq\frac{1}{|\S|}\sum_{t\in\S}\sum_{k=0}^{\ell_{\max}}\sum_{v\in\S}(d^{in}(v)+1)\mathds{1}_{\{\PR_{\vecE_v}^k[t]\geq\rmax\}}
=\frac{1}{|\S|}\sum_{v\in\S}\sum_{k=0}^{\ell_{\max}}\sum_{t\in\S}(d^{in}(v)+1)\mathds{1}_{\{\PR_{\vecE_v}^k[t]\geq\rmax\}}\\
&\leq\frac{1}{|\S|}\sum_{v\in\S}(\ell_{\max}+1)\cdot(d^{in}(v)+1)\frac{1}{\rmax}
=O\left(\frac{\ell_{\max}}{\rmax}\cdot\frac{\sum_{v\in\S}d^{in}(v)}{|\S|}\right)
=O\left(\frac{\ell_{\max}\dbar}{\rmax}\right)
\end{align*}
Finally, we choose $\rmax= \sqrt{\frac{\epsilon^2\delta}{\ell_{\max}\log(\ell_{\max}/p_{f})}}$ to balance $T_f$ and $T_r$ and get the result. 
\end{proof}

\section{Applications of MSTP estimation}
\label{sec:apps}

\begin{itemize}[leftmargin=*]
\item\textbf{Estimating the Stationary Distribution and Hitting Probabilities:} MSTP-estimation can be used in two ways to estimate stationary probabilities $\pi[t]$. First, if we know the mixing time $\tau_{mix}$ of the chain $W$, we can directly use Algorithm \ref{alg:MSTP} to approximate $\pi[t]$ by setting $\ell_{\max}=\tau_{mix}$ and using any source distribution $\vecS$. Theorem \ref{thm:runtime} then guarantees that we can estimate a stationary probability of order $\delta$ in time $O(\tau_{mix}^{3/2}\sqrt{\overline{d}/\delta})$. In comparison, Monte Carlo has $O(\tau_{mix}/\delta)$ runtime. We note that in practice, we usually do not know the mixing time -- in such a setting, our algorithm can be used to compute an estimate of $\PR_{\vecS}^{\ell}[t]$  for all values of $\ell\leq \ell_{\max}$.
%\footnote{In practice, this can be made faster using the `balancing trick' (proposed in Algorithm $4$ in~\cite{Lofgren2014}) to dynamically choose $\rmax$ at runtime -- the resulting algorithm is instance-wise faster than the Monte Carlo algorithm.}. 

An alternative is to modify Algorithm \ref{alg:MSTP} to estimate the \emph{truncated hitting time} $\widehat{p}_{\vecS}^{\ell,hit}[t]$(i.e., the probability of hitting $t$ starting from $\vecS$ for the first time in $\ell$ steps). By setting $\vecS=\vecE_t$, we get an estimate for the expected {\em truncated return time}  $\EE[T_t\mathds{1}_{\{T_t\leq\ell_{\max}\}}] = \sum_{\ell\leq\ell_{\max}}\ell\widehat{p}_{\vecE_t}^{\ell,hit}[t]$ where $T_t$ is the hitting time to target $t$. Now, using that fact that $\pi[t]=1/\EE[T_t]$, we can get a lower bound for $\pi[t]$ which converges to $\pi[t]$ as $\ell_{\max}\rightarrow\infty$. We note also that the truncated hitting time has been shown to be useful in other applications such as identifying similar documents on a document-word-author graph \cite{sarkar2008fast}.

To estimate the truncated hitting time, we modify Algorithm \ref{alg:MSTP} as follows: at each stage $i\in\{1,2,\ldots,\ell_{\max}\}$ (note: not $i=0$), instead of REVERSE-PUSH-MSTP$(t,i)$, we update $\widetilde{\p}_t^i[t]=\p_t^i[t] + \r_t^i[t]$, set $\widetilde{\r}_t^i[t]=0$ {\em and do not push back $\r_t^i[t]$ to the in-neighbors of $t$ in the $(i+1)^{th}$ stage}. The remaining algorithm remains the same. It is is plausible from the discussion in Section \ref{ssec:intuition} that the resulting quantity $\widehat{p}_{\vecS}^{\ell,hit}[t]$ is an unbiased estimate of $\PP[\mbox{Hitting time of }t = \ell|X_0\sim\vecS]$ -- we leave developing a complete algorithm to future work.

%\item \textbf{Exact Stationary Probabilities in Strong Doeblin chains}: A strong Doeblin chain~\cite{Doeblin1940} is obtained by mixing a Markov chain $W$ and a distribution $s$ as follows: at each transition, the process proceeds according to $W$ with probability $\alpha$, else samples a state from $s$. Doeblin chains are widely used in ML applications -- special cases include the celebrated PageRank metric~\cite{Page1999}, variants such as HITS and SALSA~\cite{Lempel2000}, and other algorithms for applications such as ranking~\cite{Negahban2012} and structured prediction~\cite{Steinhardt2015}. An important property of these chains is that if we sample a starting node $V_0$ from $s$ and sample a trajectory of length $Geometric(\alpha)$ starting from $V_0$, then {\em the terminal node is an unbiased sample from the stationary distribution}~\cite{Athreya2003}. 
%There are two ways in which our algorithm can be used for this purpose: one is to replace the REVERSE-PUSH-MSTP algorithm with a corresponding local update algorithm for the strong Doeblin chain (similar to the one in Andersen et al.~\cite{Andersen2007} for PageRank), and then sample random trajectories of length $Geometric(\alpha)$. 
%and average the residues of the terminal nodes (similar to~\cite{Lofgren2014}). 
%A more direct technique is to choose some $\ell_{\max}>>1/\alpha$, estimate $\{\PR_{\vecS}^{\ell}[t]\}\,\fall\ell\in[\ell_{\max}]$ and then directly compute the stationary distribution as $\PR[t]=\sum_{\ell=1}^{\ell_{\max}}\alpha^{\ell-1}(1-\alpha)\PR_{\vecS}^{\ell}[t]$. 

\item\textbf{Graph Diffusions:}
%The second approach proposed above shows that we can use MSTP-estimation as a subroutine to estimate the $t^{th}$ element of any given polynomial of the transition matrix:
If we assign a weight $\alpha_i$ to random walks of length $i$ on a (weighted) graph, the resulting scoring functions
$f(W,\vecS)[t]:=\sum_{i=0}^{\infty}\alpha_i\left(\vecS^TW^{i}\right)[t]$ are known as a \emph{graph diffusions}~\cite{Chung2007} and are used in a variety of applications. The case where $\alpha_i=\alpha^{i-1}(1-\alpha)$ corresponds to PageRank. If instead the length is drawn according to a Poisson distribution (i.e., $\alpha_i=e^{-\alpha}\alpha^i/i!$), then the resulting function is called the \emph{heat-kernel} $h(G,\alpha)$ -- this too has several applications, including finding communities (clusters) in large networks~\cite{Kloster2014}. Note that for any function $f$ as defined above, the truncated sum $f^{\ell_{\max}}=\sum_{i=0}^{\ell_{\max}}\alpha_i\left(\PR_{\vecS}^TW^{i}\right)$ obeys $||f-f^{\ell_{\max}}||_{\infty}\leq\sum_{\ell_{\max}+1}^{\infty}\alpha_i$.
Thus a guarantee on an estimate for the truncated sum directly translates to a guarantee on the estimate for the diffusion.  We can use MSTP-estimation to efficiently estimate these truncated sums.
% as long as the tail $\sum_{\ell_{\max}+1}^{\infty}\alpha_k$ is small. 
We perform numerical experiments on heat kernel estimation in the next section.

\item\textbf{Conductance Testing in Graphs:} MSTP-estimation is an essential primitive for conductance testing in large Markov chains~\cite{Goldreich2011}. In particular, in regular undirected graphs, Kale et al~\cite{Kale2008} develop a sublinear bidirectional estimator based on counting collisions between walks in order to identify `weak' nodes -- those which belong to sets with small conductance. Our algorithm can be used to extend this process to any graph, including weighted and directed graphs.

%\item\textbf{Local Algorithms:} There is a lot of interest recently on {\em local algorithms} -- those which perform computations given only a small neighborhood of a source node~\cite{Lee2013}. In this regard, we note that \texttt{Bidirectional-MSTP} gives a natural local algorithm for MSTP estimation, and thus for the applications mentioned above -- given a $k$-hop neighborhood around the source and target, we can perform \texttt{Bidirectional-MSTP} with $\ell_{\max}$ set to $k$. The proof of this follows from the fact that the invariant in Equation \eqref{eq:mstpest} holds after any sequence of REVERSE-PUSH-MSTP operations. 

\end{itemize}

\section{Experiments}
\label{sec:sims}

To demonstrate the efficiency of our algorithm on large Markov chains, we use \emph{heat kernel estimation} (cf. Section \ref{sec:apps}) as an example application. The heat kernel is a graph diffusion, defined as the probability of stopping at the target on a random walk from the source, where the walk length is sampled from a $Poisson(\ell)$ distribution. It is important in practice as it has empirically been shown to detect communities well, in the sense that the heat kernel from $s$ to $t$ tends to be larger for nodes $t$ sharing a community with $s$ than for other nodes $t$.  Note that for finding complete communities around $s$, previous algorithms which work from a single $s$ to many targets are more suitable.  Our algorithm is relevant for estimating the heat kernel from $s$ to a small number of targets.  For example, in the application of personalized search on a social network, if user $s$ is searching for nodes $t$ satisfying some query, we might find the top-$k$ results using conventional information retrieval methods for some small $k$, then compute the heat kernel to each of these $k$ targets, to allow us to place results $t$ sharing an inferred community with $s$ above less relevant results.  A related example is if a social network user $s$ is viewing a list of nodes $t$, say the set of users attending some event, our estimator can be used to rank the users $t$ in that list in decreasing order of heat kernel.  This can help $s$ quickly find the users $t$ closest to them in the network.

We use our estimator to compute heat kernels on diverse, real-world graphs that range from millions to billions of edges as described in Table \ref{table:graphs}.  
\begin{table}
\centering
\caption{Datasets used in experiments}
\label{table:graphs}
\begin{tabular}{|c|c|c|c|} \hline
Dataset & Type & \# Nodes & \# Edges\\ \hline
Pokec & directed & 1.6M & 30.6M\\ \hline
LiveJournal & undirected & 4.8M & 69M\\ \hline
Orkut & undirected & 3.1M & 117M\\ \hline
Twitter-2010 & directed & 42M & 1.5B\\ \hline
\end{tabular}
\end{table}
For each graph, for random (source, target) pairs, we compute the heat kernel.  We set average walk-length $\ell=5$ since for larger values of $\ell$, the walk mixes into the stationary distribution, ``forgetting'' where it started. %, but our qualitative running-time results are not sensitive to the choice of $\ell$.
  We set a maximum length of 10 standard deviations above the expectation $\ell + 10 \sqrt{\ell} \approx 27$; the probability of a walk being longer than this is $10^{-12}$, so we can ignore that event.

We compare Algorithm \ref{alg:MSTP} to two state-of-the-art algorithms: The natural Monte Carlo algorithm, and the push forward algorithm introduced in \cite{Kloster2014}.  All three have parameters which allow them to trade off speed and accuracy.  For a fair comparison, we choose parameters such that the mean relative error of all three algorithms is around 10\%, and for those parameters we measure the mean running time of all three algorithms.  We implemented all three algorithms in Scala (for the push forward algorithm, our implementation follows the code linked from \cite{Kloster2014}).

\begin{figure}
\begin{center}
\includegraphics[width=\textwidth]{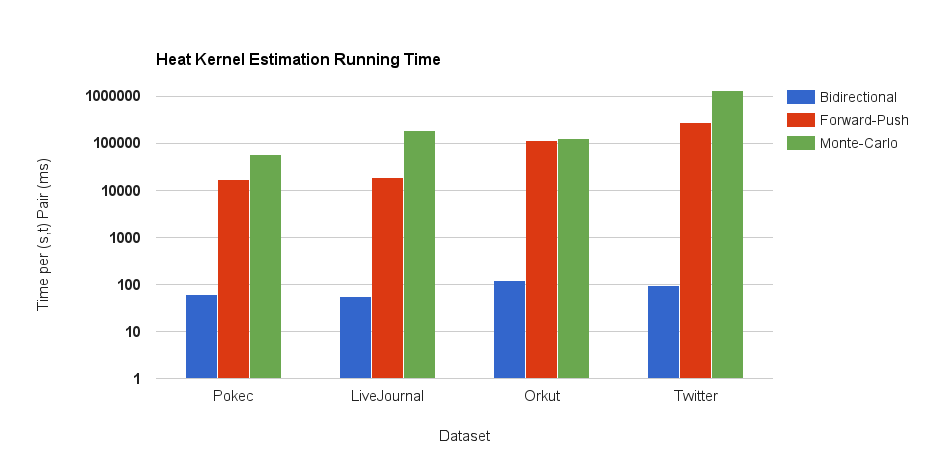}
\end{center}
\label{fig:runtime}
\caption{Estimating heat kernels: Bidirectional MSTP-estimation vs. Monte Carlo, Forward Push. For a fair comparison of running time, we choose parameters such that the mean relative error of all three algorithms is around 10\%.   Notice that our algorithm is 100 times faster per (source, target) pair than state-of-the-art algorithms.
%Visualizing a sequence of REVERSE-PUSH-MSTP operations: Given the Markov chain on the left, we choose $R$ as the target, and perform REVERSE-PUSH-MSTP operations $(R,0),(G,1),(B,1)$ and $(R,2)$. Each state has multiple virtual nodes at any level $i$, each corresponding to a unique length-$i$ path to the target $R$. Edges correspond to completed push operations; colored virtual nodes are those where either the estimate or residue is non-zero.
}
\end{figure}

Our results are presented in Figure \ref{fig:runtime}.
% \todo{or in in Table \ref{table:runtime}}  
% \begin{table}
% \centering
% \caption{Running time per (source, target) pair of our algorithm compared to two previous algorithms}
% \label{table:runtime}
% \begin{tabular}{|c|c|c|c|c|} \hline
% Dataset & Bidirectional & Forward-Push & Monte Carlo 
% & Relative Improvement\\ \hline
% Pokec & 61 ms & 16 s & 56 s & 270x\\ \hline
% LiveJournal & 56 ms & 18 s & 187 s & 330x\\ \hline
% Orkut & 120 ms & 120 s & 120 s & 960x\\ \hline
% Twitter-2010 & 93 ms & 270 s & 1317 s & 2900x\\ \hline
% \end{tabular}
% \end{table}
 We find that on these four graphs, our algorithm is 100x faster (per $(s, t)$ pair) than these state-of-the-art algorithms.  For example, on the Twitter graph, our algorithm can estimate a heat kernel score is $0.1$ seconds, while the state-of-the-art algorithms both take more than 4 minutes. We note though that the state-of-the-art algorithms were designed for community detection, and return scores from the source to all targets, rather than just one target. Our algorithm's advantage applies in applications where we want the score from a source to a single target or small set of targets.  

The Pokec \cite{takac2012data}, Live Journal \cite{mislove-2007-socialnetworks}, and Orkut \cite{mislove-2007-socialnetworks} datasets were downloaded from the Stanford SNAP project \cite{SnapProject}. The  
%DBLP-2011 \cite{BRSLLP}, 
Twitter-2010 \cite{BRSLLP}
% and UK 2007-05 Web Graph \cite{BRSLLP} were 
was downloaded from the Laboratory for Web Algorithmics \cite{WebAlgorithmics}.  For reproducibility, the source code of our experiments are available on our website\footnote{\url{http://cs.stanford.edu/~plofgren}}.

\chapter{Precomputation for PPR Estimation}
 \label{precompute_ppr}
Our experiments show that \bippr{} is efficient, especially for uniform random targets, but for worst-case targets with many in-neighbors, it can still take several seconds on Twitter-2010, which limits its usefulness for real-time search.  In this chapter we describe a simple method of precomputing forward walk and reverse residual vectors for each node in a large graph which enables very fast, real-time PPR estimation.
\section{Basic Precomputation Algorithm}
As we show in Section \ref{ppr_as_dot_product}, the estimation in \bippr{} can be expressed as a simple dot product between a vector $x^s$ depending only on $s$, and a vector $y^t$ depending only on $t$, which can both be pre-computed.  In particular, if we define  $y^t = (p^t, r^t) \in \R^{2n}$ and $x^s = (e_s, \tilde{\pi}_s) \in \R^{2n})$ where $e_s$ is the standard basis vector and $ \tilde{\pi}_s$ is a precomputed random walk vector, then $\pi_s(t) \approx x^s \cdot y^t$.  The basic idea of this section is to precompute all these vectors $x^s$ and $y^t$ in advance, then compute the dot-product at run time when $s$ and $t$ are given.  The challenge  is that the sparcity of the vectors $y^t$ vary dramatically: on average they have $O(\sqrt{m}/\epsilon)$ non-zero values for $\epsilon$ relative error and constant failure probability (Theorem \ref{thm:bidirmain}), but ``popular'' targets $t$ with high PageRank will have up to $\Theta(n)$ non-zero values.  For this reason it can be slow to access $y^t$ and compute the dot-product $x^s \cdot y^t$ at run time on a single machine.  Also, the total size of all the $x^t$ and $y^t$ is $O(n \sqrt{m} / \epsilon)$, which makes it infeasible to store them on a single machine.

Our solution is to distribute all the vectors $x^s$ and $y^t$ among a set of servers in advance and compute the dot-product in a distributed way.  Suppose we use $k$ servers and sharding function $h: V \to [k]$, so the $i$th server stores the non-zero coordinates $x^s_v$ and $y^t_v$ for all $s, t \in V$ and for all $v$ such that $h(v)=i$ .  Then we can write the dot-product as a sum of independent contributions
\[ x^s \cdot y^t = \sum_{i=1}^k E_i \]
where
\[ E_i = \sum_{v: h(v) = i} x^s(v) y^t(v) .\]
During pre-computation, we compute each vector $x^s$ and shard it among the $k$ servers, so for each $s \in [n]$, server $i$ has local access to $x^s(v)$ for all $v$ such that $h(v)=i$.  We similarly shard every vector $y^t$ among the $k$ servers.
Given a query $(s, t)$, we see that each shard can do a local dot-product based on the nodes $v$ assigned to it to compute $E_i$, send a single number $E_i$ to a central broker, and then the broker can return an estimate to $\PR_s(t)$.  

The total communication cost is $O(k)$ and the expected computation per server (which concentrates because the number of nonzero entries of the vectors is much greater than $k$ for realistic values of $k$ and $m$) is $\frac{1}{k}$ times the computation cost for a single machine, which is the best parallelism speed-up we could hope for.  In contrast, if we simply stored each vector $x^s$ and $y^r$ in a distributed key-value store, we would need to transfer at least one of the two vectors over the network, requiring communication $\Theta \pn{\sqrt{m} / \epsilon}$.

The total storage cost of this method is $O(n \sqrt{m} / \epsilon)$ because we store $2n$ vectors (a forward vector $x^v$ and reverse vector $y^v$ for each $v \in V$) and the proof of Theorem \ref{thm:fastpprtime} shows they each have $O(\sqrt{m} / \epsilon)$ non-zero coordinates on average.  The total pre-computation time is also $O(n \sqrt{m} / \epsilon)$.

\section{Combining Random Walks to Decrease Storage}
\label{sec:combine_walks}
We can decrease this storage by combining the random walk vector from multiple nodes.  A simple version of this technique (proposed in \cite{fogaras2004towards}) just uses the neighbors of the source node $s$. To illustrate the idea, suppose every node $s$ in the graph has 10 neighbors $v_1, \ldots, v_{10}$, and our accuracy analysis says we need 100,000 random walks from each $s$ to form $x^s$.  Then rather than pre-compute 100,000 random walks from each $s$, we can pre-compute 10,000 random walks from each node $v$ in the graph to form $\tilde{x}^v$, and average the walks from the neighbors of $s$ to form
\[x^s = (1-\alpha) \frac{1}{10} \sum_{i=1}^{10} \tilde{x}^{v_i}.\]
This is because if we sampled 100,000 walks from $s$, a fraction $1-\alpha$ of them would continue to a neighbor, and on average $\frac{1}{10}$ of those would continue from each neighbor, so we can combine the 10,000 walks at each neighbor to simulate 100,000 walks from $s$.
%This works because for $t \neq s$, 
%\[ \pi_s(t) = (1 - \alpha)\frac{1}{\outdegree{s}} \sum_{ u \in \outneighbors{s}} \pi_u(t). \]
Of course in real graphs some nodes $s$ will have small out-degree and others will have large out-degree, so we need a more general technique.

We can combine the walks from a larger set of nodes around $s$ using the \fpush{} algorithm from Section \ref{sec:forward_push}. Recall that this algorithm returns a vector of forward estimates $p_s$ and forward residuals $r_s$ which satisfy the invariant that for all $t \in V$
\[
 \pi_s(t) = p_s(t) + \sum_{v \in V} r_s(v) \pi_v(t).
\]
If we apply the push forward technique until all forward residuals $r_s(v)$ are less than some threshold $\rmaxf$, then we can simulate $n_f$ walks from $s$ by taking $\rmaxf n_f$ random walks from each node $u$ in the graph, letting $\tilde{x}^{v}$ be their stopping distribution, and defining
\[x^s =  \sum_{v \in V} r_s(v) \tilde{x}^{u_i}.\]
We don't need to explicitly form $x^s$ on a single machine.  Rather we pre-compute $p_s$ and $r_s$ and store them in a key-value store with key $s$.  Then given query $(s, t)$ the broker broadcasts $r_s$ to all servers, each server $i$ computes estimate
\[ E_i = \sum_{v: h(v) = i} r_s(u) \tilde{x}^u(v) y^t(v) \]
and the broker returns $p_s(t) + \sum_i E_i \approx \pi_s(t)$.

Notice that averaging walks from the out-neighbors of $s$ is a special case of this technique: after the first forward-push of their algorithm, $r_s(v) = (1-\alpha)\frac{1}{\outdegree{s}}$ for out-neighbors $v$ of $s$ and $r_s(v)=0$ for other nodes $v$, so we recover the simpler out-neighbor technique.

The above algorithm can be improved by treating high-degree nodes specially.  The issue is that if we push forward from a node with high out-degree (like Barack Obama who follows more than 500,000 people on Twitter), that greatly increases the size of the forward residual vector.  Our proposed fix is to choose some maximum degree $\dmax$ and only push forward from nodes with degree at most $\dmax$.  For nodes $v$ of degree larger than $\dmax$, we do the full number of forward walks required for an accurate estimate without random walk sharing.  Full pseudo-code is given in Algorithm \ref{alg:pprDotPrecomputation} and \ref{alg:pprDotQuery}.
\begin{algorithm}[!ht]
\caption{\texttt{PPR-Precomputation}$(G, \alpha, \delta, \dmax)$}
\label{alg:pprDotPrecomputation}
\begin{algorithmic}[1]
\REQUIRE graph $G=(V, E)$ with edge weights $w_{u,v}$, teleport probability $\alpha$, minimum PPR $\delta$ (estimates of $\pi_s(t)$ are accurate if $\pi_s(t) \geq \delta$), max degree parameter $\dmax$
\STATE Note: This algorithm runs on a cluster of large machines which can each hold the graph in RAM.  The for loop over $v \in V$ is parallelized across these large machines.
\STATE Choose parameters $\rmaxr = \sqrt[3]{\frac{c_2^2 \delta}{c_1 c_3}}$, $\rmaxf = \sqrt[3]{\frac{c_3^2 \delta}{c_1 c_2}}$, $\nwalk=\frac{c_1 \rmaxf \rmaxr}{\delta}$ (see Section \ref{pprStorageAnalysis} for analysis and details on these constants).
\STATE Choose some hash function $h: V \to [\nshard]$, where $\nshard$ is the number of servers.
\FOR{all nodes $v \in V$}
\STATE  Run algorithm \fpush{}($v$, $\alpha$, $\rmaxf$) with two modifications: First, Run until $\forall u, r_v(u) < \rmaxf$. Second, only push forward from nodes $u$ with $\outdegree{u} \leq \dmax$.  Store the resulting forward estimate and residual vectors  $(p_v, r_v)$ in a distributed key-value store with key $v$.
\STATE Run algorithm \rpush{}($G, \alpha, v, \rmaxr$).  With the resulting reverse estimate and residual vectors $(p^v, r^v)$, define $y^v=(p^v, r^v)\in \R^{2n}$ and send each nonzero entry $y^v(u)$ to server $h(u)$.
% and each nonzero entry $r^v_u$ to server $h(u)$
\STATE If $\outdegree{v} \leq \dmax$, sample $\nwalk$ random walks from $v$; otherwise sample $\frac{c_1  \rmaxr}{\delta}$ walks from $v$.  Stop each walk after each step with probability $\alpha$. Let $\tilde{\pi}_v$ be the distribution of their endpoints.  Define $\tilde{x}_v = (e_s, \tilde{\pi}_s)\in \R^{2n}$ and send each non-zero entry $\tilde{x}_v(u)$ to server $h(u)$.
\ENDFOR
\end{algorithmic}
\end{algorithm}

\begin{algorithm}[!ht]
\caption{\texttt{PPR-Query-Given-Precomputation}$(s, t)$}
\label{alg:pprDotQuery}
\begin{algorithmic}[1]
\REQUIRE source node $s$, target node $t$
\STATE Note: This is executed on an app server or broker server.
\STATE Retrieve the forward estimate and residual vector $(p_s, r_s)$ from the key-value store. 
\STATE Broadcast $(s, t, r_s)$ to all ppr storage servers.
\STATE Each server $i$ computes
\[ E_i = \sum_{v: h(v) = i} r_s(u) \tilde{x}_u(v) y^t(v) \]
and returns $E_i$ to the broker.
\RETURN $p_s(t) + \sum_i E_i \approx \pi_s(t)$.
\end{algorithmic}
\end{algorithm}    

%Note on related work section: \cite{Bahmani2010} also proposes a technique for using pre-computed walks to simulate a larger number of random walks from a smaller number of random walks which in part inspired our approach.  They propose to precompute a number of random walks starting from every node.  Then given a source node $s$, they sample a random walk until a node is reached with has pre-computed random walks that haven't been used yet in the estimation of $\pi_s$, and at the end of their algorithm they produce a large number of walks from $s$ using less communication (but not less computation) than it would take to sample the walks directly from $s$.   Their approach efficiently solves the top-k personalized PageRank problem from a single source to all targets.  However, for re-ranking applications, where we are given a single source and a small number of targets to rank by personalized PageRank, their method doesn't focus on just the targets given, and thus requires more communication and computation than our approach which uses  bidirectional methods to only estimate personalized PageRank to the relevant targets.  

\section{Storage Analysis on Twitter-2010}
\label{pprStorageAnalysis}
To estimate the amount of storage needed, both with and without combining walks, we now present a storage analysis for the Twitter-2010 graph, one of the largest publicly available social networks with 1.5 billion edges.  Without walk sharing, the only parameter to optimize is $\rmaxr$, the maximum residual for the reverse push algorithm.  From Theorem \ref{thm:bidirmain}, the number of walks needed per node is then $c_1 \frac{\rmaxr}{\delta}$ where $\delta$ is the smallest personalized PageRank value we can accurately estimate (for example $\delta=\frac{4}{n}$ as in our experiments) and $c_1$ is a constant controlling the relative error of the estimates.  Our experiments in Section \ref{sec:bippr_experimental_time} found that $c_1=7$ achieved mean relative error less than 10\%, which is a natural accuracy goal given that the personalized PageRanks vary on a wide scale.  The storage requirement of the random walk endpoint distribution is thus bounded simply by the total number of walks 
\[n c_1 \frac{\rmaxr}{\delta}.\]

\begin{figure}[tbp]
\begin{center}
\includegraphics[width=1\columnwidth]{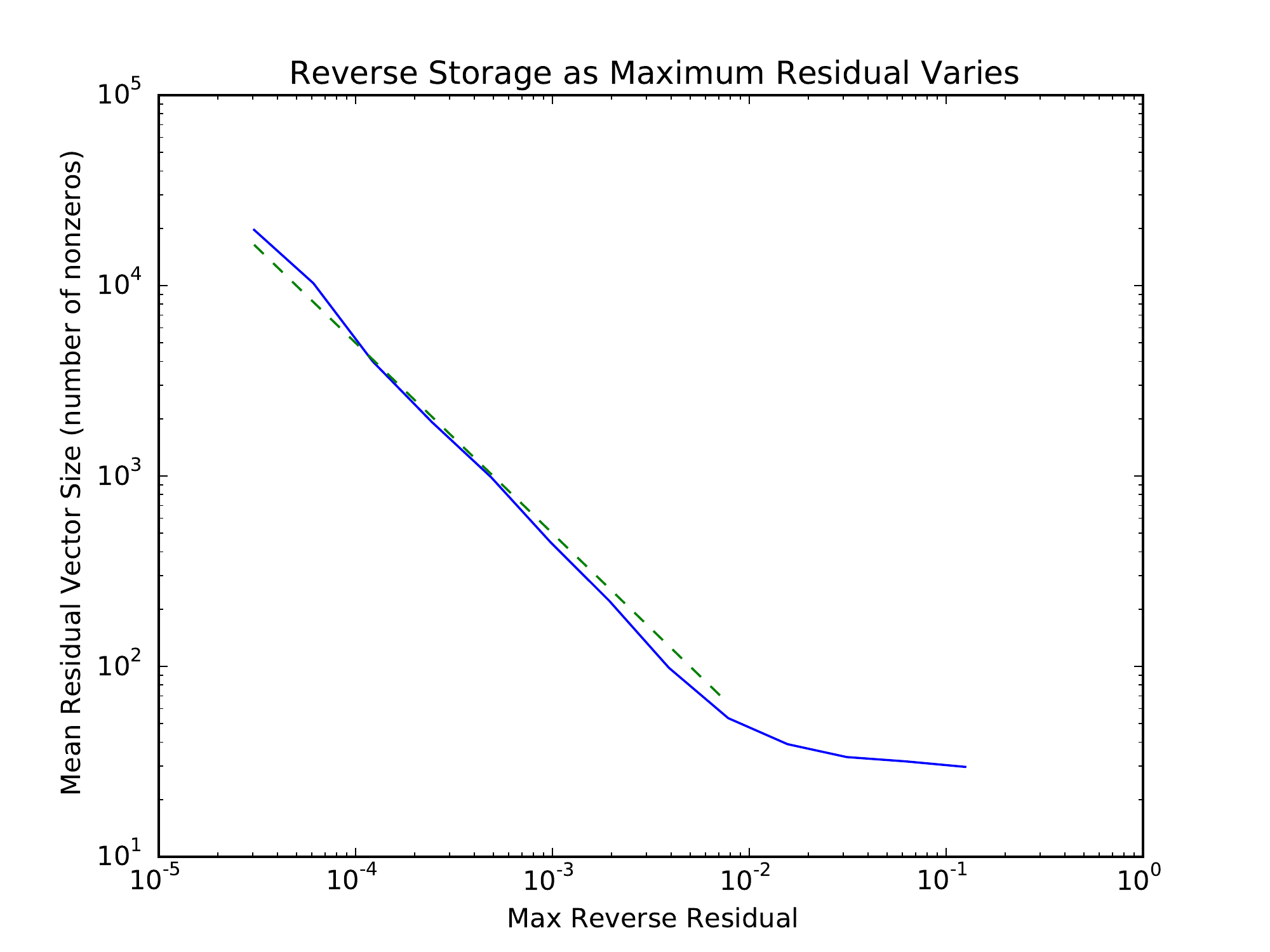}
\caption{Mean reverse residual size (number of non-zeros) as $\rmaxr$ varies, on Twitter-2010.  The dotted line is $y=\frac{0.5}{\rmaxr}$.}
\label{fig:reverse_residual_size}
\end{center}
\end{figure}

Now we consider the storage cost of the residual vectors.  A worst case bound on the total size (i.e. number of non-zeros) of the residual vectors from Theorem \ref{thm:fastpprtime} is $\frac{m}{\rmaxr}$. However, we find that in practice the total size is smaller than this.  On Twitter-2010, we sampled 2000 nodes $t$ uniformly at random, ran the algorithm \rpush{} for various values of $\rmaxr$ and measured the number of non-zero entries in the residual vector.  Figure
\ref{fig:reverse_residual_size}
shows that $\frac{c_2}{\rmaxr}$, where $c_2 \approx 0.5$ for Twitter-2010, is a good estimate of the reverse storage required per node.  Thus the total storage required is approximately
\[n c_1 \frac{\rmaxr}{\delta} + n c_2 \frac{1}{\rmaxr}.\]
If we choose $\rmaxr$ to minimize this expression, we get storage of $2 n \sqrt{\frac{c_1  c_2}{\delta}} = 1.0 \cdot 10^{12}$,
%n = 41.6e6; 2 * n * math.sqrt(0.5*7*n)
 or 8TB assuming each non-zero entry takes 8 bytes to store.  Here $n=41 $ million and we choose $\delta=\frac{4}{n}$ (meaning we can estimate PPR values 4 times the PPR of a random node).

\begin{figure}[tbp]
\begin{center}
\includegraphics[width=1\columnwidth]{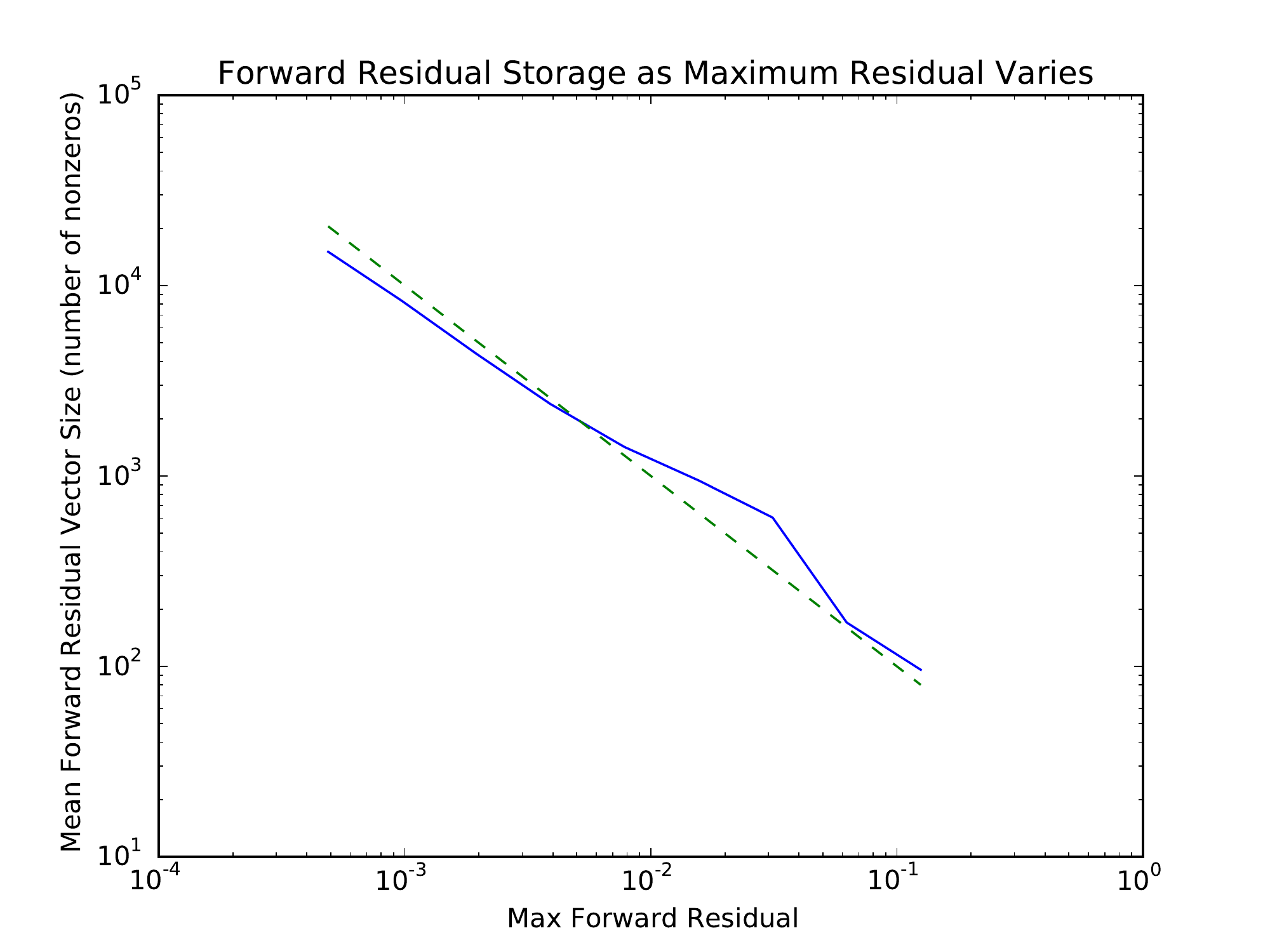}
\caption{Mean forward residual size (number of non-zeros) as $\rmaxf$ varies on Twitter-2010.  The dotted line is $y=\frac{10}{\rmaxf}$.}
\label{fig:forward_residual_size}
\end{center}
\end{figure}

Now we consider how this improves if we re-use walks.  Recall from above that we need to choose a parameter $\rmaxf$ which is the maximum forward residual value we allow, and then the number of forward walks needed is $\frac{c_1 \rmaxf \rmaxr}{\delta}$.  To estimate the size of the forward residuals, we sample 2000 uniformly random nodes on Twitter-2010, run the \fpush algorithm modified as described in Algorithm \ref{alg:pprDotPrecomputation} for each node for various values of $\rmaxf$, and measure the mean number of non-zero entries in the forward residual vectors.  As described above, we did not push forward from nodes with degree greater than $\dmax$, where we chose $\dmax = 1000$.  Figure
\ref{fig:forward_residual_size}
shows that $\frac{c_3}{\rmaxf}$, where $c_3 \approx 10$ for Twitter-2010, is a good estimate of the forward residual storage required per node.
The total storage required is then approximately
\[n \frac{c_3}{\rmaxf} + n c_1 \frac{\rmaxr \rmaxf}{\delta} + n \frac{c_2}{\rmaxr}.\]
If we choose $\rmaxr$  and $\rmaxf$ to minimize this expression, we get storage of $3 n \sqrt[3]{\frac{c_1  c_2 c_3}{\delta}} = 1.4 \cdot 10^{11}$ entries 
%n = 41.6e6; 3 * n * math.pow(0.5*7*10*n, 1.0/3.0)
under the same assumptions as above.  In addition, for nodes with outdegree greater than $\dmax$, we need to store the full number of $\frac{c_1 \rmaxr}{\delta}$ random walks.  On Twitter-2010, only 220K nodes (0.5\% of all nodes)
% n = 41.6e6; 224090 / n
 have degree greater than $\dmax=1000$, so their total storage is $220,000 \cdot \frac{c_1 \rmaxr}{\delta} \approx 3 \cdot 10^{10}$ entries.
%n = 41.6e6; c1=7; c2=0.5; c3=10; 224090 * c1**(2.0/3) * c2**(2.0/3) * n**(2.0/3) / c3**(1.0/3)
The total of these is $1.7  \cdot 10^{11}$ entries, or 1.4 TB.

 %Note, these numbers are for the correct twitter orientation (outdegree counts people you follow).  I checked @WholeFoods's (nodeId 23933986) outdegree and indegree.
\section{Varying Minimum Probability to Reduce Storage}
One more optimization that can reduce storage is to vary the minimum probability $\delta$ for each target $t$.  In particular, we can set $\delta' = \max(\delta, \pi[t])$, since, as we alluded to in Section \ref{sec:choosing_delta}, if $\pi_s[t] < \pi[t]$, then $s$ is less interested in $t$ than a uniform random source is, so it is not useful to measure how (un)interested $s$ is in $t$.  Since we pre-compute the walks for each source $s$ independently of $t$, we cannot vary the number of walks simulated $w$, but for a given number of simulated walks we can solve the equation $w = c_1 \frac{\rmaxr}{\delta'} = c_1 \frac{\rmaxr}{\max(\delta, \pi[t])} $ for $\rmaxr$ to get
\[ \rmaxr = \frac{w \max(\delta, \pi[t])}{c_1}.\]
We leave the analysis of this optimization to future work.

\chapter{Personalized PageRank Search}
\label{sec:search_chapter}
In this chapter, based on \cite{bippr}, we extend our method of estimating PPR between a single (source, target) pair to doing full searches from a source $s$ to a set of candidate targets.

%!TEX root = main.tex
\section{Personalized PageRank Search}

We now turn from Personalized PageRank estimation to the Personalized PageRank search problem: 
%Recall from the introduction that this is the following problem:
\begin{center}
\emph{Given a start node $s$ (or distribution $\sigma$) and a query $q$ which filters the set of all targets to some list $T = \{t_i\} \subseteq V$, return the  top-$k$ targets ranked by $\pi_s[t_i]$.}
\end{center}

We consider as baselines two algorithms which require no pre-computation.  They are efficient for certain ranges of $\size{T}$, but our experiments show they are too slow for real-time search across most values of $\size{T}$:
\begin{itemize}[nosep,leftmargin=*]
\item \noindent\textbf{\mc{}}~\cite{Avrachenkov2007,Fogaras2005}:  Sample random walks from $s$, and filter out any walk whose endpoint is not in $T$.  If we desire $\nsample$ samples, this takes time $O\pn{\nsample/\pi_s[T]}$, where $\pi_s[T] := \sum_{t \in T} \pi_s[t]$ is the probability that a random walk terminates in $T$.  
This method works well if $T$ is large, but in our experiments on Twitter-2010 it takes minutes per query for $\size{T}=1000$ (and hours per query for $\size{T}=10$).

\item \noindent\textbf{\bipprbasic{}}: On the other hand, we can estimate $\pi_s[t]$ to each $t \in T$ separately using \bippr.  
This has an average-case running time $O\pn{\size{T} \sqrt{\bar{d}/\delta_k}}$ where $\delta_k$ is the PPR of the $k^{th}$ best target.  
This method works well if $T$ is small, but is too slow for large $T$; 
in our experiments, it takes on the order of seconds for $\size{T} \leq 100$, but more than a minute for $\size{T}=1000$.
\end{itemize}

%On the other hand, 
If we are allowed pre-computation, then we can improve upon \bippr{} %(\bipprbasic{}) 
by precomputing and storing a reverse vector from all target nodes. 
\label{ppr_as_dot_product} To this end, we first observe that the estimate $\hat{\pi}_s[t]$  can be written as a dot-product.
Let $\tilde{\pi}_s$ be the empirical distribution over terminal nodes due to $w$ random walks from $s$ (with $w$ chosen as in Theorem \ref{thm:bidirmain}); we define the \emph{forward vector} $x_s \in \R^{2n}$ to be the concatenation of the basis vector $e_s$ and the random-walk terminal node distribution.
On the other hand, we define the \emph{reverse vector} $y^t \in \R^{2n}$, to be the concatenation of the estimates $p^t$ and the residuals $r^t$.
Formally, define
\begin{equation}
\label{eq:frvectors}
x_s = (e_s, \tilde{\pi}_s) \in \R^{2n},\quad\quad y^t = (p^t, r^t) \in \R^{2n}.
\end{equation}  
Now, from Algorithm \ref{alg:BIPPR}, we have
\begin{equation}
\label{eq:dotprod}
	\hat{\pi}_s[t] = \langle x_s,y^t\rangle .
\end{equation}

The above observation motivates the following algorithm:
\begin{itemize}[nosep,leftmargin=*]
\item \noindent\textbf{\bipprprecomp{}}: In this approach, we first use \rpush{} to pre-compute and store a reverse vector $y^t$ for each $t \in V$. At query time, we generate random walks to form the forward vector $x_s$; now, given any set of targets $T$, we compute $\size{T}$ dot-products $\langle x_s, y^t\rangle$, and use these to rank the targets.  
This method now has an \emph{worst-case} running time $O\pn{\size{T} \sqrt{\bar{d}/\delta_k}}$. 
In practice, it works well if $T$ is small, but is too slow for large $T$. In our experiments (doing 100,000 random walks at runtime) this approach takes around a second for $\size{T} \leq 30$, but this climbs to a minute for $\size{T}=10,000$.
\end{itemize} 

The \bipprprecomp{} approach is faster than \bipprbasic{} (at the cost of additional precomputation and storage), and also faster than \mc{} for small sets $T$, but it is still not efficient enough for real-time personalized search.  
%\todo{Crucially, it still has running time $\Omega(\size{T})$ as it computes a score for each target.}
This motivates us to find a more efficient algorithm that scales better than \bipprbasic{} for large $T$, yet is fast for small $|T|$. 
In the following sections, we propose two different approaches for this -- the first based on pre-grouping the precomputed reverse-vectors, and the second based on sampling target nodes from $T$ according to PPR.  
For convenience, we first summarize the two approaches: 

\begin{itemize}[nosep,leftmargin=*]
\item \noindent\textbf{\bipprgrouped{}}: Here, as in \bipprprecomp{}, we compute an estimate to each $t \in T$ using \bippr{}. 
However, we leverage the sparsity of the reverse vectors $y^t = (p^t, r^t)$ by first grouping them in a way we will describe. This makes the dot-product more efficient.   
This method has a \emph{worst-case} running time of $O\pn{\size{T} \sqrt{\bar{d}/\delta_k}}$, and in experiments we find it is much faster than \bipprprecomp{}.  
%It works well if $T$ is small or moderately sized, but its running time is $\Omega(\size{T})$ so it is slow for the largest target sets.
For our parameter choices its running time is less than 250ms across the range of $\size{T}$ we tried.

\item \noindent\textbf{\bipprsampling{}}: We again first pre-compute the reverse vectors $y^t$. 
Next, for a given target $t$, we define the \emph{expanded target-set} $T_t=\{v \in [2n]|y^t[v]\neq 0\}$, i.e., the set of nodes with non-zero reverse vectors from $t$. 
At run-time, we now sample random walks forward from $s$ to nodes in the expanded target sets. 
Using these, we create a sampler in average time $O\pn{\epr/\delta_k}$ (where as before $\delta_k$ is the $k^{th}$ largest PPR value $\pi_s[t_k]$), which samples nodes $t \in T$ with probability proportional to the PPR $\pi_s[t]$.  We describe this in detail in Section \ref{sec:sampler}. 
Once the sampler has been created, it can be sampled in $O(1)$ time per sample.  
The algorithm works well for any size of $T$, and has the unique property that in can identify the top-$k$ target nodes without computing a score for all $\size{T}$ of them.  For our parameter choice its running time is less than 250ms across the range of $\size{T}$ we tried.
\end{itemize}

We note here that the case $k=1$ (i.e., for finding the top PPR node) corresponds to solving a Maximum Inner Product Problem.  
In a recent line of work, Shrivastava and Li~ \cite{shrivastava2014asymmetric,shrivastava2014improved} propose a sublinear time algorithm for this problem based on Locality Sensitive Hashing; however, their method assumes that there is some bound $U$ on $\norm{y^t}_2$ and that $\max_t \langle x_s , y^t\rangle$ is a large fraction of $U$. %For example, in Figure 1 of \cite{shrivastava2014improved} the smallest fraction of $U$ they consider is $\max_t x_s \cdot y^t > 0.1 U$.  
In personalized search, we usually encounter small values of $\max_t \langle x_s , y^t\rangle$ relative to $\max \norm{y^t}_2$ -- finding an LSH for Maximum Inner Product Search in this regime is an interesting open problem for future research.  
Our two approaches bypass this by exploiting particular structural features of the problem -- \bipprgrouped{} exploits the sparsity of the reverse vectors to speed up the dot-product, and \bipprsampling{} exploits the skewed distribution of PPR scores to find the top targets without even computing full dot-products.

%!TEX root = main.tex

\subsection{Bidirectional-PPR with Grouping} 
\label{sec:bippr_dot_product} 

In this method we improve the running-time of \bipprprecomp{} by pre-grouping the reverse vectors corresponding to each target set $T$.
Recall that in \bipprprecomp{}, we first pre-compute reverse vectors $y^t = (p^t, r^t) \in \R^{2n}$ using \rpush{} for each $t$.  At run-time, given $s$, we compute forward vector $x_s = (e_s, \tilde{\pi}_s)$ by generating sufficient random-walks, and then compute the scores $\langle x_s , y^t\rangle$ for $t \in T$.  
Our main observation is that \emph{we can decrease the running time of the dot-products by pre-grouping the vectors $y^t$ by coordinate}.  
The intuition behind this is that in each dot product $\sum_v x_s[v] y^t[v]$, the nodes $v$ where $x_s[v] \neq 0$ often don't have $y^t[v] \neq 0$, and most of the product terms are 0.   
Hence, we can improve the running time by grouping the vectors $y^t$ in advance by coordinate $v$.  Now, at run-time, for each  $v$ such that $x_s[v] \neq 0$, we can efficiently iterate over the set of targets $t$ such that $y^t[v] \neq 0$.  

An alternative way to think about this is as a sparse matrix-vector multiplication $Y^{T} x_s$ after we form a matrix $Y^T$ whose rows are $y^t$ for $t \in T$.  
This optimization can then be seen as a sparse column representation of that matrix.

\begin{algorithm}[ht]
\caption{BiPPRGroupedPrecomputation$(T, \epr)$}
\label{alg:approxContributionsToSet}
\begin{algorithmic}[1]
\REQUIRE Graph $G$, teleport probability $\alpha$, target nodes $T$, maximum residual $\epr$
%\STATE // We will call ApproxContributions on each target, and group the resulting vectors by coordinate
\STATE $z\leftarrow$ empty hash map of vectors such that for any $v$, $z[v]$ defaults to an empty (sparse) vector in $\R^{2|V|}$
\FOR{$t \in T$}
  \STATE Compute $y^t = (p_t, r_t)\in \R^{2|V|}$ via \rpush{}$(G, \alpha, t, \epr)$
%  \STATE $y_t = (p_t, r_t) \in \R^{2n}$ %// E.g. $y_t$ can store the values of $p_t$ in negative coordinates
  \FOR{$v \in [2 \size{V}] \text{ such that } y_t[v] > 0$}
    \STATE $z[v][t] = y^t[v]$
  \ENDFOR
\ENDFOR
\RETURN $z$
\end{algorithmic}
\end{algorithm}
\vspace{-0.2cm}

\begin{algorithm}[ht]
\caption{BiPPRGroupedRankTargets$(s, \epr, z)$}
\label{alg:RankTargets}
\begin{algorithmic}[1] 
\REQUIRE Graph $G$, teleport probability $\alpha$, start node $s$, maximum residual $\epr$, $z$: hash map of reverse vectors grouped by coordinate 
\STATE Set number of walks $w = c \frac{\epr}{\delta}$  (In experiments we found $c=20$ achieved precision@3 above 90\%.)
\STATE Sample $w$ random-walks of length $Geometric(\alpha)$ from $s$; compute $\tilde{\pi}_s[v]=$ fraction of walks ending at node $v$
% = \frac{1}{w}\size{\{\text{walks terminating at $v$}\}}$.
\STATE Compute $x_s = (e_s, \tilde{\pi}_s) \in \R^{2|V|}$
\STATE Initialize empty map score from $V$ to $\R$
%\STATE // We will set score($t$) to be $\langle x_s , y_t\rangle$ efficiently by only iterating over non-zero terms
\FOR{$v$ such that $x_s[v] > 0$}
  \FOR{$t$ such that $z[v][t] > 0$}
    \STATE  score($t$) $\pluseq x_s[v] z_v[t]$
  \ENDFOR
\ENDFOR
\STATE Return $T$ sorted in decreasing order of score
\end{algorithmic}
\end{algorithm}

We refer to this method as \bipprgrouped{}; the complete pseudo-code is given in Algorithm \ref{alg:RankTargets}.  
The correctness of this method follows again from Theorem \ref{thm:bidirmain}.  
In experiments, this method is efficient for $T$ across all the sizes of $T$ we tried, taking less than 250 ms even for $\size{T} = 10,000$.  
The improved running time of \bipprgrouped{} comes at the cost of more storage compared to \bipprprecomp{}. 
In the case of name search, where each target typically only has a first and last name, each vector $y^t$ only appears in two of these pre-grouped structures, so the storage is only twice the storage of \bipprprecomp{}.
On the other hand if a target $t$ contains many keywords, $y^t$ will be included in many of these pre-grouped data structures, and storage cost will be significantly greater than for \bipprprecomp{}.

%\todo{delete?  In some application, there may be an efficient way to prune the set of candidate targets to a smaller set $T'$.  (For example, if a user searches for ``Football'' on Twitter,  we could simply take the top-100 candidates mentioning football sorted by global relevance metrics.)  In this case, we can efficiently do $\abs{T'}$ \fasterppr queries, and take the top-$k$.  But for full personalization, or for situations where it is impossible to create a short candidate list, we also propose a sampling approach.}

%!TEX root = main.tex

\subsection{Sampling from Targets Matching a Query}
\label{sec:sampler}

The key idea behind this alternate method for PPR-search is that by sampling a target $t$ in proportion to its PageRank we can quickly find the top targets without iterating over all of them.  
After drawing many samples, the targets can be ranked  according to the number of times they are sampled.  
Alternatively a full \bippr{} query can be issued for some subset of the targets before ranking. 
This approach exploits the skewed distribution of PPR scores in order to find the top targets.  
In particular, prior empirical work has shown that on the Twitter graph, for each fixed $s$, the values $\pi_s[t]$ follow a power law \cite{Bahmani2010}.

We define the PPR-Search Sampling Problem as follows:
\begin{center}
\emph{Given a source distribution $s$ and a query $q$ which filters the set of all targets to some list $T = \{t_i\} \subseteq V$, sample a target $t_i$ with probability $p[t_i] = \frac{\pi_s[t_i]}{\sum_{t \in T} \pi_s[t]}$.}
\end{center}
We develop two solutions to this sampling problem.  The first, in $O(w) = O(\epr/\delta_k)$ time, generates a data structure which can generate an arbitrary number of independent samples from a distribution which approximates the correct distribution. 
The second can generate samples from the exact distribution $\pi_s[t_i]$, and generates complete paths from $s$ to some $t \in T$, but requires time $O(\epr/\pi_s[T])$ per sample.  Because the approximate sampler is more efficient, we present that here and defer the exact sampler to Section \ref{sec:path_chapter}.

\noindent\textbf{The \bipprsampling{} Algorithm}

\begin{figure}[thb]
\centering
\includegraphics[width=1\columnwidth]{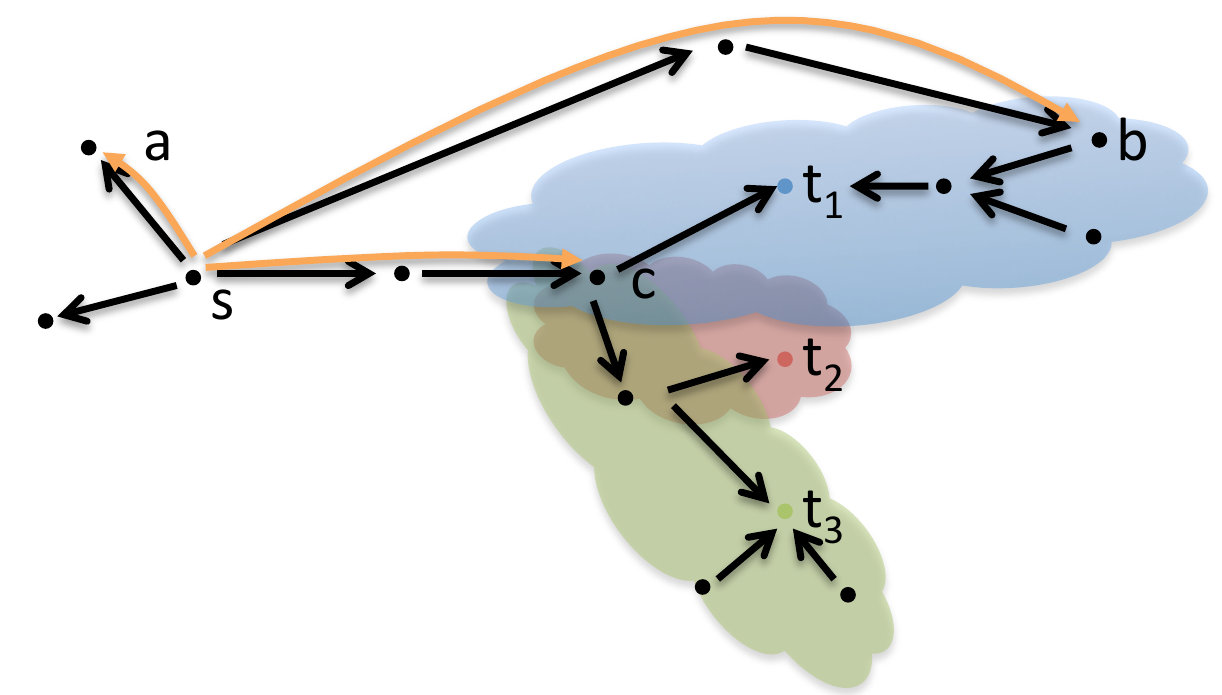}
\caption[Example]{Search Example: Given target set $T=\{t_1, t_2, t_3\}$, for each target $t_i$ we have drawn the expanded target-set, i.e., nodes $v$ with positive residual $y^{t_i}[v]$.  
From source $s$, we sample three random walks, ending at nodes $a$, $b$, and $c$.  
Now suppose $y^{t_1}(b)=0.64, y^{t_1}(c) = 0.4, y^{t_2}(c)=0.16$, and $y^{t_3}(c)=0.16$ -- note that the remaining residuals are $0$.
Then we have $y^T(a)=0, y^T(b)=0.64$ and $y^T(c)=0.72$, and consequently, the sampling weights of $(a,b,c)$ are $(0,0.213,0.24)$.    
Now, to sample a target, we first sample from $\{a,b,c\}$ in proportion to its weight. Then if we sample $b$, we always return $t_1$; if we sample $c$, we sample $(t_1,t_2,t_3)$ with probability $(5/9,2/9,2/9)$.}
\label{fig:example1}
\end{figure}

The high level idea behind our method is \emph{hierarchical sampling}.  
Recall that the start node $s$ has an associated forward vector $x_s = (e_s, \pi_s)$, and from each target $t$ we have a reverse vector $y^t$; the PPR-estimate is given by $\pi_s[t] \approx \langle x_s , y^t\rangle$.  
Thus we want to sample $t \in T$ with probability:
\[ p[t] = \frac{\langle x_s , y^t\rangle}{\sum_{j \in T} \langle x_s , y^{j}\rangle}. \]
We will sample $t$ in two stages: first we sample an intermediate node $v \in V$ with probability:
\[ p'_s[v] = \frac{x_s[v]  \sum_{j \in T}  y^{j}[v]}{\sum_{j \in T} \langle x_s ,  y^{j}\rangle}. \]
Following this, we sample $t\in T$ with probability:
\[ p''_v[t] = \frac{y^t[v]}{\sum_{j \in T} y^{j}[v]} . \]
It is easy to verify that $p[t]=\sum_{v\in V}p'_s[v]p''_v[t]$. 
Figure \ref{fig:example1} shows how the sampling algorithm works on an example graph.  The pseudo-code is given in Algorithm \ref{alg:samplerPrecompute} and Algorithm \ref{alg:SampleApprox}.

Note that we assume that some set of supported queries is known in advance, and we first pre-compute and store a separate data-structure for each query $Q$ (i.e., for each target-set $T= \{t \in V: t \text{ is relevant to } Q\}$).  In addition, we can optionally pre-compute $w$ random walks from each start-node $s$, and store the forward vector $x_s$, or we can compute $x_s$ at query time by sampling random walks.

%\noindent\textbf{Precomputation}: 
%Given a target-set $T$, for each $t \in T$, we compute residuals $r^t$ and estimates $p^t$ using \rpush{}$(G,\alpha,t,\epr)$ (Algorithm \ref{alg:invPPR}); the choice of $\epr$ is discussed in the analysis section.  
%For each $T$ and each $v$, we also pre-compute $ \sum_{j \in T}  y^{j}_v$ and a constant-time sampler for the distribution: 
%\[p''_v[T]= \frac{y^t[v]}{\sum_{j \in T} y^{j}[v]} .\]
%
\begin{algorithm}[ht]
\caption{SamplerPrecomputationForSet$(T, \epr)$}
\label{alg:samplerPrecompute}
\begin{algorithmic}[1]
\REQUIRE Graph $G$, teleport probability $\alpha$, target-set $T$, maximum residual $\epr$
\FOR{$t \in T$}
  \STATE Compute $y^t = (p^t, r^t)\in \R^{2\size{V}}$ via \rpush{}$(G, \alpha, t, \epr)$
  %// E.g. $y^t$ can store the values of $p^t$ in negative coordinates
\ENDFOR
\STATE Compute $y^{T} = \sum_{t \in T} y^t$
\FOR{$v \in V \text{ such that } y^T[v] > 0$}
  \STATE Create sampler$_v$ which samples $t$ with probability $p''_v[t]$  (For example, using the alias sampling method  \cite{Walker:1977:EMG:355744.355749}, \cite[section~3.4.1]{knuth1998vol2}).
\ENDFOR
\RETURN ($y^T, \{\text{sampler}_v\}$)
\end{algorithmic}
\end{algorithm}

%\noindent\textbf{Sampling Algorithm}: 
%Given start node $s$ and target-set $T$, we load the relevant target data structures for $T$ and either load $x_s$ or compute it at run-time.  Next, in $O(w)$ time, we create a sampler for the distribution $p'_s[v]$ over $v$:
%\[p'_s[v] = \frac{x_s[v]  \sum_{j \in T}  y^{j}[v]}{\sum_{j \in T}\langle x_s ,  y^{j} \rangle} = \frac{x_s[v] y^{T}[v]}{\langle x_s , y^T \rangle}. \]
%Now we can generate independent samples in constant time per sample. We sample $v$ from this distribution, and then sample $t$ from $p''_v[t]$ using the pre-computed sampler. 

\begin{algorithm}[ht]
\caption{SampleAndRankTargets$(s, \epr, y^T, \{\text{sampler}_v\})$}
\label{alg:SampleApprox}
\begin{algorithmic}[1] 
\REQUIRE Graph $G$, teleport probability $\alpha$, start node $s$, maximum residual $\epr$, reverse vectors $y^T$, intermediate-node-to-target samplers $\{\text{sampler}_v\}$.
\STATE Set number of walks $w = c \frac{\epr}{\delta}$. In experiments we found $c=20$ achieved precision@5 above 90\% on Twitter-2010.
\STATE Set number of samples $n_{s}$ (We use $n_{s} = w$)
%\FOR{$i \in [0, w-1]$}
\STATE Sample $w$ random walks from $s$ and let $\tilde{\pi}_s$ be the empirical distribution of their endpoints; compute forward vector $x_s = (e_s, \tilde{\pi}_s) \in \R^{2\size{V}}$
\STATE Create $\text{sampler}_s$ to sample $v \in [2 \size{V}]$ with probability $p'_s[v]$, i.e., proportional to $x_s[v] y^T[v]$
\STATE Initialize empty map score from $V$ to $\mathbb{N}$
\FOR{$j \in [0,n_{\text{s}}-1]$}
  \STATE Sample $v$ from $\text{sampler}_s$
  \STATE Sample $t$ from $\text{sampler}_v$
  \STATE Increment score($t$)
\ENDFOR
\STATE Return $T$ sorted in decreasing order of score
\end{algorithmic}
\end{algorithm}

\noindent\textbf{Running Time}: 
For a small relative error for targets  with $\pi_s[t] > \delta$, we use $w = c\epr/\delta$ walks, where $c$ is chosen as in Theorem \ref{thm:bidirmain}.  
The support of our forward sampler is at most $w$ so its construction time is $O(w)$ using the alias method of sampling from a discrete distribution \cite{Walker:1977:EMG:355744.355749},  \cite[section~3.4.1]{knuth1998vol2}. 
Once constructed, we can get independent samples in $O(1)$ time.  
Thus the query time to generate $\nsample$ samples is $O \pn{c \epr/\delta + \nsample}$. 

\noindent\textbf{Accuracy}:
\bipprsampling{} does not sample exactly in proportion to $\pi_s$; instead, the sample probabilities are proportional to a distribution $\hat{\pi}_s$ satisfying the guarantee of Theorem \ref{thm:bidirmain}. %Formally, with probability at least $1- \pfail$, we have:
%\[ \abs{\pi_s[t]-\hat{\pi}_s[t]} \leq \max\left( \epsilon \pi_s[t], \frac{\delta}{2}\right). \]
In particular, for all targets $t$ with $\pi_s[t] \geq \delta$, this will have a small relative error $\epsilon$, while targets with $\pi_s[t] < \delta$ will likely be sampled rarely enough that they won't appear in the set of top-$k$ most sampled nodes.

\noindent\textbf{Storage Required}: 
The storage requirements for \bipprsampling{}  (and for \bipprgrouped{}) depends on the distribution of keywords and how $\rmax$ is chosen for each target set.  For simplicity, here we assume a single maximum residual $\epr$ across all target sets, and assume each target is relevant to at most $\gamma$ keywords.  For example, in the case of name search, each user typically has a first and last name, so $\gamma = 2$.
\begin{theorem}
  \label{thm:storage_all_targets}
Let graph $G$, minimum-PPR value $\delta$ and time-space trade-off parameter $\rmax$ be given, and suppose every node contains at most $\gamma$ keywords. Then the total storage needed for \bipprsampling{} to construct a sampler for any source node (or distribution) $s$ and any set of targets $T$ corresponding to a single keyword is $O\pn{\frac{\gamma m}{\alpha \epr}}$.
\end{theorem}
We can choose $\epr$ to trade-off this storage requirement with the running time requirement of $O\pn{c\epr/\delta}$ -- for example, we can set both the query running-time and per-node storage to $\sqrt{c \gamma\bar{d}/\delta}$ where $\bar{d}=m/n$ is the average degree.
Now for name search $\gamma=2$, and if we choose $\delta= \frac{1}{n}$ and $\alpha=\Theta(1)$, the per-query running time and per-node storage is  $O(\sqrt{m})$.
\begin{proof}
For each set $T$ corresponding to a keyword, and each $t \in T$, we push from nodes $v$ until for each $v$, $r^t[v] < \epr$.  Each time we push from a node $v$, we add an entry to the residual vector of each node $u \in \inneighbors{v}$, so the space cost is $\indegree{v}$.  Each time we push from a node $v$, we increase the estimate $p^t[v]$ by $\alpha r^t[v] \geq \alpha \epr$, and $\sum_{t \in T} p^t[v] \leq \sum_{t \in T} \pi_v[t] = \pi_v[T]$ so $v$ can be pushed from at most $\frac{\pi_v[t]}{\alpha \epr}$ times. Thus the total storage required is 
\begin{equation}
\label{eq:storage}
 \sum_{v \in V}  \indegree{v} (\#\text{ of times $v$ pushed}) \leq  \sum_{v \in V}  \indegree{v} \frac{\pi_v[T]}{\alpha \epr}
\end{equation}

Let $\mathcal{T}$ be the set of all target sets (one target set per keyword).  Then the total storage over all keywords is 
\begin{align*}
\sum_{T \in \mathcal{T}}\sum_{t \in T} \sum_{v \in V}  \indegree{v} \frac{\pi_v[t]}{\alpha \epr} 
& \leq \gamma \sum_{v \in V} \sum_{t \in V}  \indegree{v} \frac{\pi_v[t]}{\alpha \epr} \\
& \leq \gamma \sum_{v \in V}   \indegree{v} \frac{1}{\alpha \epr}
 \leq \gamma \frac{m}{\alpha \epr} .
\end{align*}
\end{proof}

%Similar bounds can be obtained under other assumptions -- for example, if an average target is relevant to $\gamma$ keywords, and the number of keywords per target is independent of that target's PageRank.
  
\noindent\textbf{Adaptive Maximum Residual}: 
One way to improve the storage requirement is by using larger values of $\epr$ for target sets $T$ with larger global PageRank.  
Intuitively, if $T$ is large, then it's easier for random walks to get close to $T$, so we don't need to push back from $T$ as much as we would for a small $T$.  
We now formalize this scheme, and outline the savings in storage via a heuristic analysis, based on a model of personalized PageRank values introduced by Bahmani et al. \cite{Bahmani2010}.
%\todo{remove? Formally, Bidirectional PPR returns a result with small relative error as long as the number of walks is at least $c \frac{\epr}{\pi_s[t]}$.  Let $t_k \in T$ be the $k$th target sorted in decreasing order of $\pi_s[t]$ and let $\delta_k = \pi_s[t_k]$.  Then if we have fixed running time requirement (implying a fixed number of random walks) we should set $\epr$ to be proportional to $\delta_k$. }

For a fixed $s$, we assume the values $\pi_s[v]$ for all $v \in V$ approximately follow a power law with exponent $\beta$.
%\[ \pi_s[t_i] = \frac{1 - \beta}{n^{1 - \beta}} i^{-\beta}. \] 
Empirically, this is known to be an accurate model for the Twitter graph -- Bahmani et al.~\cite{Bahmani2010} find that the mean exponent for a user is $\beta = 0.77$ with standard deviation $0.08$.  
To analyze our algorithm, 
%we need to know the distribution of $\pi_s$ when restricted to some set $T$.  \todo{remove? One natural model would be to assume that $\pi_s$ follows a power law and $T$ is uniformly chosen from all subsets of size $\size{T}$.  The analysis becomes unnecessarily complicated however, so} for simplicity 
we further assume that $\pi_s$ restricted to $T$ also follows a power law, i.e.:
\begin{equation}
  \label{eq:power_law}
   \pi_s[t_i] = \frac{1 - \beta}{\size{T}^{1 - \beta}} i^{-\beta} \pi_s[T].
\end{equation}

Suppose we want an accurate estimate of $\pi_s[t_i]$ for the top-$k$ results within $T$, so we set $\delta_k = \pi_s[t_k]$. 
%= \frac{1 - \beta}{\size{T}^{1 - \beta}} k^{-\beta}
From Theorem \ref{thm:bidirmain}, the number of walks required is:
$$w = c \frac{\epr}{\delta_k} = c_2 \frac{\epr \size{T}^{1 - \beta} }{\pi_s[T]}$$
where  $c_2 = k^{\beta} c/(1 - \beta)$.
If we fix the number of walks as $w$, then we must set $\epr = w \pi_s[T]/ (c_2 \size{T}^{1-\beta})$. Also, for a uniformly random start node $s$, we have $\EE[\pi_s[T]]=\pi[T]$ (the global PageRank of $T$). This suggests we choose $\epr(T)$ for set $T$ as:
\begin{equation}
	\label{eq:rmax}
\epr(T) = \frac{w \pi[T]}{c_2 \size{T}^{1-\beta}}
\end{equation}

Going back to equation \eqref{eq:storage}, suppose for simplicity that the average $\indegree{v}$ encountered is $\bar{d}$.  
Then the storage required for this keyword is bounded by:
\[ \sum_{v \in V}  \indegree{v} \frac{\pi_v[T]}{\epr} = \bar{d} \frac{n \pi[T]}{\epr} 
= \frac{m c_2 \size{T}^{1-\beta}}{w} . \]
Note that this is independent of $\pi[T]$.  There is still a dependence on $\size{T}$, which is natural since for larger $T$ there are more nodes which make it harder to find the top-$k$.
For $\beta=0.77$ %(for the Twitter network, from \cite{Bahmani2010})
, the rate of growth, $\size{T}^{0.23}$ is fairly small, and in particular is sublinear in $\size{T}$.

\noindent\textbf{Dynamic Graphs}: So far we have assumed that the graph and keywords are static, but in practice they change over time.  When a keyword is added to some node $T$, the node's reverse vector $y^t$ needs to be added to the sampling data structure for that keyword.  When an edge is added, the residual values need to be updated.  We leave the extension to dynamic graphs to future work.  
%!TEX root = main.tex

\subsection{Experiments}

We conduct experiments to test the efficiency of these personalized search algorithms as the size of the target set varies.  We use one of the largest publicly available social networks, Twitter-2010 \cite{WebAlgorithmics} with 40 million nodes and 1.5 billion edges.  For various values of $\size{T}$, we select a target set $T$ uniformly among all sets with that size, and compare the running times of the four algorithms we propose in this work, as well as the \texttt{Monte Carlo} algorithm.  We repeat this using 10 random target sets and 10 random sources $s$ per target set, and report the median running time for all algorithms.  We use the same target sets and sources for all algorithms.

\begin{figure}[t]
\centering
\includegraphics[width=0.66\columnwidth]{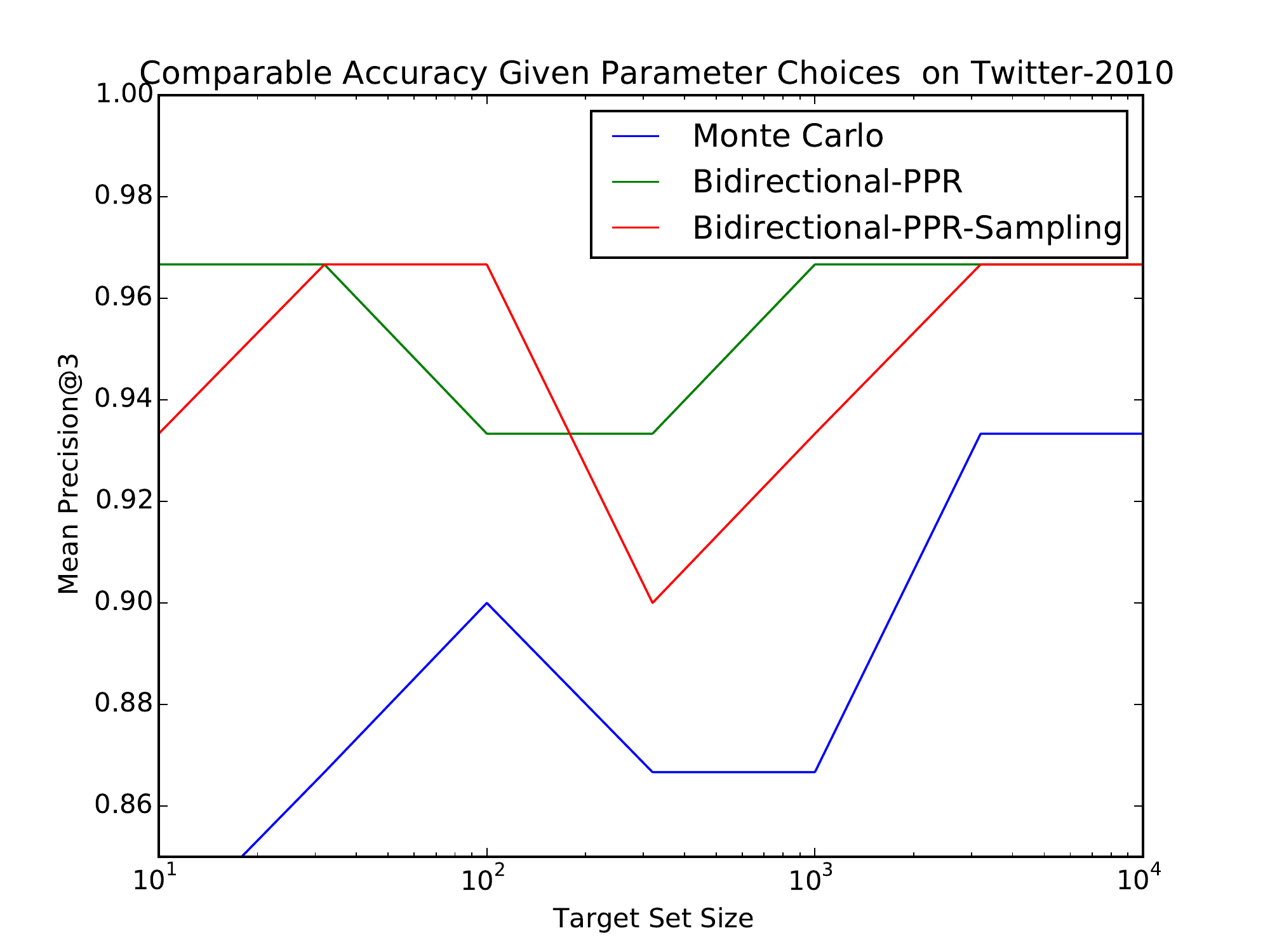}
\caption[Accuracy]{Median precision@3 for the search algorithms we compare.  Notice that the Precision@3 of our algorithms exceeds $90\%$ and exceeds the precision@3 of \mc.}
\label{fig:search_accuracy}
\end{figure}

\noindent\textbf{Parameter Choices}:
Because all five algorithms have parameters that trade-off running time and accuracy, we choose parameters such that the accuracy is comparable so we can compare running time on a level playing field.  To choose a concrete benchmark, we chose parameters such that the precision@3 of the four algorithms we propose are consistently greater than 90\% for the range of $\size{T}$ we used in experiment. We chose parameters for \mc\ so that our algorithms are consistently more accurate than it, and its precision@3 is greater than 85\%.  In Figure \ref{fig:search_accuracy} we plot the precision@3 of the algorithms for the parameters we use when comparing running time when $w=100,000$.  Note that \bipprprecomp\ and \bipprgrouped\ compute the same estimates, and these estimates are similar to those of \bippr, so we plot a single line for their accuracy.

We used $\delta = \pi_s(t_k)$ where $\pi_s(t_k)$ is estimated using Eqn. \ref{eq:power_law}, using $k=3$, power law exponent $\beta=0.77$ (the mean value found empirically on Twitter), and assuming $\pi_s(T) = \frac{\size{T}}{n}$ (the expected value of $\pi_s(T)$ since $T$ is chosen uniformly at random).  Then we use Equation \ref{eq:rmax} to set $\epr$, using $c=20$ and two values of $w$, 10,000 and 100,000. We used the same value of $\epr$ for \bipprprecomp{}, \bipprgrouped{}, and \bipprsampling{}.  For \mc, we sampled $\frac{40}{\delta}$ walks\footnote{Note that \mc\ was too slow to finish in a reasonable amount of time, so we measured the average time required to take 10 million walks, then multiplied by the number of walks needed.  When measuring precision, we simulated the target weights \mc\ would generate, by sampling $t_i$ with probability $\pi_s(t_i)$; this produces exactly the same distribution of weights as \mc\ would.}.  
% \todo{is that an embarrassing amount? One prior work reports 60MB total PPR pre-computation storage for citeseer (for all keywords).  Should we multiply $w$ by 10?}.  The best choice for $w$ depends on the latency requirement or amount of storage available.

%d in section \ref{sec:four_algorithms}.  
\begin{figure*}
\centering
\subfigure[Running Time, More Precomputation]{
\label{fig:search_runtime10K}
\includegraphics[width=0.66\columnwidth]{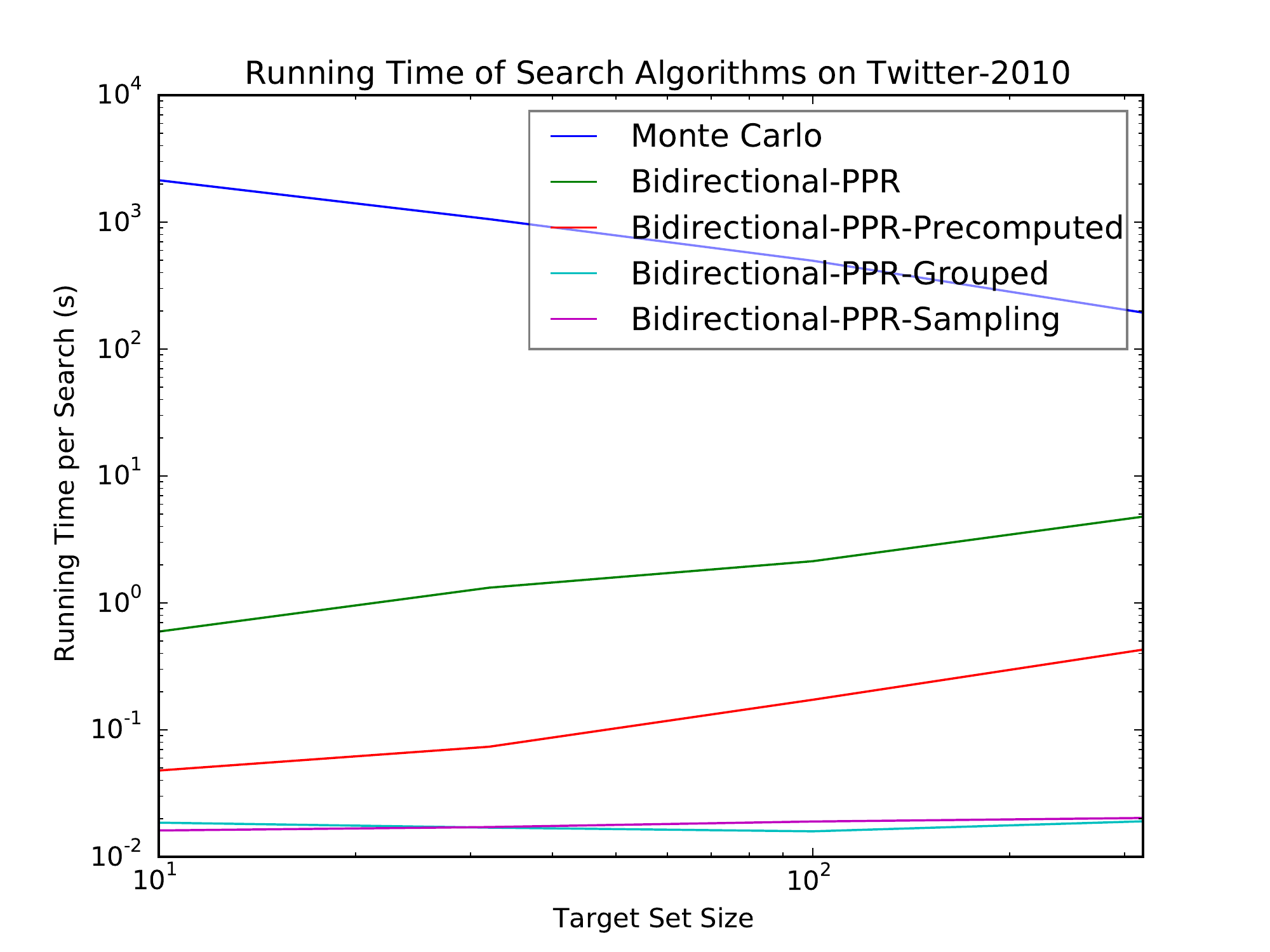}
}
\hfill
\subfigure[Running Time, Less Precomputation]{
\label{fig:search_runtime100K}
\includegraphics[width=0.66\columnwidth]{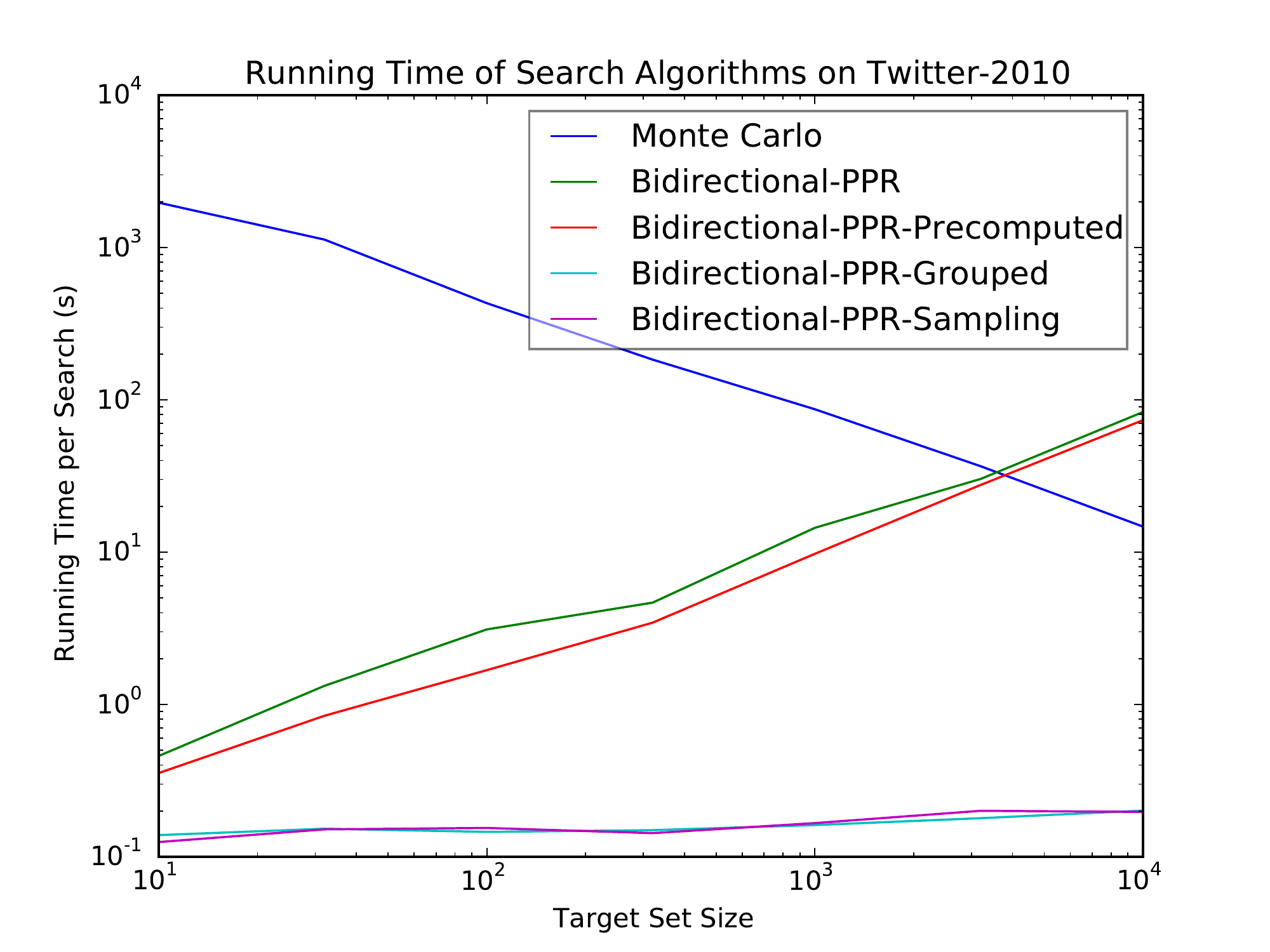}
}
%\subfigure[Other stuff]{
%\includegraphics[width=.65\textwidth]{somethingelse.png}}
\caption[Comparing PPR-Search Algorithms]{%In Figure \ref{fig:search_accuracy}, we plot the Precision@3 for several search algorithms, while in figure \ref{fig:search_runtime}, 
Running time on Twitter-2010 (1.5 billion edges), with parameters chosen such that the Precision@3 of our algorithms exceeds $90\%$ and exceeds the precision@3 of \mc.\ The two plots demonstrate the storage-runtime tradeoff: Figure \ref{fig:search_runtime10K} (which performs $10K$ walks at runtime) uses more pre-computation and storage compared to Figure \ref{fig:search_runtime100K} (with $100K$ walks).}
\label{fig:search_runtime}
\end{figure*}

\noindent\textbf{Results}:
Figure \ref{fig:search_runtime} shows the running time of the five algorithms as $\size{T}$ varies for two different parameter settings in the trade-off between running time and precomputed storage requirement.   Notice that \mc\ is very slow on this large graph for small target set sizes, but gets faster as the size of the target set increases.  For example when $\size{T}=10$ Monte Carlo takes half an hour, and even for $\size{T}=1000$ it takes more than a minute.
\bippr{} is fast for small $T$, but slow for larger $T$, taking more than a second when $\size{T} \geq 100$.  In contrast, \bipprgrouped{} and \bipprsampling{} are both fast for all sizes of $T$, taking less than 250 ms when $w=10,000$ and less than 25 ms when $w = 100,000$.  

The improved running time of \bipprgrouped\ and \bipprsampling{}, however, comes at the cost of pre-computation and storage. With these parameter choices, for $w=10,000$ the pre-computation size per target set in our experiments ranged from 8 MB (for $\size{T}=10$) to 200MB (for $\size{T}=1000$) per keyword.  For $w=100,000$, the storage per keyword ranges from 3 MB (for $\size{T}=10$) to 30MB (for $\size{T}=10,000$).

\begin{figure}
\centering
\includegraphics[width=0.66\columnwidth]{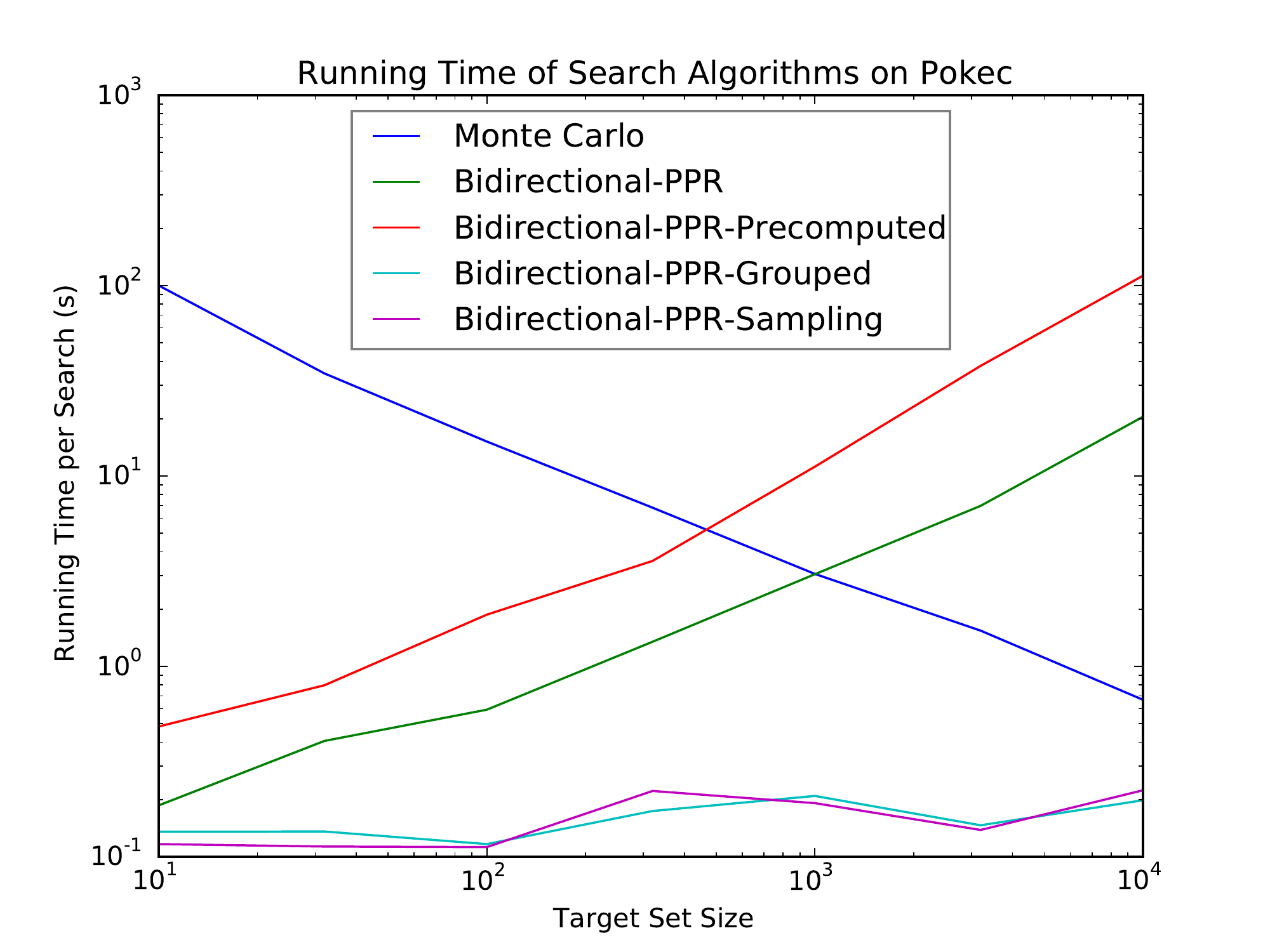}
\caption[Pokec Running Time]{Running time on Pokec (30 million edges) performing 100K walks at runtime.  Notice that \mc\ is slow for small $\size{T}$, \bippr\ is slow for large $\size{T}$, and \bipprgrouped\ and \bipprsampling\ are fast across the entire range of $\size{T}$.}
\label{fig:runtime_pokec}
\end{figure}

To get a larger range of $\size{T}$ relative to $\size{V}$, we also perform experiments on the Pokec graph \cite{SnapProject} which has 1.6 million nodes and 30 million edges. Figure \ref{fig:runtime_pokec} shows the results on Pokec for $w=100,000$.  Here we clearly see the cross-over point where \mc\ becomes more efficient than \bippr, while \bipprgrouped\ and \bipprsampling\ consistently take less than 250 milliseconds.  On Pokec, the storage used ranges from 800KB for $\size{T}=10$ to 3MB for $\size{T}=10,000$.

We implement our algorithms in Scala and report running times for Scala, but in preliminary experiments \bipprgrouped{} is 3x faster when re-implemented in C++, we expect the running time would improve comparably for all five algorithms.  Also, we ran each experiment on a single thread, but the algorithms parallelize naturally, so the latency could be improved by a multi-threaded implementation.
We ran our experiments on a machine with a 3.33 GHz 12-core Intel Xeon X5680 processor, 12MB cache, and 192 GB of 1066 MHz Registered ECC DDR3 RAM.  %http://snap.stanford.edu/moin/Hulk
We measured the running time of the tread running each experiment to exclude garbage collector time.  
We loaded the graph used into memory and completed any pre-computation in RAM before measuring the running time of the algorithms.

\section{Sampling  Paths Conditioned on Endpoint}
\label{sec:path_chapter}
\subsection{Introduction}
In this chapter we apply our bidirectional ideas to create an algorithm for sampling a random path from a given source $s \in V$ conditioned on the path ending  in a given set of targets $T \subseteq V$ (which of course might be a single target $T=\{t\}$).  Our results generalize to any source distribution and Markov Chain based the methods of Chapter \ref{sec:mc_chapter}, and in fact the algorithm is likely simpler for a fixed path length, but in this chapter we present the algorithm for the case of a single source node and for Personalized PageRank (i.e. a geometrically distributed length).  

This is a general problem which could be used as a subroutine in other graph algorithms and could have diverse applications.  Our motivation was in personalized search.  The method in this chapter can be viewed as an exact version of the approximate target sampler presented in section \ref{sec:sampler}.  In contrast to that sampler, given source $s$ this method samples a target $t$ from a given set $T$ with probability exactly proportional to $\pi_s(t)$.  A related application to personalized search is showing a user why a search result came up.  For example, if a user searches on Twitter for ``Adam'' and finds Adam Messinger, this algorithm could generate a few representative paths through the Twitter network from the user to Adam Messinger.  

We are not aware of prior work on this problem.  Our baseline of comparison is simply Monte Carlo with rejection sampling: we repeatedly sample paths from $s$ until we find one that happens to end in $T$. The running time of that algorithm is geometrically distributed with expected time $\frac{1}{\pi_s[T]}$ where $\pi_s[T]$ is the probability that a path from $s$ happens to end in $T$. 

Our approach, similar to that of \bippr{} (Section \ref{sec:bippr}), is to use the \rpush{} algorithm from the target set to find a larger set of nodes with a significant probability of probability of reaching $T$ on a random walk.  However, we modify the \rpush{} algorithm such that each node $v$ also keeps track of where it receives its residual probabilities from, so from $v$ we can sample a neighbor in proportion to the amount of residual $v$ received from that neighbor.  

The running time of our sampler is similar to the running time of \bippr{}.  Like \bippr{} our algorithm has a parameter $\rmax$ which lets it trade off reverse work and forward work. The reverse running time is the same as for \bippr{} , $O \pn{\size{T} \frac{\bar{d}}{\rmax}}$ for an average target set $T$ (see Section \ref{sec:bippr_running_time} for more detail).  After executing \rpush{}, the time per independent path sampled is $O \pn{\frac{\rmax}{\pi_s[T]}}$ or $ O(\frac{n \rmax}{\size{T}} )$ for a random target set.  If we chose $\rmax$ to minimize the running time for average target sets, we get $O(\sqrt{m})$ running time to sample a random path, or $O(\sqrt{k m})$ time to sample $k$ independent paths.

%The main limitation of our algorithm is simply its running time (?).
\subsection{Path Sampling Algorithm}
\subsubsection{Algorithm Overview}
If we run the \rpush{} algorithm starting with one unit of residual on each element of $T$, as in equation \eqref{eq:ppr_dot_product} we get
\[
  \pi_s(T) = p^T[s] + \sum_{v \in V} \pi_s[v] r^T[v].
\]
Our basic approach will to be sample $v \sim \pi_s$ using a random walk, then accept that walk with probability $\frac{r^T[v]}{\rmax}$.  (If we reject, we keep sampling until we accept.) Once we accept, we sample the remaining path from $v$ to a random target $t \in T$ using information about where each node received residual from to trace the residual back to its source in $T$.

To enable that, we modify the \rpush{} algorithm so that each node stores where it received its residual from, as shown in Algorithm \ref{alg:contributionsToSet}.  Then we can sample a path forward from an intermediate node $u$ towards $T$ in proportion to where residual came from.  For example, if a node $u$ received 0.1 units of residual from its out-neighbor $v_1$ and 0.2 units of residual from its outneighbor $v_2$, then from $u$ we would choose $v_1 $ with probability $\frac{0.1}{0.1 + 0.2}$.  We continue sampling like this until we reach one of the initial residuals that started at each target in $T$.  

One complication in this approach is that a node $u$ might receive residual mass from some node $v$, push that mass backwards, then later receive new residual mass from another node $v'$. If we reach $u$ its then unclear with what probability we should go to $v$ vs $v'$.  Our solution to this problem is to store multiple residual vector samplers at $u$, one for each time $u$ pushes back, plus an additional sampler for any residual mass remaining at $u$.  Then if we reach $u$ when sampling a path, we can transition to the next node based on where $u$ had recieved residual at the relevant time when $u$ pushed back.

\subsubsection{An Edge Case}
Typically,  $p^T[s]=0$, since $p^T[s]$ is non-zero only for $s$ ``close'' to $t$.  If $p^T[s] > 0$, then $\pi_s(T) > \alpha \rmax$, so simply sampling random walks without using any of these techniques will find a path to $T$ in time $O \pn{\frac{1}{\rmax}}$.  

However, if $\rmax$ is very small the running time of that method could be significant.  To fully handle the case $p^T[s] > 0$ we make two changes to the algorithm.  First, when selecting intermediate node $v$, instead of sampling $v \sim \pi_s$ and accepting with probability $\frac{r^T[v]}{\rmax}$, we accept $v=s$ (without doing a walk) with probability 
\[\frac{\pEst^T[s]}{\pEst^T[s] + \rmax}. \]
We also take a walk to $v \sim \pi_s$ and accept it with probability
\[\frac{r^T[v]}{\pEst^T[s] + \rmax}.\]
(As shown in the pseudo-code, we do this in such a way that both events cannot happen.)  With the remaining probability we try again for both options.  Once we accept, the probability of accepting $v=s$ without a walk is 
\[\frac{\pEst^T[s]}{\pEst^T[s] + \sum_v \pi_s[v] r^T[v]} = \frac{\pEst^T[s]}{\pi_s(T)}.\]
Similarly the probability of accepting node $v$ after a walk is $\frac{\pi_s[v] r^T[v]}{\pi_s(T)}$.
The second change we make is to allow us to sample residuals of $v$ in the case that we choose $v = s$ without a walk.  Since $v$ may push multiple times, and the nodes which gave $v$ residual may be different for different pushes, we store a sampler $P^T[v]$ at each node $v$ which pushes, and when $v$ pushes we add its current residual vector $R^T[v]$ to $P^T[v]$ with weight $r^T[v]$.  The complete algorithm is shown as Algorithm \ref{alg:SampleExact}.

\begin{algorithm}[ht]
\caption{\precompSamplers{}$(T, \rmax)$~\cite{Andersen2007}}
\label{alg:contributionsToSet}
\begin{algorithmic}[1]
\REQUIRE Graph $G$, teleport probability $\alpha$, target nodes $T$, maximum residual $\rmax$

\STATE Initialize (sparse) estimate-vector $\pEst^T = \vec{0}$ and (sparse) residual-vector 
\[ r^T[v] = \indicator[v \in T] =  \begin{cases}
   1 &\text{ if } v \in T \\
   0 &\text{else}
  \end{cases} \]
\STATE We will also store a map of samplers $R^T$ to keep track of where each node received its residual from.
For each $u$, $R_t[u]$ is map from samplers to their current weight, where the weight of the sampler associated with node $v$ equals the \emph{current} residual amount $u$ has received from $v$.  Equivalently, $R_t[u]$ is a discrete probability distribution that samples a sampler associated with $v$ with probability proportional to the current residual $u$ has received from $v$.  
Note that over the course of the algorithm, if a node $u$ is pushed multiple times, it will create a new sampler each time, and any previous references to its sampler will refer to a previous sampler.  Initially each $t \in T$ receives one unit of residual from a constant sampler which always returns $t$. The current sampler represents the current residual, so we maintain the invariant $r^T[u] = \sum_v R^T[u][v]$.

We also store a map of samplers $P^T$ based on estimates received.  For each node $v$ which has pushed at least once, $P^T[v]$ will be a sampler over previous samplers associated with $v$, with weight equal to the amount of estimate $v$ accumulated during that push.  Initially $P^T$ is an empty map.  We maintain the invariant $\pEst[u] = \sum_z P^T[u](z)$. %(todo better explanation)

\WHILE{$\exists v\in V\,s.t.\,r^T[v]>\rmax$}
	\STATE $\text{sampler}_v$ = makeImmutableSampler($R^T[v]$)
	\STATE $\pEst^T[v] = \pEst^T[v] + \alpha r^T[v]$
	\STATE $P^T[v]\text{.increaseValue}(\text{sampler}_v, \alpha r^T[v])$
	\FOR{$u\in\mathcal{N}^{in}[v]$}
		\STATE      $\Delta_u =  (1-\alpha) \frac{r^T[v]}{d^{out}[u]}$
		\STATE     $ r^T[u] = r^T[u] + \Delta_u$
		\STATE     $ R^T[u]\text{.increaseValue}(\text{sampler}_v, \Delta_u)$
	\ENDFOR
	\STATE  Update $r^T[v]=0$
	\STATE  Update $R^T[v] = \vec{0}$ (An empty map/sampler)
\ENDWHILE

\RETURN $(r^T, \pEst^T, R^T, P^T)$
\end{algorithmic}
\end{algorithm}

\begin{algorithm}[ht]
\caption{\samplePath{}$(s,T, \rmax, \pEst^T, r^T, R^T)$}
\label{alg:SampleExact}
\begin{algorithmic}[1] 
\REQUIRE Graph $G$, teleport probability $\alpha$, start node $s$, target set $T$, maximum residual $\rmax$, residuals, estimates, and residual and estimate maps $(r^T, \pEst^T, R^T, P^T)$ from \precompSamplers{}$(T, \rmax)$
%\IF{$\hat{\pi}^T[s] \geq \rmax$}
% \STATE Simply sample random walks from $s$ until the last node is some node $t \in T$, and return $t$.
%\ENDIF
\WHILE{currentSampler has not been set}
\STATE Sample $x \sim \text{Uniform}[0,1]$
\IF {$ x < \frac{\pEst^T[s]}{\pEst^T[s] + \rmax}$}
\STATE  Set currentSampler = P[s].\text{sample}().
\label{SampleTargetAcceptStartNodeLine}
\STATE Set walk $W=[s]$ (i.e.~a length-1 walk)
\ELSE
\STATE Sample a walk $W$ from $s$, and let $u$ be its last node.
\IF {$ x - \frac{\pEst^T[s]}{\pEst^T[s] + \rmax} < \frac{r^T[u]}{\pEst^T[s] + \rmax}$}
\STATE  Set $currentSampler = R[u]$ \label{SampleTargetAcceptLine}
\ENDIF
\ENDIF
\ENDWHILE
%\STATE currentSampler = $R[v]$
\WHILE {currentSampler is not a ConstantSampler}
  \STATE currentSampler = currentSampler.sample()
    \label{SampleTargetConditionalNeighborLine}
  \STATE Append node of currentSampler to $W$
  %That is, set $v$ to $v'$ with probability 
  %\[\frac{R^T[v](v')}{r^T[v]} \]
\ENDWHILE
\RETURN $W$ %the $t$ associated with currentSampler
\end{algorithmic}
\end{algorithm}

\subsubsection{Running Time} Our probability of accepting $v$ on each iteration of the while loop in SampleTarget is
\[ \frac{\pEst^T[s] + \sum_v \pi_s[v]  r^T[v]}{\pEst[s] + \rmax}  = \frac{\pi_s(T)}{\pEst^T[s] + \rmax}\]
so on average we expect to do
\[ \frac{\pEst^T[s] + \rmax}{\pi_s(T)} \leq 1 + \frac{\rmax}{\pi_s(T)} \]
walks before accepting, and each walk takes constant expected time $O\pn{\frac{1}{\alpha}}$.  %In the typical case when $\pEst[s]=0$ (i.e. the source node $s$ is not close to $T$), the expected number of walks is
%\[ \frac{\rmax}{\pi_s(T)}. \]
%(Note that if $T$ contains $\gamma$ nodes $t$ with $\pi_s(t) > \delta$, then the running time is less than $\frac{\rmax}{\delta \gamma}. $)

\subsubsection{Correctness}
We now show that this method is correct by induction on the reverse push operation in \precompSamplers{}.
\begin{theorem}
\label{thm:sampling_exact} For any weighted graph $G$, source node (or distribution) $s$, target set $T$, and maximum residual $\rmax$, \samplePath{} (executed on the output of \precompSamplers{}$(T, \rmax)$) %returns a target $t$ with probability exactly
%\[ \frac{ \pi_s(t)}{\sum_{t' \in T} \pi_s(t')} .\]
returns a random walk from $s$ conditioned on ending in $T$.
\end{theorem}

\begin{proof} 
(Sketch)
We prove this by induction on the number of iterations of the while loop of \precompSamplers{}.  As our base case, before the while loop executes, we have $r^T[v] = \indicator[ v \in T]$, so \samplePath{} will simply do rejection sampling until it finds a walk ending at some node $v \in T$, which is clearly sampled from the right distribution. 

%(todo: it probably is clearest if we assume $ \pEst^T[s]=0$ here, and move this proof to the appendix.  For now, the full proof is here.)

Now suppose by induction that we have some $(r^T, \pEst^T, R^T, P^T)$ such that \samplePath{}$(s, T, \rmax, r^T, \pEst^T, R^T, P^T)$ returns a node $v$ from the correct distribution. Let $(r'^T, \pEst'^T, P'^T, R'^T)$ be the residuals and estimates after one more iteration of the while loop in \precompSamplers{} applied to some node $v$ (i.e. one more reverse push operation on $v$).  Then we have
\[ \pEst'^T[v] = \pEst^T[v] + \alpha r^T[v],\]
\[ P'^T[v] = P^T[v] \text{ with increased value }  \alpha r^T[v] \text{ on } R^T[v], \]
\[ r'^T[v] = 0, \]
and for $u\in \inneighbors{v}$
\begin{align*}
\Delta_u &=  (1-\alpha)\frac{r^T[v]}{d^{out}[u]} \\
r'^T[u] &= r^T[u] + \Delta_u \\
R'^T[u][v] &= R^T[u][v] + \Delta_u
\end{align*}
All other entries of $r^T$, $\pEst$, $R^T$, and $P^T$ are unchanged.  Consider how the behavior of \samplePath{} has changed with this iteration.  Before this iteration, there was a probability $\frac{(\pi_s[v] r^t[v]) / (p^T[s] + \rmax)}{ (\sum_v \pi_s[v] r^t[v]) / (p^T[s] + \rmax) + p^T[s] / (p^T[s] + \rmax)} =  \pi_s[v] \frac{r^T[v]}{\pi_s(T)}$  that on line \ref{SampleTargetAcceptLine} we would accept $v$, and that probability is now 0.  However, after this iteration, there is an increased probability of picking some in-neighbor $u \in \inneighbors{v}$ on line \ref{SampleTargetAcceptLine} and transitioning to $v$ on the first iteration of line \ref{SampleTargetConditionalNeighborLine}.  The change $\Delta(R[v]) $ in the probability of choosing $R[v]$ either on line \ref{SampleTargetAcceptLine} or line \ref{SampleTargetConditionalNeighborLine} is exactly 
\begin{align*}
\Delta(R[v]) &= \frac{\indicator[s=v] \pEst'[v]  + \pi_s[v] r'^T[v]}{\pi_s(T)} 
    + \sum_{u \in \inneighbors{v}} \pn{\pi_s[u] \frac{r'^T[u]}{\pi_s(T)} } \pn{ \frac{R'^T[u][v]}{r'^T[u]}} \\
  &\phantom{=} - \frac{\indicator[s=v] \pEst[v] + \pi_s[v] r^T[v]}{\pi_s(T)} - \sum_{u \in \inneighbors{v}} \pn{\pi_s[u] \frac{ r^T[u]}{\pi_s(T)} } \pn{ \frac{ R^T[u][v]}{ r^T[u]}} \\
  &= \frac{ \indicator[s=v] \alpha r^T[v] - \pi_s[v] r^T[v] + \sum_{u \in \inneighbors{v}} \pi_s[u] \Delta_u}{\pi_s(T)} \\
  &= \frac{r^T[v]}{\pi_s(T)} \pn{\indicator[s=v] \alpha - \pi_s[v] 
    + \sum_{u \in \inneighbors{v}} \pi_s[u] \frac{1-\alpha}{d^{out}[u]}} \\
  &= 0
\end{align*}
The last line follows from a recursive equation for PPR (Equation (11) in \cite{Andersen2006}):
\[ \pi_s[v] = \sum_{u\in \inneighbors{v}} \pi_s[u]\frac{1-\alpha}{d^{out}[u]}  + \alpha \indicator[s=v]. \]

%The term $\frac{r^T[v]}{\rmax}\pi_s[v]$ is balanced by the fact that $r'^T[v]=0$, and the term $\frac{r^T[v]}{\rmax} \alpha \indicator[s=v]$ is balanced by the fact that $\pEst'[v]=\pEst[v] + \alpha r^T[v]$ and line \ref{SampleTargetAcceptStartNodeLine}.  
Thus the probability of reaching the sampler $R[v]$ on line \ref{SampleTargetConditionalNeighborLine} is unchanged by this iteration.  Furthermore, the distribution of paths occurring before reaching the sampler $R[v]$ is unchanged.  If $v=s$, then before the push at $v$ with probability $\alpha r^T[v]/\pi_s(T)$ we reach $R[v]$ directly, and after the push the increase in probability of reaching $R[v]$ directly is $p'^T[v] - p^T[v]=\alpha r^T[v]/\pi_s(T)$.  %(todo: Should I make the proof more formal?)

Finally note that the probability of reaching any other sampler $R(v')$ %on line \ref{SampleTargetAcceptLine} 
is also unchanged, since before increasing $r_T[u]$ this probability is 
\[ \frac{r^T[u]}{\pi_s(t)} \frac{R(u, v')}{r^T(u)} \]
which is equal to the probability of reaching sampler $R(v')$ after increasing $r_t[u]$:
\[ \frac{r'^T[u]}{\pi_s(t)} \frac{R(u, v')}{r'^T(u)} \]
%when we increase $r^T[u]$ to $r'^T[u]$, the denominator on line \ref{SampleTargetConditionalNeighborLine} also increases from $r^T[u]$ to $r'^T[u]$, so the overall probability of reaching $v'$ is also unchanged. 
Thus the distribution of paths returned using $(r'^T, \pEst'^T, R'^T, P'^T)$ is identical to the distribution of paths returned using $(r^T, p^T, R^T, P^T)$.  Our induction hypothesis was that distribution of paths returned using $(r^T, p^T, R^T, P^T)$ is the correct distribution%, $\frac{\pi_s(t)}{\sum_{t' \in T} \pi_s(t')}$
, so our proof is complete.
%we have changed the probabilities of paths not going through $v$.  However, since we also increased $R'^T[v]$, the effect of increasing $r'^T[u]$ cancels out for other $v' \in \outneighbors{u}$.  For $v' \in \outneighbors{u} \backslash \{v\}$, the probability of sampling $u$ on line \ref{SampleTargetAcceptLine} and then sampling $v$ on line  \ref{SampleTargetConditionalNeighborLine} is
\[\]
\end{proof}

\chapter{Conclusion and Future Directions}
In this chapter, we briefly recap the contributions in this thesis, and then describe a number of interesting research problems that could lead to nice publications.

\section{Recap of Thesis}
In this thesis we presented new algorithms for estimating random walk scores, including personalized PageRank.  While past algorithms either use linear algebra or sample random walks, we achieve dramatic speed improvements by combining local liner algebra operations from the target node backwards with random walks from the source node forwards.  Our algorithm gives provably accurate estimates, and on six diverse graphs it is 70x faster than any past algorithm at estimating individual PPR scores.  For estimating a single value $\pi_s[t]$, \bippr{} takes average time $O(\sqrt{m} / \epsilon)$ for constant failure probability, and \cite{fastppr} includes a lower bound of $\Omega(\sqrt{n})$, so there is little asymptotic improvement to be made on sparse graphs for random targets.  However, on the Twitter-2010 graph, \bippr{} still takes several seconds to compute scores for the most popular targets, and this motivates pre-computing the reverse linear algebra (residuals) and forward walks.  That pre-computation motivates several problems described in the next section.

We then observed that on undirected graphs, rather than do linear algebra backwards from the target and random walks forwards, it is possible to do random walks backwards from the target and linear algebra forwards. This alternative algorithm allows for cleaner worst-case bounds and may be useful in future applications.  There is also an interesting symmetry between \bippr{} and this alternative algorithm, and there may be applications which can exploit that symmetry.

We also observed that our bidirectional algorithm does not require the memoryless property of Personalized Pagerank walks, but generalizes naturally to estimating random walk probabilities of any given length on any weighted graph (or Markov Chain).  This algorithm can be used for example, to estimate the heat kernel score, and we hope future applications of Markov Chains find use for this algorithm.  Natural extensions of this algorithm to other random walk quantities like hitting time would be interesting.

We then applied these algorithms to describe new personalized search ranking algorithms.  In particular, we described an algorithm which pre-groups targets by name and allows the scores of all targets matching a given name query to be efficiently computed, even when the set of nodes matching the name query is large.  We also described a sampling algorithm which can find the most relevant nodes without needing to score them all.  Note that these two algorithms are not restricted to estimating PPR, but could be applied to personalized search with any scoring function which can be written as a dot-product between some vector associated with the searching user and some vector associated with each document.  For example, the vectors could arise from some machine learning or latent semantic analysis algorithm.

This is an exciting new capability, but we believe there is still significant improvements to be made.  In particular, it may be possible to decrease the amount of storage needed, especially in the case of text search when each target is relevant to multiple keywords.  Our methods also require the potential queries to be known in advance.

\section{Open Problems in PPR Estimation}
\subsection{The Running-Time Dependence on $\bar{d}$}
The average-case running time of \bippr{} to detect a PPR value of size $\delta=\frac{1}{n}$ to constant relative error with constant failure probability is $O(\sqrt{m})$.  However, the average-case lower bound presented in \cite{fastppr} is $\Omega(\sqrt{n})$.  This leaves a gap of $\sqrt{\dbar} = \sqrt{m/n}$ between the achieved running time and the lower bound.  Can the average-case running time be improved to $O(\sqrt{n})$, or alternatively does there exist a graph where for random $(s, t)$ pairs, $\Omega(\sqrt{m})$ edges must be looked at on average to estimate a PPR value of size $\frac{1}{n}$?

\subsection{Decreasing Pre-computation Storage or Running Time}
When we precompute all the residual vectors $r^t$ from every target in order to make real-time queries faster, the storage is significant, $O(n \sqrt{m} / \epsilon)$.  Is there some way to decrease the amount of storage required without increasing the running time at query time?  

In Section \ref{sec:combine_walks} we presented a method to to combine random walks from the out-neighbors of $s$ to decrease the number of stored walks. One approach to this problem is to find an analogous way to combine the residual vectors from different nodes to create the residual vector for the given target $t$ at runtime.

%One potential approach is to use the relationship between the residual vectors of different nodes: if the residuals $r^v$ are known for all in-neighbors $v \in \inneighbors{t}$, then
%\[ \sum_{v \in \inneighbors{t}} (1-\alpha) w_{v,t} r^v \]
%might be useful for forming a residual vector for $t$, but it is unclear.  The issue is that even if each entry $r^v[u] < \rmax$ that does not imply that the above sum is less than $\rmax$. %is likely a valid residual vector for $t$.  If the residual vectors $r^v$ have maximum value $\rmax$, then with effective maximum residual of $\max_{v \in \inneighbors{t}} (1-\alpha) w{v, t} \rmax$.  

\subsection{Maintaining Pre-computed Vectors on Dynamic graphs}
\label{sec:dynamic_graph}
Suppose we precompute the residual vectors $r^t$  for every target $t$ to enable real-time PPR estimation or for personalized search.  How can we update these residual vectors as edges are added to the graph?

 For the forward walks, there is a natural algorithm for maintaining the walks as the graph evolves \cite{Bahmani2010}. Something similar might work for maintaining the residuals. In particular, when an edge $(u, v)$ is added, we can compute how much excess residual $u$ has received from other out-neighbors, as well as how much residual $u$ should have received from $v$, and from this restore the loop invariant of \rpush{} (Section \ref{sec:reverse_push}), at which point push operations can restore the requirement that for all $t, u \in V$, $\abs{r^t[u]} \leq \rmax$.

\subsection{Parameterized Worst-Case Analysis for Global PageRank}
%In Section \ref{sec:choosing_delta}, we proposed choosing delta based on $t$, since if we had a lower bound for $\pi_s[t]$ we only n
For a fixed choice of $\delta$, we give an average-case running time analysis for \bippr{}, but no worst-case running time analysis. %Our algorithm takes as input a lower bound $\delta$ for the quantity being estimated.  However, in practice we may not have a lower bound.  For global PageRank, we can use $\delta = \alpha / n$, since every node has $\pi[t] \geq \alpha / n$, but if we could choose a larger $\delta$ \bippr{} would run faster.  
Is there a variation of \bippr{} or some assumption on the graph for which there is a worst-case running time bound, perhaps parameterized by some property of the target like its global PageRank?

%Is there an algorithm for choosing $\delta$ which leads to a rigerous worst-case running time for estimating the PageRank of a node $t$?

One approach is to use a sequence of $\delta$ values, say $\delta_i = 2^{-i}$ and try progressively smaller values of $\delta$ until the estimate returned is greater than $(1 + 2e) \delta$, as needed by Theorem \ref{thm:bidirmain} to guarantee a small relative error.  The heuristic parameterized estimate in Section \ref{sec:parameterized_bippr} may also be relevant.%Simply using a value for $\pfail$ which is $\log(n)$ smaller than before and applying a union bound lets us prove accuracy.

\subsection{Computing Effective Resistance or Hitting Times}
As briefly sketched in Section \ref{sec:apps}, our random walk methods can be used to estimate other random walk scores.  As defined for example in these lecture notes\footnote{\url{http://web.stanford.edu/class/cme305/Notes/6.pdf}}, the hitting time between two nodes $s$ and $t$ is the expected number of steps before a random walk from $s$ first reaches $t$.  On an undirected graph, the effective conductance between $a$ and $b$ is $d_a$ times the probability that a random walk from $a$ reaches $b$ before returning to $a$, and the effective resistance between $s$ and $t$ is the inverse of this.  These scores have applications in the analysis of networks.  It would be interesting to apply bidirectional techniques to estimating them and compare the running time of the resulting algorithm with the best previous algorithms.

\section{Open Problems in Personalized Search}
\subsection{Decreasing Storage for Text Search}
For text search, for each target $t$, the algorithm in Section \ref{sec:sampler} requires the associated residual $r^t$ to be included in a data-structure for every word $t$ is related to. For example, there is a data structure incorporating $\{r^t: t \text{ is relevant to ``computer''}\}$ and one incorporating $\{r^t: t \text{ is relevant to ``science''}\}$, so a target relevant to both ``computer'' and ``science'' will have its entire residual vector included in both.  Is there some way to not include $r^t$ in the data structure associated with every keyword for which $t$ is relevant.

\subsection{Handling Multi-Word Queries}
In Chapter \ref{sec:search_chapter}, we assume that a single keyword is given, for example ``Adam'' or ``computer'' but not both.  If two keywords are given, our algorithm is useful in some cases.  In particular, if one keyword is likely conditioned on the other, for example if 10\% of all people named ``Adam'' are relevant to ``computer,'' we can use rejection sampling.  In this example, we could find the top 30 people named Adam, reject those not relevant to ``computer,'' and we would have the top 3 people matching ``Adam'' and ``computer.''  However, there is some efficiency loss in this algorithm, and it requires the keywords to be correlated, so an interesting open problem is finding an algorithm for PPR search which supports multiple keywords or more general queries.

\subsection{User Study of Personalized Search Algorithms for Social Networks}
While there has been past empirical work measuring the improvement in relevance for topic-based Personalized PageRank \cite{haveliwala2002topic}, history based personalized search \cite{dou2007large}, and social network search based on shortest path \cite{vieira2007}, we are not aware of any published evaluation of personalized PageRank for ranking name or text search queries on a social network.  This experiment for example could be done on Twitter, Facebook, or Google+.  Participants could choose common names to search for (or be given common names to search for), shown results using different algorithms (either interleaved or side-by-side), and note which results are most relevant.  Or a live A/B test could be done on a social network site, and implicit feedback based on clicks could be used (see the chapter on ``Evaluation in information retrieval'' in \cite{manning2008introduction} for more background).  Natural ranking strategies to compare could include global PageRank (or in-degree), shortest path distance, and personalized PageRank.

%\subsection{(optional) Using PPR to Measure Publication Impact}
%  LocalWords:  impactful

%\appendix
%\include{appendix1}
%\include{appendix2}
%\include{appendix3}

\bibliographystyle{abbrv}
\bibliography{PPR-refs,PPR-Search-refs,PPR-generalized-refs}

%\onlinesignature

\end{document}